\documentclass[prd,twocolumn,superscriptaddress,showpacs,aps,amsmath,amssymb,nofootinbib]{revtex4-2}

\usepackage{natbib}
\usepackage{color}
\usepackage{ragged2e}
\usepackage{graphicx}
\usepackage{amsmath}
\usepackage{amssymb}
\usepackage{verbatim}
\usepackage{float}
\usepackage{epsfig}
\usepackage{psfrag}
\usepackage{dsfont}
\usepackage{amsfonts}
\usepackage{mathrsfs}
\usepackage{multirow}
\usepackage{bm}
\usepackage{hyperref}
\usepackage{xspace}
\usepackage{tikz}
\usepackage{afterpage}
\usetikzlibrary{matrix}
\usepackage[acronym]{glossaries} 
\usepackage{IEEEtrantools}

\usepackage[normalem]{ulem}
\usepackage{mwe}
\usepackage{aas_macros}
\hypersetup{
  colorlinks=true,        % false: boxed links; true: colored links
  linkcolor=blue,         % color of internal links                     
  citecolor=cyan          % color of links to bibliography              
}
\usepackage{pifont}% http://ctan.org/pkg/pifont     

\graphicspath{{./plots/}}                         

\usepackage{ragged2e}
\allowdisplaybreaks                                                   
\newcommand{\eg}{e.g.,~}
\newcommand{\changes}[1]{{\textcolor{red}{#1}}}
\begin{document}

\title{\boldmath On the nature of oscillating modes of proto-neutron stars\unboldmath}

\author{Dimitra Tseneklidou$^1$, Alejandro Torres-Forné$^{1,2}$, Pablo Cerdá-Durán$^{1,2}$, Martin Obergaulinger$^{1,2}$ and Jos\'e A. Font}

\affiliation{Departamento de Astronom\'ia y Astrof\'isica, Universitat de Val\`encia, Av. Vicent Andrés Estellés 19, 46100, Burjassot (Valencia), Spain}
\affiliation{Observatori Astron\`omic, Universitat de Val\`encia, Catedr\'atico Jos\'e Beltr\'an 2, 46980, Paterna, Spain}

\begin{abstract}
Newborn proto-neutron stars (PNS) are potential targets of gravitational wave asteroseismology, the study of the (inner) structure of stars via their oscillation modes. To prepare for the eventual detection of such modes by current and future gravitational wave observatories, theoretical studies have obtained the possible spectrum of oscillations. However, there has been disagreement when it comes to identifying and classifying specific modes according to the physical mechanism that excites them. In this paper, we present a novel scheme to classify the oscillation modes of a newly born PNS surrounded by a stalled accretion shock in core-collapse supernovae (CCSNe). Our classification is physically motivated, as it is based on the energy of the restoring forces of the mode. We investigate the nature of the modes by considering the different regions of the CCSN that they stem from. We apply this scheme to a set of 28 non-rotating 1D, 2D and 3D CCSN simulations performed with two different numerical codes and using different progenitors and equations of state. In that way, we find that there are different families of f-, p- and g-modes in the system, living in different areas. More specifically, we identify two families of f- and p-modes associated with the PNS and the shock, respectively, and two families of g-modes originating from the two convectively stable regions, one in the PNS core and one near its surface. We also find that the dominant high-frequency emission mode is the f-mode of the PNS, associated with the strong density gradient at its surface. Our classification procedure is automatic, performs consistently for all the models considered, and paves the way to systematic studies of the dependence of the mode frequencies on the PNS properties, with direct application to parameter estimation from future observations of gravitational wave signals.

\end{abstract}

\maketitle
%%%%%%%%%%%%%%%%%%%%%%%%%%%%%%%%%%%%%%%
\section{Introduction}
%%%%%%%%%%%%%%%%%%%%%%%%%%%%%%%%%%%%%%%
Core-collapse supernovae (CCSNe) are among the most violent and energetic events in the Universe. 
Information about these events is expected to be collected through multiple messengers, including electromagnetic radiation, 
neutrinos, and gravitational waves (GWs). As such, they represent some of the most promising GW sources for 
the current and next generation of interferometers \cite{Szczepanczyk2022,Srivastava2019,Kotake:2017,
Gossan:2016, Mezzacappa2025}. Understanding 
the physics that drives the explosion, and decoding the information encoded within these signals, is 
therefore essential.

At the end of their lives, massive stars, typically those with masses between $8$ and $100~\text{M}_\odot$, 
develop an onion-like structure, where different elements burn in concentric shells surrounding an iron core. 
When the core mass surpasses the Chandrasekhar limit, electron degeneracy pressure can no longer counteract 
gravity, triggering a rapid core collapse, amplified by electron capture and photodisintegration. During collapse, 
the innermost region reaches nuclear saturation density, at which point the nuclear equation of state (EoS) 
abruptly stiffens. The resulting pressure increase causes the innermost region to rebound at core bounce, forming 
a hot, compact proto-neutron star (PNS) and launching a powerful hydrodynamic shock wave outwards. As the shock 
propagates through the rest of the infalling iron core, it loses energy to the dissociation of heavy nuclei and a 
burst of neutrino emission, stalling within a few tens of milliseconds into a standing accretion shock at a 
typical distance of $\sim 100$ km, behind which matter moves subsonically towards the center 
\cite[see e.g.][]{Janka:2006fh, Janka2012,Janka2016, 2020_Mueller_review,2021Nature_CCSN}.

In non-rotating progenitors, which form the primary focus of this paper, the post-shock region may develop prompt convection shortly after bounce \cite{Burrows1993, Bruenn1994}, driven by the unstable negative 
entropy and lepton gradients left behind by the shock. This early overturn, when present, is followed by 
neutrino-driven convection and, in some cases, the standing accretion shock instability (SASI) around the stalled 
shock \citep{Blondin2003,Foglizzo:2006fu}. Two processes may 
excite oscillation modes of the newborn PNS, setting it into the non-spherical, time-dependent motion responsible for GW emission: on the one hand, the turbulent accretion downflows produced by post-shock instabilities 
striking the PNS surface \citep{Murphy:2009};
on the other hand, sustained Ledoux convection developing deep inside the PNS \cite{Andresen:2017,Murphy2025}. 
Whether the stalled shock 
is revived depends heavily on the explosion mechanism: in the neutrino-driven scenario, neutrinos emitted 
by the hot PNS deposit enough energy to relaunch it, while in rapidly rotating, strongly magnetized cores, 
the magneto-rotational mechanism can drive the explosion instead
\cite{2021Nature_CCSN,Obergaulinger:2017qno,Mosta2014,
Bugli2021,Powell2023, Yamada2024}. Magnetic-field amplification may also
contribute to neutrino-driven explosions in more slowly rotating
progenitors \cite{Muller2020a}.
If the shock is not revived, accretion onto the PNS continues until it collapses into a black hole (BH), 
producing a failed SN \cite{2017handbook_SN.1095Janka, 2020_Mueller_review, Cerda-Duran:2013}. Because the 
explosion is intrinsically multidimensional, it is widely studied through 2D and 3D numerical 
simulations \cite{Cerda-Duran:2013,Mueller:2013,Melson2015,Kuroda:2016,
Andresen:2017,OConnor2018,Powell:2018,Radice:2018,
Burrows2020,Bollig2021,Mezzacappa2023}.
These simulations employ different approximations for gravity and
neutrino transport \cite{Janka2012,Janka2016,2020_Mueller_review,Mezzacappa2020}.
Fully general-relativistic multidimensional calculations remain
comparatively rare, and long-term 3D models that follow the explosion
for the better part of a second or longer do not combine full general
relativity with multi-angle neutrino transport
\cite{2020_Mueller_review,Mezzacappa2025,Janka2025}.
These highly sophisticated 
computational models are crucial to capture the complex, non-linear coupling between general-relativistic 
hydrodynamics, multi-angle neutrino transport, and multi-dimensional hydrodynamic instabilities that 
govern the post-bounce evolution \cite{Janka:2006fh, Foglizzo:2006fu}.

The GW emission carries crucial information about the explosion dynamics, the stellar progenitor, 
and the properties of hot nuclear-density matter. However, extracting it requires connecting specific features 
of an otherwise stochastic waveform to the physics of the source. The strongest part of the signal typically lasts 
$200\text{--}500\text{ ms}$, until a successful explosion sets in \cite{Mueller:2013}, or extends 
to $1\text{ s}$ or longer when a BH forms \cite{Cerda-Duran:2013}. Weaker GW emission may persist for several seconds, as long as the PNS survives \cite{Choi2024}. In the low-frequency band ($\sim 50\text{--}250\text{ Hz}$), a SASI-related GW signal may appear as a prominent stochastic feature tied to the large-scale 
sloshing of the stalled shock \cite{Blondin2003,Foglizzo:2006fu}. The high-frequency emission, by contrast, 
is dominated by a monotonically rising track in the time-frequency plane, referred to in the literature as the dominant emission mode \cite{Cerda-Duran2025} or high-frequency feature (HFF) \cite{Casallas-Lagos2023, CasallasLagos2026}. The dominant feature starts near $\sim 100\text{ Hz}$ and rises to $1\text{--}2\text{ kHz}$, driven by the steady 
contraction and accretion of the newborn compact object, which shifts its characteristic frequencies upward over time. 
In rapidly rotating cores, the spectrum is further enriched by quasi-radial modes, whose frequency decreases 
towards zero as a BH forms \cite{Cerda-Duran:2013}.

Connecting these spectroscopic features in the GW emission to the underlying stellar oscillations is the primary aim of asteroseismology \citep{Mezzacappa2025}, which relates the normal modes of the star to its 
interior structure and thus offers a direct way to probe the newborn remnant. Earlier works in neutron-star (NS) asteroseismology relied on idealized setups, \eg treating linear perturbations of non-rotating stars in hydrostatic equilibrium  \cite{McDermott1983,Reisenegger1992,Ferrari2003,Passamonti2005,Dimmelmeier2006,Kruger:2014pva,Camelio:2017nka}. 
Previous applications to PNSs mainly targeted the late cooling phase between the CCSN explosion and the formation of the solid crust, during which the dominant process is the cooling of the PNS by neutrino emission. In this phase, the spectrum hosts acoustic pressure-supported p-modes, buoyancy-supported g-modes, and the fundamental f-mode.
Crucially, each class of modes is physically defined by the dominant restoring force 
acting on fluid elements within distinct layers of the star.

Although historically the main focus had been on the late phases of a CCSN, this changed when asteroseismology calculations
were extended to the early post-bounce evolution preceding the explosion \cite{Sotani:2016uwn}. This 
highly dynamic regime requires evaluating perturbations on realistic backgrounds extracted directly from 
multidimensional numerical simulations. Initial efforts to model these complex environments utilized the 
relativistic Cowling approximation, where spacetime perturbations are completely neglected 
\cite{Torres_et_al_I,Sotani2020}. This framework was progressively refined by gradually restoring the 
general-relativistic (GR) degrees of freedom,  first incorporating perturbations of 
the lapse function \cite{Morozova:2018}, and subsequently accounting for conformal factor 
perturbations \cite{Torres_et_al_II}. These additions captured the dominant GR metric effects and allowed the PNS and the post-shock region to 
be modeled as a single, hydrodynamically coupled system rather than treating the core as an isolated 
object in vacuum. 

More recently, this perturbative line of 
work has been extended by incorporating a steady, 
subsonic accretion flow alongside a consistent treatment of the boundary conditions (BCs) at the stalled 
shock location \cite{Tseneklidou:2025}. The present study, however, does not include the background accretion flow. While the exact treatment of the outer boundary conditions 
exerts minimal influence on deeply buried internal modes, like core g-modes \cite{Sotani2019}, it represents 
a critical prerequisite for accurately capturing the p-modes and shock-centered instabilities that 
directly interact with the accretion layer \cite{Torres_et_al_I,Torres_et_al_II}.

The oscillation modes are typically computed by solving an eigenvalue problem numerically using a shooting method, such as that implemented in
the publicly available linear perturbation code 
\texttt{GREAT}\footnote{Available at \url{https://www.uv.es/cerdupa/codes/GREAT/}} (General Relativistic 
Eigenmode Analysis Tool) \cite{Torres_et_al_I,Torres_et_al_II}, which we use in this work. Alternatively, there have been efforts to solve the eigenvalue problem using a spectral collocation framework \cite{Tseneklidou:2025} or physics-informed neural networks \cite{Tseneklidou2025b}.

These modern frameworks have firmly established that the tracks seen in GW spectrograms correlate directly 
with specific f-, p- and g-modes of the PNS \cite{Torres_et_al_I,Morozova:2018,Torres_et_al_II,Sotani2019,Westernacher-Schneider2019,Westernacher-Schneider2020,Sotani2020}. 
This correspondence has led to the formulation of EoS-independent universal relations between observable GW mode 
frequencies and PNS properties, expressed as combinations of the system's mass and radius \cite{Sotani2017,Torres_et_al_letter,Sotani2019}. Those relations may 
allow the inference of PNS parameters in the case of a nearby galactic CCSN \cite{Bizouard2021,Bruel2023}. 
The core g-modes, in turn, have been linked to the EoS itself \cite{Jakobus2023,Wolfe2023,Jakobus2024}, 
opening the possibility of constraining its parameters. These universal relations should be studied systematically in order to assess all possible uncertainties affecting the inference process.
Recent work has also highlighted the need for caution when
using universal relations, since the inferred PNS properties can depend
on the adopted relation, mode identification, and modeling assumptions
\cite{Murphy2026}.
Unfortunately, the available universal relations are based on limited sets of simulations (the largest set being $25$ simulations \cite{Torres_et_al_letter}), due to the difficulty of automatically processing large simulation ensembles (see discussion below).

Other approaches extract source information directly from CCSN GW signals without identifying the underlying oscillation modes. For rapidly rotating progenitors, waveform catalogues, Bayesian inference and machine-learning methods have been used to constrain rotation and the EoS \cite{Richers2017,Rover2009,Pastor-Marcos2024,Nunes2024,Abylkairov2025,Akhmetali2026}.

Despite being the prime target for GW detection, the exact physical nature of the dominant emission mode remains heavily debated. 
Some studies have attributed the origin of its emission to
buoyancy-supported g-modes excited in the convectively stable layer around the surface of the PNS \cite{Mueller:2013,Cerda-Duran:2013,Torres_et_al_I,Torres_et_al_II,Torres_et_al_letter}, whereas other authors instead identify it with the global fundamental 
(f) mode \cite{Morozova:2018,Sotani2019,Sotani2020} or some combination of both \cite{Vartanyan2023,Murphy2025}. 
Further insight into this problem has been obtained by studying the
spatial origin of the GW emission in 3D simulations. Andresen et
al.~\cite{Andresen:2017} showed that high-frequency PNS motions are
predominantly excited by PNS convection in their 3D models, whereas
Murphy et al.~\cite{Murphy2025} combined a decomposition of the GW
emission into different regions with a modal analysis. They found that
the dominant emission (HFF) follows successively the modes labeled as $g_2$, $g_1$, and
$f$ according to the Generalized Cowling Nomenclature, with the
transitions taking place close to avoided crossings. These results
established an important contribution of the $f$-mode to the HFF, but
also illustrate that the modal identification depends on the adopted
classification scheme.
This ambiguity is severely compounded by the fact that the computed
eigenmodes do not possess, a priori, any intrinsic
information regarding their physical character. Consequently, post-processing classification schemes are needed to extract any information. 
This task is traditionally handled by counting radial nodes; however, 
in the highly dynamic, non-uniform stratification of a young PNS, avoided crossings cause severe mixing of 
the eigenfunctions, rendering mode tracking based on geometric node counting highly unreliable (see discussion in~\cite{Torres_et_al_II}).  
Alternative schemes have attempted to map modes through these 
avoided crossings, e.g., using eigenvalue matching classification procedures \cite{Torres_et_al_II}, or the Classification Based on Modal Properties (CBMP) procedure \cite{Rodriguez2023}.
However, a systematic determination of the physical character of
each mode based directly on its underlying restoring forces is still
lacking. Such a physically grounded classification scheme would enable the automated analysis of large sets of simulations (beyond what is currently possible) and systematically relate mode frequencies to PNS properties.
This remains an essential step towards reliably inferring PNS structural 
properties from the observed GW signals.

In this paper, we investigate the physical nature of the normal oscillation modes of the PNS-shock system by shifting 
the diagnostic focus from geometric node structures to its local thermodynamics. We develop a formalism to compute the contribution of the different restoring forces (compression, buoyancy, gravity) to the mode energy (Section \ref{sec:linear_perts}), and apply it to a set of numerical simulations of CCSNe (Section \ref{sec:CCSNeSimulations}), using the eigenmodes computed with \texttt{GREAT}. By systematically studying the behavior of the mode frequencies, the different energy contributions, the impact of boundary conditions and the spatial structure of the eigenfunctions, we build up evidence supporting the existence of several classes of f-, p- and g-modes coexisting in the system (Section \ref{sec:ModeClassificationFramework}). Finally, we present a new automated procedure, capable of classifying the modes in their different families (Section~\ref{sec:ModeClass}), paving the way for a systematic analysis of large simulation ensembles and for future robust asteroseismic inference of PNS properties (Section~\ref{sec:conclusions}). Given that this study addresses fundamental questions about the nature of PNS oscillation modes, which will set the basis for our future work, we have tried to be comprehensive in the details, derivations and bibliography presented. Additional details of the thorough derivations and numerical procedures are provided in the appendices.

Throughout this paper, we use a space-like metric signature $(-, +, +, +)$ and, if not explicitly mentioned otherwise, geometrized units with 
$c = G = 1$, where $c$ stands for the speed of light and $G$ is Newton's gravitational constant. 
Greek indices run from $0$ to $3$, Latin indices run from $1$ to $3$, and we adopt Einstein's 
summation convention for repeated indices. 

%%%%%%%%%%%%%%%%%%%%%%%%%%%%%%%%%%%%%%%
\section{Linear perturbations of a spherically symmetric background}
\label{sec:linear_perts}
%%%%%%%%%%%%%%%%%%%%%%%%%%%%%%%%%%%%%%%

We consider three-dimensional linear adiabatic perturbations of a spherically symmetric, self-gravitating hydrostatic equilibrium configuration (background hereafter, for which $\partial_t =0$ and velocity is zero).
We follow the procedure presented in \cite{Torres_et_al_II} to obtain the eigenmodes of the system formed by the PNS and the stalled accretion shock. Here, we outline the procedure and introduce our methodology for calculating the energy of the modes. 

%%%%%%%%%%%%%%%%%%%%%%%%%%%%%%%%%%%%%%%
\subsection{Metric}\label{sec:Metric}
%%%%%%%%%%%%%%%%%%%%%%%%%%%%%%%%%%%%%%%

The metric can be described in terms of the 3+1 decomposition as
\begin{equation}\label{eq:metric}
    ds^{2} 
    = g_{\mu\nu}dx^{\mu}dx^{\nu} 
    = -\alpha^{2}dt^{2} +  \gamma_{ij}(dx^{i} + \beta^i dt)(dx^{j}+\beta^j dt), 
\end{equation}
where $\beta^i$ is the shift 3-vector, $\alpha$ is the lapse function, and $\gamma_{ij}$ is the spatial 3-metric.
We perform a conformal decomposition of the latter, $\gamma_{ij} = \psi^4 (f_{ij} + h_{ij})$, with $\psi$ being the conformal factor, $f_{ij}$ the flat spatial 3-metric and $h_{ij}$ the deviation of the 3-metric from conformal flatness. The conformal factor is defined such that the determinant of the 3-metric, $\gamma$, satisfies $\sqrt{\gamma} = \psi^6 \sqrt{f}$, with $f$ the determinant of the flat 3-metric, in spherical coordinates, $\sqrt{f} = r^2 \sin\theta$. We denote the covariant derivative with respect to $\gamma_{ij}$ and $f_{ij}$ as $\nabla_i$ and $\bar \nabla_i$, respectively. 

This metric allows us to study the perturbations of any spherically symmetric background metric in isotropic coordinates, for which $\beta^i=0$ and $h_{ij}=0$. Instead of analyzing the general case for the perturbations of the metric, we assume the following two approximations: i) we follow the CFC approximation \citep{Isenberg:2007zg,Wilson1996,CC:2009}, which assumes that the 3-metric is conformally flat, i.e. $h_{ij}=0$ (for the perturbed metric) in a quasi-isotropic gauge with maximal slicing. This approximation has shown excellent results in the non-linear dynamics of compact objects such as neutron stars \citep{Shibata2004,Ott2007a,Ott2007b}, with deviations from GR, in the context of CCSNe, smaller than $1\%$~\citep{Cerda-Duran2005, Dimmelmeier:2002II, Dimmelmeier:2002I}. This approximation removes the propagating GW degrees of freedom from the spacetime dynamics. As a consequence, purely space-time features, such as w-modes \cite{Kokkotas1992}, disappear from the resulting solutions. Additionally, the approximation removes a channel of energy losses. For the system considered here, it means that the oscillatory solutions will be purely adiabatic and undamped, constituting a set of normal oscillation modes of the system. ii) We disregard perturbations of the shift, i.e., $\beta^i=0$, for the perturbed metric. For weak gravitational fields (with compactness $M/R \ll 1$), the shift behaves as a first-order post-Newtonian correction \citep{Blanchet1990}, so its removal is expected to produce errors of the order of $M/R$ ($\sim 10\%$ error for PNSs). 
The results in \cite{Torres_et_al_II} showed that the oscillation modes computed with these two approximations accurately matched the GW frequencies of multidimensional CCSNe simulations. Furthermore, including 1PN corrections introduced differences under 10\%, confirming that these approximations are appropriate for this scenario.

%%%%%%%%%%%%%%%%%%%%%%%%%%%%%%%%%%%%%%%
\subsection{Fluid}\label{sec:Fluid}
%%%%%%%%%%%%%%%%%%%%%%%%%%%%%%%%%%%%%%%

We consider a perfect fluid, whose energy-momentum tensor is described by ${T}^{\mu\nu} = \rho h {u}^{\mu}{u}^{\nu} + p {g}^{\mu\nu}$,
where $\rho$ stands for the rest-mass density, $p$ for the pressure, ${u}^{\mu}$ is the $4-$velocity, $h \equiv 1 + \epsilon +p/\rho$ is the specific enthalpy and $\epsilon$ the specific internal energy. The energy density can be described as $e\equiv \rho(1 + \epsilon)$. On the hypersurface, we define two velocities: the Eulerian velocity $\upsilon^i = u^i/W$ and the advective (coordinate) velocity $\upsilon^{*i}=\alpha\upsilon^i$.

The condition of non-rotating hydrostatic equilibrium implies that the background satisfies  $v^{i}=0$ and $\partial_r p / (\rho h) = \mathcal{G}$, where $\mathcal{G}\equiv- \partial_r \ln{\alpha}$ is the gravitational acceleration. At this point, let us introduce two characteristic frequencies of the background: the square of the relativistic Brunt-Väisälä (BV) frequency,
\begin{equation}\label{eq:BVfreq}
    \mathcal{N}^{2} \equiv \frac{\alpha^{2}}{\psi^{4}} \mathcal{G} \mathcal{B}_r
\end{equation}
and the relativistic Lamb frequency $\mathcal{L}$,
\begin{equation}\label{eq:Lamb_freq}
    \mathcal{L}^{2} \equiv \frac{\alpha^{2}}{\psi^{4}} c_{s}^{2} \frac{l(l+1)}{r^{2}},
\end{equation}
where $c_s$ is the sound speed, $\mathcal{B}_i\equiv \frac{\partial_i e}{\rho h} - \frac{\partial_i p}{p\Gamma_1}$ represents the relativistic version of the Ledoux discriminant, and $l$ is the angular degree of the perturbation (see the following paragraph). For a detailed description of the aforementioned quantities, the interested reader is referred to \cite{Torres_et_al_I, Torres_et_al_II, Tseneklidou:2025} and references therein.

The Eulerian perturbation of any scalar quantity (including the metric functions $\alpha$ and $\psi$) is of the form $q \rightarrow q + \delta q$, where $q$ refers to the background quantity. $\delta q$ is a linear combination of the normal mode solutions, which can be decomposed into spherical harmonics and a harmonic time dependence, $\delta q_{lm\pm} = \delta\hat{q}(r)Y_{lm}(\theta, \varphi){\rm e}^{\pm i\sigma t}$,
where $l$ is the degree of the spherical harmonic function, $m$ is the azimuthal number, while the $\pm$ sign corresponds to the one of the frequency $\sigma$.

Similarly, any vector quantity, e.g. the Lagrangian displacement $\vec \xi$, can be expressed as a linear combination of normal mode solutions, $\vec \xi_{lm\pm}$,
whose components in the coordinate basis are given by\footnote{Note that the radial functions $\eta_1$ and $\eta_2$ are the same used in \cite{Tseneklidou:2025} and correspond to those used in \cite{Torres_et_al_I,Torres_et_al_II} as, $\eta_r=\eta_1$ and $\eta_\perp=r \, \eta_2$. Also note that \cite{Torres_et_al_I,Torres_et_al_II} use the coordinate basis to express the components of this vector.},

\begin{align}
    \xi^{r}_{lm\pm} &= \eta_{_{1}} (r)\, Y_{lm} (\theta,\varphi)\,  e^{\pm i\sigma t}, \label{eq:xi_r} \\
    \xi^{\theta}_{lm\pm} &= \eta_{_{2}}(r)\,  \frac{1}{r} \partial_{\theta} Y_{lm}(\theta,\varphi)\, e^{\pm i\sigma t}, \label{eq:xi_theta} \\
    \xi^{\varphi}_{lm\pm} &= \eta_{_{2}} (r)\,\frac{1}{r \sin^{2} \theta} \partial_{\varphi} Y_{lm} (\theta,\varphi)\, e^{\pm i\sigma t}. \label{eq:xi_phi}
\end{align}

The Lagrangian displacement $\xi^{i}$ is related to the advective velocity perturbation $\delta v^{*i}$ by $\partial_{t}\xi^{i}=\delta v^{*i}$. 
Lagrangian perturbations of any scalar quantity $q$ are related to the Eulerian perturbations as $\Delta q = \delta q +\xi^r\partial_r q$. 

The assumption that the perturbations are adiabatic implies that
\begin{equation}\label{eq:adiabaticity}
    \frac{\Delta p}{\Delta\rho} = hc_s^2.
\end{equation}

All the definitions and approximations for the metric and fluid variables above were used by \cite{Torres_et_al_II} to perturb the general relativistic hydrodynamics equations to the linear order. The resulting equations, together with an appropriate set of boundary conditions, form a one-dimensional (radial) eigenvalue problem (Eqs. (33)-(34) and (37)-(38) in \cite{Torres_et_al_II}), whose solutions are the eigenfrequencies, $\sigma$, and eigenfunctions $(\eta_{_{1}}, \eta_{_{2}}, \delta \hat \alpha, \delta \hat \psi)$. For spherically symmetric backgrounds: i) modes with different $l$ and $m$ decouple, ii) the eigenfrequencies and the radial part of the eigenfunctions depend only on $l$. For simplicity, we avoid an explicit labeling of these quantities with $l$, since we work with fixed $l$ (typically $l=2$) for the rest of the work. iii) $\sigma$ is purely real or imaginary, and if $\sigma$ is an eigenfrequency, $-\sigma$ is as well, and with the same eigenfunction. For a detailed derivation of the eigenvalue problem and its solutions, the interested reader is referred to \cite{Torres_et_al_II}.

%%%%%%%%%%%%%%%%%%%%%%%%%%%%%%%%%%%%%%%%%%%%%%%%%%%%%%%%%%
\subsection{Standing wave solutions}
\label{subsec:StandingWave}
%%%%%%%%%%%%%%%%%%%%%%%%%%%%%%%%%%%%%%%%%%%%%%%%%%%%%%%%%%

In general, the solutions of the eigenvalue problem (e.g. $\vec \xi_{lm\pm}$) are complex functions.  However, the physical solutions for the normal oscillation modes (e.g. $\vec \xi$) should be real functions, constructed as linear combinations of the former, that correspond to standing wave solutions. Using these physical solutions is particularly important for computing the energy of the perturbations.

Instead of using the decomposition of $\vec \xi_{lm\pm}$ given by Eqs.~\eqref{eq:xi_r}-\eqref{eq:xi_phi}, which is used in the numerical calculation of the eigenmodes, for the calculation of the standing wave solutions, it is convenient to use a description in terms of vector spherical harmonics.
According to \cite{Thorne:1980}, any vector can be expanded into ``pure-spin" vector harmonics. In the case of the Lagrangian displacement, it can be written as
\begin{equation}\label{eq:xi_vector_harmonics}
    \vec{\xi}_{lm\pm} = \bigg[ \eta_{_{R}} \,\vec{Y}^{R,lm} + \eta_{_{E}} \,\vec{Y}^{E,lm} + \eta_{_{B}} \,\vec{Y}^{B,lm}
    \bigg] {\rm e}^{\pm i \sigma t},
\end{equation}
where the $\eta$-variables are radial functions, $\vec{Y}^{R,lm}$ and $\vec{Y}^{E,lm}$ have electric-type parity (polar), while $\vec{Y}^{B,lm}$ has ``magnetic-type parity" (axial). $\vec{Y}^{R,lm}$ is purely radial, while the other two are purely transverse (angular).
Since our background is a spherically symmetric ideal fluid in hydrostatic equilibrium, this implies that the axial perturbations will vanish (i.e., $\eta_{{}_{B}} = 0$). 
By comparison of the two relations for the displacement vector, Eqs.~\eqref{eq:xi_r}-\eqref{eq:xi_phi} and \eqref{eq:xi_vector_harmonics}, we get that $\eta_R = \eta_1$ and $\eta_E = \sqrt{l\left( l+1\right)}\, \eta_2$. The derivation of these expressions is shown in Appendix \ref{app:LagrangianDispl}.
These vector spherical harmonics are orthogonal,
\begin{equation}\label{eq:Ylm_orthonormal}
    \int \vec{Y}^{J,lm}\vec{Y}^{J',l'm'*}d\Omega = \delta_{JJ'}\delta_{ll'}\delta_{mm'},
\end{equation}
where $d\Omega= \sin{\theta}d\theta d\phi$, $\delta_{ij}$ is the Kronecker delta, $J=R, E, B$ and their complex conjugates are given by $Y^{J,lm*} = (-1)^m Y^{J,l-m}$.

Although both $\pm \sigma$ are solutions of the eigenvalue problem, we are searching for those $\vec{\xi}_{lm\pm}$, that result in standing waves for the physical displacement. To do so, we can construct a linear combination of these complex solutions such that all perturbation variables are real (in particular, the displacement). A possible combination having the form of a real standing wave solution is the following
\begin{align}\label{eq:xi_combination}
    \vec{\xi} = \frac{1}{4}\Bigg[ \vec{\xi}_{lm\,+} + \vec{\xi}_{lm\,-} + (-1)^m\left( \vec{\xi}_{l-m\,+} + \vec{\xi}_{l-m\,-} \right)
    \Bigg].
\end{align}

If we expand the above expression in terms of the harmonic time dependence, use the properties of the spherical harmonics above, and define $\vec\xi_{lm}$ such that $\vec\xi_{lm\pm} = \vec\xi_{lm} e^{\pm i\sigma t}$, then we arrive at
\begin{equation}\label{eq:xi_real}
    \vec{\xi} = {\rm{Re}} \left( \vec{\xi}_{lm} \right) \cos{(\sigma t )} ,
\end{equation}
showing that the vector $\vec{\xi}$ is indeed purely real and corresponds to a standing wave. For a given $l$ and $m$, this linear combination is sufficiently general for the purpose of this work. Other choices or linear combinations can introduce a phase shift in the time dependence or rescale the amplitude. Neither affects our analysis, since they correspond to arbitrary choices of the oscillation phase and the normalization of the eigenfunction, respectively and the normalization of the eigenfunction. The same linear combination holds for all the first-order perturbations to be real, e.g., for a generic scalar quantity $q$, we get

\begin{eqnarray}
\label{eq:dq_real}
\delta q &=& \frac{1}{4} \left [
\delta q_{lm\, +} + \delta q_{lm\, -} + (-1)^m (\delta q_{l-m\, +} + \delta q_{l-m\, -})  
\right ] \nonumber \\
&=& {\rm Re} (\delta q_{lm}) \cos (\sigma t)
\end{eqnarray}
where $\delta q_{lm}$ is such that $\delta q_{lm\pm} = \delta q_{lm} e^{\pm i \sigma t}$.

%%%%%%%%%%%%%%%%%%%%%%%%%%%%%%%%%%%%%%%%%%%%%%%%%%%%%%%%%%
\subsection{Acceleration}
\label{subsec:acceleration}
%%%%%%%%%%%%%%%%%%%%%%%%%%%%%%%%%%%%%%%%%%%%%%%%%%%%%%%%%%
In this section, we present a new analytical procedure to calculate the energy of the modes. Our starting point is the computation of the acceleration $\vec{a} = \partial_{tt} \vec\xi$. Similarly to Eqs.~\eqref{eq:xi_r}–\eqref{eq:xi_phi}, this vector can be decomposed into two independent components, $a_i = -\sigma^2 \eta_i$ $(i=1,2)$, which can be derived from the momentum equations (30)–(31) of \cite{Torres_et_al_II}:

\begin{widetext}
\begin{IEEEeqnarray}{lllllrrlrrlcllclllll}
    \label{eq:mom1_Drho}
    -\sigma^2 \eta_{_{1}} 
    &= \alpha^2\psi^{-4}
    &\Bigg\{ 
    &-
    &\frac{1}{\rho h}
    &\bigg[ \partial_r - \mathcal{G}\left(1 + \frac{1}{c_s^2} \right)\bigg] 
    &\,\left(p\Gamma_1\frac{\Delta\hat{\rho}}{\rho}\right)
    &\,+
    &\frac{1}{\rho h}\bigg[ \partial_r - \mathcal{G}\left(1 + \frac{1}{c_s^2} \right)\Bigg]
    &\,\left( \eta_{_{1}}  \partial_r p \right) 
    &\,-\partial_r \,
    &\left( \frac{\delta\hat{\alpha}}{\alpha} \right) 
    &\Bigg\}
    &- \mathcal{N}^2 
    &\,\eta_{_{1}}&, 
    \\
    \label{eq:mom2_Drho}
    -\sigma^2 \eta_{_{2}} 
    &= \frac{\alpha^2\psi^{-4}}{r} 
    & \Bigg \{ 
    &-
    &
    &\frac{1}{\rho h} 
    &\,\left(p\Gamma_1\frac{\Delta\hat{\rho}}{\rho}\right)
    &\,+  \,
    &\frac{1}{\rho h}
    & \left ( \eta_{_{1}} \partial_r p \right )  
    &\,-
    &\frac{\delta\hat{\alpha}}{\alpha} &\Bigg\}&&&.
\end{IEEEeqnarray}
\vspace{-1cm}
\[
\hspace{0.21\textwidth}
\underbrace{\hspace{0.29\textwidth}}_{\text{compression (p)}}
\underbrace{\hspace{0.26\textwidth}}_{\text{free surface (f)}}
\underbrace{\hspace{0.08\textwidth}}_{\text{ \shortstack{gravitational  \\ potential ($\alpha$)}}}
\hspace{0.02\textwidth}
\underbrace{\hspace{0.07\textwidth}}_{\text{buoyancy (g)}}
\hspace{0.07\textwidth}
\]
\end{widetext}

The original equations of \cite{Torres_et_al_II} are written in terms of the Eulerian perturbation of the pressure, $\delta \hat p$. However, here it is possible to express them in terms of the Lagrangian perturbation of the
density, $\Delta \hat \rho$. 
Through the adiabaticity condition, 
Eq.~\eqref{eq:adiabaticity}, $\delta\hat{p}$ translates into 
\begin{equation}\label{eq:dP_as_Drho}
    \delta\hat{p} = hc_s^2\Delta\hat{\rho} - \eta_1\partial_r p .
\end{equation}
The advantage of expressing the momentum equations in this form is that the contribution of each restoring force is easier to identify. The terms scaling with the perturbation of density, $\Delta\hat{\rho}$, represent the action of compression, those containing the Brunt-Väisälä frequency, $\mathcal{N}^2$, are the result of buoyancy, while the action of the gravitational potential appears through the perturbations of the metric (here entering through the lapse function, $\delta\hat{\alpha}$). Lastly, the term scaling with $\eta_1 \partial_r p$ indicates the restoring force associated with a free surface.

Let us elaborate on our last statement by considering several limits:
\begin{itemize}
    \item {\it Incompressible fluid localized in a thin layer}:
    In the case of an incompressible fluid, $c_s^{-2} = 0$, adiabatic perturbations have constant density, $\Delta\hat{\rho}=0$ (see Eq.~\eqref{eq:adiabaticity}, i.e., a pressure perturbation will not induce a density change, and thus the fluid is incompressible). In that case, metric perturbations are negligible (i.e.~Cowling approximation, approximately true for a relativistic incompressible fluid), hence $\delta \hat \alpha=0$. 
    The condition of a localized thin layer means that buoyancy is only relevant in a region of size much smaller than the pressure scale height, $H_p = p/\partial_r p$. Under this condition it is possible to have buoyant regions where the pressure gradient term in Eq.~(\ref{eq:dP_as_Drho}), $\eta_1 \partial_r p$, is negligible compared to the buoyancy term, $\mathcal{N}^2 \eta_1$, which is the only one remaining. In this limit, only g-modes can exist, so we call the term $\mathcal{N}^2 \eta_1$ the {\it buoyant (g) contribution} to the acceleration. This limit was actually used in \cite{Torres_et_al_I} (g-mode limit in that work, without the assumption of vanishing pressure gradients) to estimate the frequency of g-modes. 
     
    \item {\it Non-stratified fluid with no pressure gradients in the Cowling approximation:}
    In this limit there is no buoyancy, ${\mathcal B} = 0$, and therefore $\mathcal{N}^2 = 0$. When combined with the Cowling approximation ($\delta \hat \alpha=0$), the only surviving terms in the acceleration are those with $\Delta \hat \rho$. In this limit, only p-modes are possible; therefore, we call those terms the {\it contribution of compression (p)} to the acceleration. This limit was used in \cite{Torres_et_al_I} to estimate the frequency of p-modes.)
    
    \item {\it Incompressible non-stratified fluid in the Cowling approximation:} in this case $\Delta \hat \rho = \mathcal{N}^2=\delta\hat \alpha =0$, so the only surviving term is the one with $\eta_1 \partial_r p$. 
    Under such conditions, the only possible mode is the fundamental (f-)mode, which represents a global surface deformation \cite{Cowling1941}. Interface modes are also possible if density discontinuities (or strong gradients) are present.
    Since this term is the minimum requirement to have an f-mode, we call it the {\it contribution of the free surface} to the acceleration or simply the {\it f-contribution} (for its relation to the f-mode).   
\end{itemize}

In the CFC approximation, the metric terms $\delta\hat{\alpha}$ do not generate any family of modes by themselves. It can be shown from Eqs.~(24) and (25) in \cite{Torres_et_al_II}, that if the other terms are zero (incompressible, no buoyancy, and no free surface) then we also have $\delta\hat\alpha=0$. However, this term contributes to modes with large density deformations, in particular to the f-mode (see discussion in Section~\ref{subsec:Class_p_g_modes}). In fact, the first derivation of the f-modes by Kelvin addressed the case of an incompressible fluid ball with no stratification and including perturbations of the gravitational potential \cite{Thompson1863}.

%%%%%%%%%%%%%%%%%%%%%%%%%%%%%%%%%%%%%%%%%%%%%%%%%%%%%%%%%%
\subsection{Energy of the modes}
\label{subsec:energyCalculation}
%%%%%%%%%%%%%%%%%%%%%%%%%%%%%%%%%%%%%%%%%%%%%%%%%%%%%%%%%%

Our next step is the computation of the mode energy as a perturbation of the ADM mass, which in the case of a vanishing extrinsic curvature (because we are considering $\beta^i=0$), a quasi-isotropic gauge, and an asymptotically flat space-time, takes the form
\begin{equation}\label{eq_ADM_mass}
    M = \int_{\Sigma_t} \psi^{-1} E \sqrt{\gamma} d^3x,
\end{equation}
where the relativistic energy density is
\begin{equation}\label{eq:total_E}
    E = \rho h W^2 -p.
\end{equation}
For further details on the derivation of the relation above, the interested reader is referred to the literature for General Relativity in the 3+1 formalism, e.g., \cite{Gourgoulhon:2012}. 

The energy of the mode can be computed by perturbing the ADM mass
\begin{equation}
    M = M_0 + \delta M,
\end{equation}
for which we need to perturb the relativistic energy density, $E=E_0 +\delta E$. Here we use the number subscript to indicate the order of the perturbation ($0$ for the background), whenever it is not clear.
According to the extremal mass-energy theorem (see section 3.5.3 in \cite{Thorne:1965lecture} and Theorem 3 in \cite{Harrison:1965}) Eq. \eqref{eq_ADM_mass} is an extremum with respect to adiabatic perturbations if the background configuration is in hydrostatic equilibrium. In other words, the background configuration is in hydrostatic equilibrium if and only if the first order perturbations of the energy vanish. 
This means that we have to keep quadratic terms in the expansion to compute the perturbation energy, $\delta M = M_2$ (and $\delta E = E_2$).

The relativistic energy density of the mode contains in general, terms involving the Lagrangian displacement, $\xi^i$, its time derivative, $\dot \xi^i$, and perturbations of scalar quantities of the fluid ($\delta\rho$, $\delta  p$...) and metric ($\delta\alpha$, $\delta\psi$...). Since we are looking for real standing wave solutions, Eqs.~\eqref{eq:xi_real} and \eqref{eq:dq_real}, the time dependence 
of all these quantities is as follows:
\begin{align}\label{eq:TimeDependence}
    \xi^i &\propto \cos \left(\sigma t\right), \nonumber \\
    \dot \xi^i &\propto \sin \left(\sigma t\right), \nonumber \\
    \delta\rho, \delta p, \dots &\propto \cos \left(\sigma t\right), \nonumber \\
    \delta\alpha, \delta\psi\dots &\propto \cos \left(\sigma t\right). 
\end{align}
Since the energy perturbation is quadratic in the perturbation equations, it contains {\it kinetic terms} that behave as 
$v^{*2}$, i.e. proportional to $\sin^2\left(\sigma t\right)$, and {\it potential terms}, proportional to $\cos^2\left(\sigma t\right)$.
Given that the system is adiabatic, no terms with other phases appear (e.g. terms with $\sin\left(\sigma t\right)\cos\left(\sigma t\right)$).

As a consequence, the total perturbation energy is the sum of the kinetic and the potential energy, which have the same amplitude but opposite phase:
\begin{equation}\label{eq:M2_E+P_ampl}
    \delta M = M_2 = \tilde{K}\sin^2{\left(\sigma t\right)} + \tilde{P}\cos^2{\left(\sigma t\right)} + \tilde{C},
\end{equation}
where $\tilde{K}$ and $\tilde{P}$ are the amplitudes of the kinetic and potential energies, respectively. Those two energies will have the same amplitude, $\tilde{K}=\tilde{P}$, but they will be pulsating out of phase. $\tilde C$ is a constant term that can generally appear and corresponds to a static deformation of the background induced by the perturbation, which is discussed later.
Since the total energy is constant, it will be related to the amplitude of the maximum kinetic (or potential) energy, $M_2=\tilde{K} + \tilde{C}=\tilde{P} + \tilde{C}$. For further discussion, the passionate reader is referred to, e.g. \cite{Thorne1967}. 

In Appendix \ref{app:CalcKineticEnergy}, we present the details of the calculation of the relativistic energy density perturbation, $E_2$, and the corresponding perturbation energy, $M_2$. Notice that, since both are equal, it is sufficient to compute the kinetic or the potential component. We choose to calculate the kinetic component, which is significantly simpler than the derivation of the potential one.
The result is remarkably simple (see Eq.~\eqref{eq:E2_dpsi2}): 
\begin{equation}\label{eq:E2_final_main_text}
    E_2 = \frac{1}{2}\alpha^{-2}\psi^4\left( \rho h \right)_0 \dot{\xi}^2  \; + \; 
    {\text{$\psi_2$ terms}} \; + \; \substack{\text{potential}\\\text{terms}}.
\end{equation}
The leading terms of the equation are qualitatively in agreement with the equivalent expression for the kinetic energy derived in \cite{Harrison:1965} and \cite{Thorne:1965lecture, Thorne1967} (where the calculations have been performed in a different gauge). 

If metric perturbations were excluded (Cowling approximation), only the first term (proportional to $\dot\xi^2$) would contribute to the kinetic energy. However, if metric perturbations are considered (see Appendix \ref{app:CalcKineticEnergy}), all the additional terms contribute to the potential energy except for part of the second order perturbation of the conformal factor, $\psi_2$, that contributes to the constant term, $\tilde C$. This term corresponds to a static non-spherical contribution to the background metric due to the presence of the perturbation.

The computation of the non-linear term $\psi_2$ is generally not trivial. It involves solving an elliptic equation for $\psi_2$, which significantly complicates the analysis.
For weak gravitational fields, $M/R \ll 1$, the conformal factor $\psi$ is a first post-Newtonian correction (order $M/R$, at most a $10\%$ contribution for proto-neutron stars).
This contribution is nominally of the same post-Newtonian order as the $\dot\xi^2$ terms, so it cannot be in general neglected. As we show in the next sections, for most of the modes considered in this work, the inclusion of the metric perturbations has a minor impact on their energy. Therefore, one could argue that for those modes the contribution of $\psi_2$ could be equally small. The only mode that would be affected significantly is the f-mode, and even in that case, this contribution is minimal or negligible for most of the data. Thus, for the sake of simplicity, we omit this term in our analysis.

The perturbed ADM mass, keeping only second-order terms, is
\begin{equation}\label{eq:dM_ADM}
M_{2} =\int_{\Sigma_t} E_2  \psi^{-1} \sqrt{\gamma}\,  d^3 x  
\; + \;
\text{$\psi_2$ terms} \; + \; \substack{\text{potential}\\\text{terms}},
\end{equation}
where the additional terms arise from the perturbation of the metric factor $\psi^{-1}\sqrt{\gamma}$ in the integral.

Substituting Eq. \eqref{eq:E2_final_main_text} into Eq. \eqref{eq:dM_ADM} and taking into account the definition of the Lagrangian displacement, Eq. \eqref{eq:xi_vector_harmonics}, the spherical harmonics properties, Eq. \eqref{eq:Ylm_orthonormal}, and that $M_2 \simeq \tilde K$, after some calculations, we arrive at the following expression for the total energy of the mode,
\begin{equation}\label{eq:M2_final}
    M_2 \simeq \frac{1}{2}\sigma^2 \int_{r} \psi^9 \alpha^{-2}  \rho h \left( \eta_R^2 +  \eta_E^2  \right) \;  r^2\; dr \,.
\end{equation}
We note that this is an approximate expression because we have neglected the terms coming from $\psi_2$ (both in $M_2$ and $E_2$), as explained above.

%%%%%%%%%%%%%%%%%%%%%%%
\subsection{Work associated with the perturbation }
\label{subsec:Work}
%%%%%%%%%%%%%%%%%%%%%%%

Up to this point, we have derived the expression for the total energy by means of the maximum kinetic energy. The terms appearing in Eq.~\eqref{eq:M2_final} are known, making it possible to compute the total energy of the mode. However, our task does not end here, as we wish to calculate the contribution to the energy of each of the restoring forces discussed in Section~\ref{subsec:acceleration}. 

The force density acting on a fluid element can be computed as $\vec f = \varrho\,  \vec a$, 
where $\varrho$ is an appropriate energy density that will be determined later. The acceleration itself, and hence $\vec f$, is a functional of $\vec \xi$. If $\vec f$ is a conservative force, then it is possible to compute the work density done to displace a fluid element by an amount $\vec \xi$ as
\begin{equation}
\label{eq:work1}
w = \int \vec f (\vec \xi) \cdot d\vec\xi = \frac{1}{2} \vec f (\vec \xi) \cdot \vec\xi.
\end{equation}
For the harmonic time dependence of the oscillation modes considered in this study, it can be proved that this equation indeed holds (see Appendix~\ref{app:ConservativeForce}).
The total work can then be computed by integrating over the system
\begin{equation} \label{eq:work}
    \mathcal{W} = \frac{1}{2}\int \vec f \cdot \vec\xi \, \sqrt{\gamma} \, d^3x = \frac{1}{2}\int \varrho \, \vec a \cdot \vec\xi \, \sqrt{\gamma} \, d^3x, 
\end{equation}
which is a quadratic functional of $\vec\xi$ (see Appendix~\ref{app:ConservativeForce}) and hence it can be written as 
\begin{equation}
    \mathcal{W} = \tilde{\mathcal{W}} \cos^2 \sigma t,
\end{equation}
where $\tilde{\mathcal{W}}$ is the work at maximum displacement.

As we discuss in Section~\ref{subsec:energyCalculation}, for conservative forces, the work at the maximum displacement should correspond to minus the maximum potential energy, $\tilde P = - \tilde{\mathcal W}$. But we already established in the same section that this is equal to the maximum of the kinetic energy and hence to the total energy of the mode, $ \tilde{\mathcal{W}} =-\tilde P =-\tilde K\simeq -M_2$. The only remaining question is to determine the appropriate energy density $\varrho$ relating the force density and the acceleration. This quantity can be obtained by comparing Eq.~\eqref{eq:work} with Eq.~\eqref{eq:dM_ADM}, and taking into account that $\xi^i = \xi^i_0 \cos\sigma t$, we have that  
\begin{align}
    \dot \xi^2 &= \sigma^2 \xi_0^2 \, \sin^2 {\left(\sigma t\right)}, \nonumber \\
    \vec a \cdot \vec \xi &= -\sigma^2 \xi_0^2 \cos^2 {\left(\sigma t\right)},
\end{align}
are equal except for the time dependence. 
As a result, the force density fulfilling the condition above should be
\begin{equation}
    \vec f= \alpha^{-2}\psi^{3} \rho h \, \vec a,
\end{equation}
i.e. $\varrho = \alpha^{-2}\psi^{3} \rho h $. Note that this only holds approximately since we neglected the kinetic terms coming from $\psi_2$.

In that manner, we can rewrite the  mode energy as the work done by the different forces,
\begin{equation}\label{eq:M2_fR_fE}
    M_2 \simeq -  \frac{1}{2}\int \vec{f}_0\cdot\vec{\xi}_0\, \sqrt{\gamma}\, d^3x = - \frac{1}{2}\int_r \left( f_{_R}\eta_{_R} + f_{_E}\eta_{_E} \right) r^2 dr,
\end{equation}
where $f_{_R}=-\psi^9 \alpha^{-2}  \rho h \sigma^2 \eta_{_R}$ is the effective radial force and $f_{_E}= -\psi^9 \alpha^{-2}  \rho h \sigma^2 \eta_{_E}$ the effective electric-type force. Notice, first, that Eq. \eqref{eq:M2_fR_fE} is generic, meaning that the energy can be expressed as the sum of the work produced by each of the forces acting in the system. And second, the effective forces $f_{_R}$ and $f_{_E}$ are proportional to the acceleration given by Eqs.~\eqref{eq:mom1_Drho}-\eqref{eq:mom2_Drho}. Combining those equations with Eq.~\eqref{eq:M2_fR_fE}, we arrive at the following relation
\begin{equation}\label{eq:M2_sum_forces}
    M_2 = - \frac{1}{2}\int \left( \vec{f}_p + \vec{f}_f + \vec{f}_{\alpha} +\vec{f}_g  \right)\cdot \vec{\xi} \; \sqrt{\gamma}\, d^3x,
\end{equation}
where the corresponding restoring forces are associated with compression ($ \vec{f}_p$), a free surface ($ \vec{f}_f$), the gravitational potential ($ \vec{f}_{\alpha}$), and buoyancy ($ \vec{f}_g$).
If we isolate the contribution of each of the forces to the total energy of the mode, we arrive at the following expressions

\begin{widetext}
\begin{align}
    \label{eq:Mp}
    M_p = &\frac{1}{2} \int_{r} \psi^5 r^2\bigg\{ \eta_1\bigg[ \partial_r - \mathcal{G}\left( 1 + \frac{1}{c_s^2}\right)\bigg]
    + \eta_2 \frac{l(l+1)}{r}
    \bigg\} \left( p\Gamma_1 \frac{\Delta \hat{\rho}}{\rho} \right) dr,
    \\ \label{eq:Mf}
    M_f = &\frac{1}{2} \int_{r} \psi^5 r^2\bigg\{ \eta_1\bigg[ \partial_r - \mathcal{G}\left( 1 + \frac{1}{c_s^2}\right)\bigg] 
    + \eta_2 \frac{l(l+1)}{r}
    \bigg\} \left( -\eta_1 \partial_r p \right) dr, 
    \\ \label{eq:Malpha}
    M_{\alpha} = &\frac{1}{2} \int_{r} \psi^5  \rho h r^2\bigg[
    \eta_1  \partial_r + \eta_2 \frac{l(l+1)}{r}
    \bigg] \left( \frac{\delta\hat{\alpha}}{\alpha} \right)dr,
    \\ \label{eq:Mg}
    M_g =& \frac{1}{2} \int_{r} \psi^9 \alpha^{-2} \rho h r^2 \mathcal{N}^2 \eta_1^2 dr ,
\end{align}
\end{widetext}
corresponding to the energy of compression, the free surface, the gravitational potential and buoyancy, respectively.

%%%%%%%%%%%%%%%%%%%%%%%%%%%%%%%%%%%%%%%%%%%%%%%%%%%%%%
\subsection{Numerical computation of the eigenmodes}
\label{subsec:NumerSolution}
%%%%%%%%%%%%%%%%%%%%%%%%%%%%%%%%%%%%%%%%%%%%%%%%%%%%%%

The system of equations describing linear oscillations of a spherically symmetric hydrostatic equilibrium configuration, under the assumptions considered in this paper, was derived in \cite{Torres_et_al_II}. For the linear perturbation analysis, we construct an effective 1D background by performing surface-area-weighted angular averages of the simulation quantities. To account for the non-spherical deformation of the shock, particularly due to the SASI, we first determine its average radius and rescale the radial profiles accordingly before averaging.
This system of equations, supplemented with suitable boundary conditions, forms an eigenvalue problem that can be solved to obtain the eigenfrequencies, $\sigma$, and the eigenfunctions $(\eta_1, \eta_2, \delta \hat \alpha, \delta \hat \psi)$. 

For the case of a PNS surrounded by a stalled accretion shock, boundary conditions are imposed at the origin of the radial coordinate (regularity condition) and at the shock location, as explained in Sec.~2.3 of \cite{Torres_et_al_II}. Here, we also consider an alternative outer boundary condition when accounting for computational domains not extending to the shock; in those cases, since the density decreases radially outwards, we treat the outer boundary as a free surface by imposing a vanishing Lagrangian pressure perturbation,
\begin{equation}\label{eq:DP=0}
    \Delta p = 0.
\end{equation}
This case is particularly useful when we impose boundary conditions at the PNS surface (see Section~\ref{subsec:BuoyancyRegions}).
If we expand the $\theta$-component of the momentum equation, Eq. (22) of \cite{Torres_et_al_II}, then the above boundary condition can be expressed as
\begin{equation}\label{eq:BC_surface}
    \Delta \hat{p} = \rho h \alpha^{-2}\psi^4 r \sigma^2 \eta_2 +  \eta_1\partial_r p - \rho h \alpha^{-1}\delta\hat{\alpha} = 0  .
\end{equation}

The numerical solution of the eigenvalue problem is performed using the open-source library \texttt{GREAT}. As described in \cite{Torres_et_al_II}, the code computes the eigenmodes using a shooting method. For this purpose, we integrate a first-order system of six coupled ODEs outwards from the center to the outer boundary radius using a second-order implicit trapezoidal rule. For this work, we have modified the code to allow imposing the outer boundary condition at a specified radius for the cases mentioned above.
Each snapshot is resampled onto a fixed radial grid that remains unchanged throughout the simulation using $C^2$-continuous cubic splines, ensuring smooth first derivatives of the background fields. This is particularly important for accurately computing the Brunt--Väisälä frequency, which depends on radial derivatives evaluated numerically. Strictly positive quantities are interpolated in logarithmic space to preserve positivity over their wide dynamic range. For the \textsc{Aenus-ALCAR} simulations, we use a grid of 800 points extending to a fixed outer radius of $300$ km. For the 2D Aenus-ALCAR simulation of a ($20\,M_\odot$) progenitor using the LS220 EoS, the median eigenfrequency differences between grids of 400 and 800 points and between 800 and 1600 points are 0.6\% and 0.3\%, respectively. This confirms convergence to better than 1\% at the adopted resolution of 800 points.
This domain contains the shock during the stalled-shock phase, although the shock may move beyond it after the onset of an explosion.

%%%%%%%%%%%%%%%%%%%%%%%%%%%%%%%%%%%%%%%%%%%%%%%%%%%%%%%
\section{CCSN simulations}\label{sec:CCSNeSimulations}
%%%%%%%%%%%%%%%%%%%%%%%%%%%%%%%%%%%%%%%%%%%%%%%%%%%%%%%
For this work, we use a diverse dataset of 1D, 2D, and 3D CCSN models
with varying combinations of 
progenitor masses, gravity
treatments, and finite-temperature EoS. The simulations
are performed with the \textsc{Aenus-ALCAR} code \citep{Just:2015} and the
\textsc{CoCoNuT} code \citep{Dimmelmeier2012}. Both are multidimensional
Godunov-based, Eulerian hydrodynamics codes formulated in spherical polar
coordinates and include multigroup neutrino transport for three species, i.e., electron neutrinos, electron antineutrinos, and heavy-flavor neutrinos.

In \textsc{Aenus-ALCAR}, neutrino transport is solved using an algebraic
Eddington-factor method with an M1 closure. The neutrino-matter coupling
includes charged-current reactions of electron neutrinos and antineutrinos with nucleons and nuclei, scattering off nucleons and nuclei, and nucleon-nucleon bremsstrahlung, together with electron--positron pair
processes and inelastic scattering off electrons. The \textsc{CoCoNuT} simulations use a more approximate neutrino treatment based on a stationary transport solution with a one-moment closure \citep{Mueller:2015}. In this
case, the included neutrino interactions comprise charged-current reactions with nucleons and nuclei, neutrino scattering off nucleons and nuclei, and neutrino production by nucleon-nucleon bremsstrahlung.

The treatment of gravity depends on the code. In our simulations using \textsc{Aenus-ALCAR}, gravity is modeled with the effective relativistic TOV-A potential \citep{Marek:2006} and redshift corrections in the time evolution \cite{ObergaulingerEtAl2026}. In contrast, \textsc{CoCoNuT} solves the hydrodynamics in general relativity using the XCFC approximation \citep{CC:2009}. 

% Radius for 1.4 Msun, maximun mass (from compose):
% SFHo: 11.89 km, 2.06 Msun
% LS220: 12.71 km, 2.06 Msun
% BHB-Lambda: 13.22 km, 2.1 Msun
% GShen-NL3 (HS NL3): 14.8 km, 2.29 Msun
% HShen-Lambda (STOS TM1 with hyperons): 14.4 km, 1.9 Msun
% HShen (STOS TM1): 14.5 km, 2.23 Msun
We explore six different finite-temperature EoS shown in Table~\ref{tab:eos} in the format provided by \cite{OConnor2010}\footnote{Available at \url{https://stellarcollapse.org}}. Mass and radius in the table are extracted from the \textsc{CompOSE} database \cite{Compose}\footnote{ \url{https://compose.obspm.fr}}. EoS are ordered according to their radii for a $1.4\, M_\odot$ NS, which approximately corresponds to increasing stiffness at saturation density. Stiffer EoS also have higher maximum mass, except for the two EoS including $\Lambda$ hyperons (BHB-$\Lambda$ and HShen-$\Lambda$) whose presence at high densities decreases it. $\Lambda$ hyperons are not accounted for in the neutrino interaction rates.

\begin{table}
\caption{EoS used in this work, including their radius for a $1.4\,M_\odot$ star and maximum mass. \label{tab:eos}}
\begin{tabular}{llll}
     EoS name        & $R_{1.4 M_\odot}$ [km] & $M_{\rm max} [M_\odot]$ & Reference \\ \hline\hline
     SFHo            & 11.89 & 2.06 & \cite{Steiner:2013}\\
     LS220           & 12.71 & 2.06 & \cite{Lattimer:1991}\\
     BHB-$\Lambda$   & 13.22 & 2.1  & \cite{Banik:2014}\\
     HShen-$\Lambda$ & 14.4  & 1.9  & \cite{HShen:2011}\\
     HShen           & 14.5  & 2.23 & \cite{HShen:2011}\\
     GShen-NL3       & 14.8  & 2.29 & \cite{GShen:2011}
\end{tabular}
\end{table}

\begin{figure}[t]
  \centering
  \includegraphics[width=\linewidth]{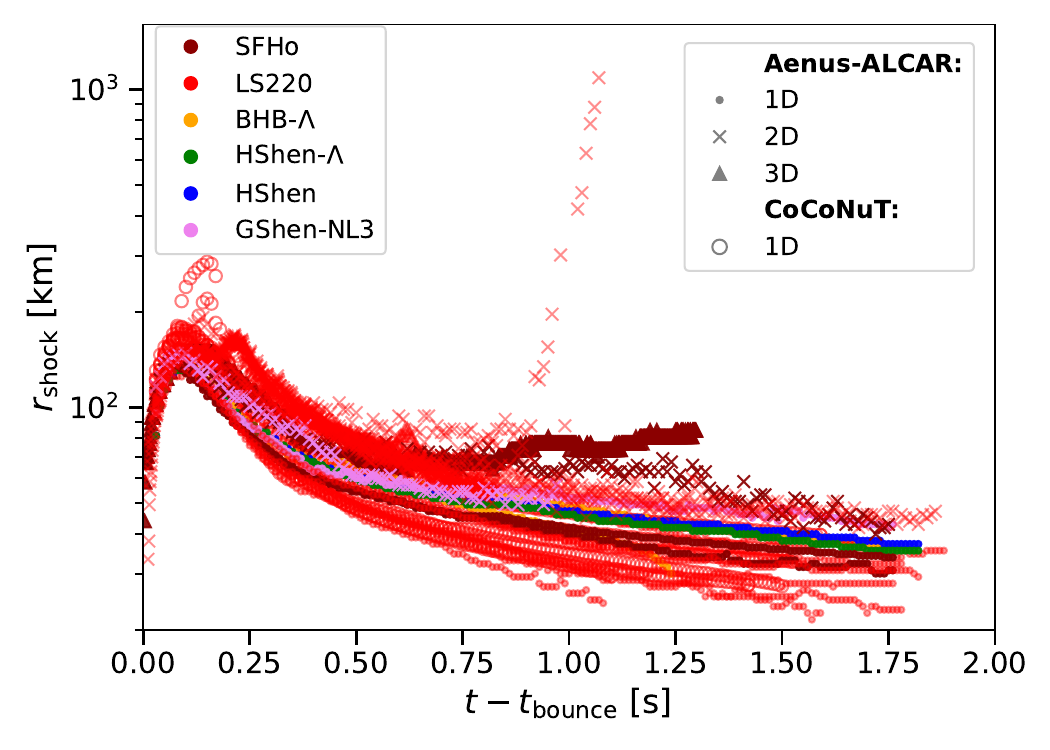}
  \caption{Time evolution of the shock location for all models. }
  \label{fig:shockposition}
\end{figure}

As initial data, we employ pre-supernova progenitors consisting of
solar metallicity stars with masses between $11$ and $75\,M_{\odot}$ \citep{Woosley2007}.
In the spherically symmetric (1D) cases, the core does not launch a supernova explosion. The mass of the PNS increases continuously via accretion, though it remains below the threshold for black-hole formation during the first second after bounce. Fig.~\ref{fig:shockposition} shows the temporal evolution of the shock position for all models. The bounce time, $t_{\rm bounce}$, is defined as the instant when the shock wave is first detected. One can easily identify the exploding model 2D s25 LS220, as its shock keeps expanding.

Table \ref{tab:simulations} summarizes the 28 simulations used in the final analysis. Each column shows the code used, the progenitor mass and metallicity, the EoS, the type of gravity treatment, and the dimensionality.
\begin{table}[t]
\caption{\label{tab:simulations}
List of simulations performed in this work. The name of the progenitor
indicates its metallicity (\texttt{s} for solar) and its zero-age main sequence mass. The symbol $*$ denotes the exploding model. The reference model appears in bold letters.}
\begin{ruledtabular}
\begin{tabular}{lclll}
Code & Progenitor & EoS & Gravity & Dim. \\
\hline
Aenus-ALCAR & s11.2 & LS220 & TOV-A & 1D, 2D \\
Aenus-ALCAR & s15   & LS220 & TOV-A & 1D, 2D \\
Aenus-ALCAR & s15   & BHB-$\Lambda$ & TOV-A & 1D \\
Aenus-ALCAR & s15   & GShen-NL3 & TOV-A & 1D, 2D \\
Aenus-ALCAR & s15   & HShen & TOV-A & 1D \\
Aenus-ALCAR & s15   & HShen-$\Lambda$ & TOV-A & 1D \\
Aenus-ALCAR & s15   & SFHo & TOV-A & 1D, 2D, 3D \\
\textbf{Aenus-ALCAR} & \textbf{s20}   & \textbf{LS220} & \textbf{TOV-A} & 1D, \textbf{2D}, 3D \\
Aenus-ALCAR & s25   & LS220 & TOV-A & 1D, 2D* \\
Aenus-ALCAR & s25   & BHB-$\Lambda$ & TOV-A & 1D \\
Aenus-ALCAR & s30   & LS220 & TOV-A & 1D \\
Aenus-ALCAR & s40   & LS220 & TOV-A & 1D \\
Aenus-ALCAR & s75   & LS220 & TOV-A & 1D \\
%Aenus-ALCAR & u20   & LS220 & TOV-A & 1D \\
CoCoNuT     & s11.2 & LS220 & XCFC & 1D \\
CoCoNuT     & s15   & LS220 & XCFC & 1D \\
CoCoNuT     & s20   & LS220 & XCFC & 1D \\
CoCoNuT     & s25   & LS220 & XCFC & 1D \\
CoCoNuT     & s30   & LS220 & XCFC & 1D \\
CoCoNuT     & s40   & LS220 & XCFC & 1D \\
CoCoNuT     & s75   & LS220 & XCFC & 1D \\
%Aenus-ALCAR & s25   & SFHo & TOV-A & 3D* \\
\end{tabular}
\end{ruledtabular}
\end{table}
\begin{figure}[t]
\centering
\includegraphics[width=\linewidth]{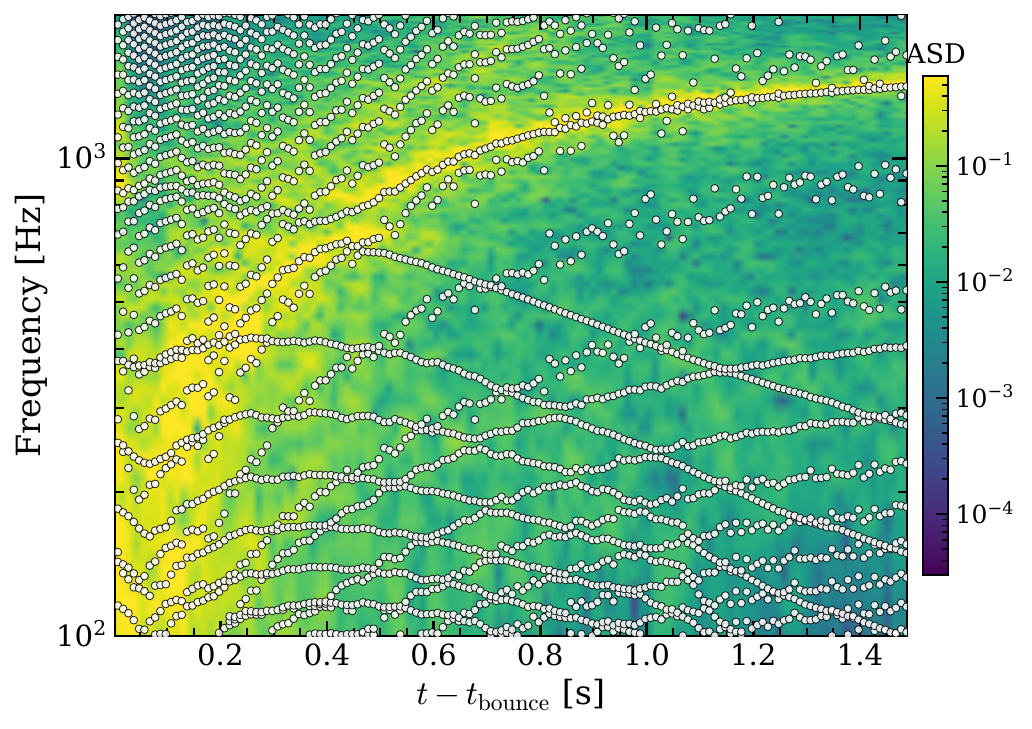}
\caption{Time-frequency diagram of the GW signal from the reference model superposed with the spectrum of the modes computed by imposing $\eta_1=0$ boundary condition at the shock. The bounce time is defined as the time of shock formation.}
\label{fig:GW_modes}
\end{figure}
For most of the plots shown in this paper and the corresponding analysis, we focus on the axisymmetric (2D) core-collapse simulation of a $20\,M_{\odot}$ progenitor with solar metallicity employing the LS220 EoS. We refer to this model as the ``reference model'' hereafter. Nevertheless, we have performed the same analysis for all models, and the comparison among them is presented in Section \ref{sec:ModeClass}.

%%%%%%%%%%%%%%%%%%%%%%%%%%%%%%%%%%%%%%%%%%%%%%%%%%%%%%%
\section{Identification of the physical origin of the modes}\label{sec:ModeClassificationFramework}
%%%%%%%%%%%%%%%%%%%%%%%%%%%%%%%%%%%%%%%%%%%%%%%%%%%%%%%

Fig.~\ref{fig:GW_modes} shows the time-frequency representation (spectrogram) of the GW signal of the reference model. The spectrum of normal modes of oscillation, obtained using \texttt{GREAT} and imposing boundary conditions ($\eta_1=0$) at the shock position (same setup as in \cite{Torres_et_al_II}), is superposed on the spectrogram. 
The main track of the GW emission closely follows the frequency of one of the modes.
However, the computed eigenmodes do not have, a priori, any information about their character (see discussion about the classification problem in the introduction).

In this section, we describe the various diagnostic analyses performed to determine the physical origin of the eigenmodes obtained.

%%%%%%%%%%%%%%%%%%%%%%%%%%%%%%%%%%%%%%%%%%%%%%%%%%%%%%%%%%%%%
\subsection{Energy-based mode classification}\label{subsec:Class_p_g_modes}
%%%%%%%%%%%%%%%%%%%%%%%%%%%%%%%%%%%%%%%%%%%%%%%%%%%%%%%%%%%%%

\begin{figure}[t]
  \centering
  \includegraphics[width=\linewidth]{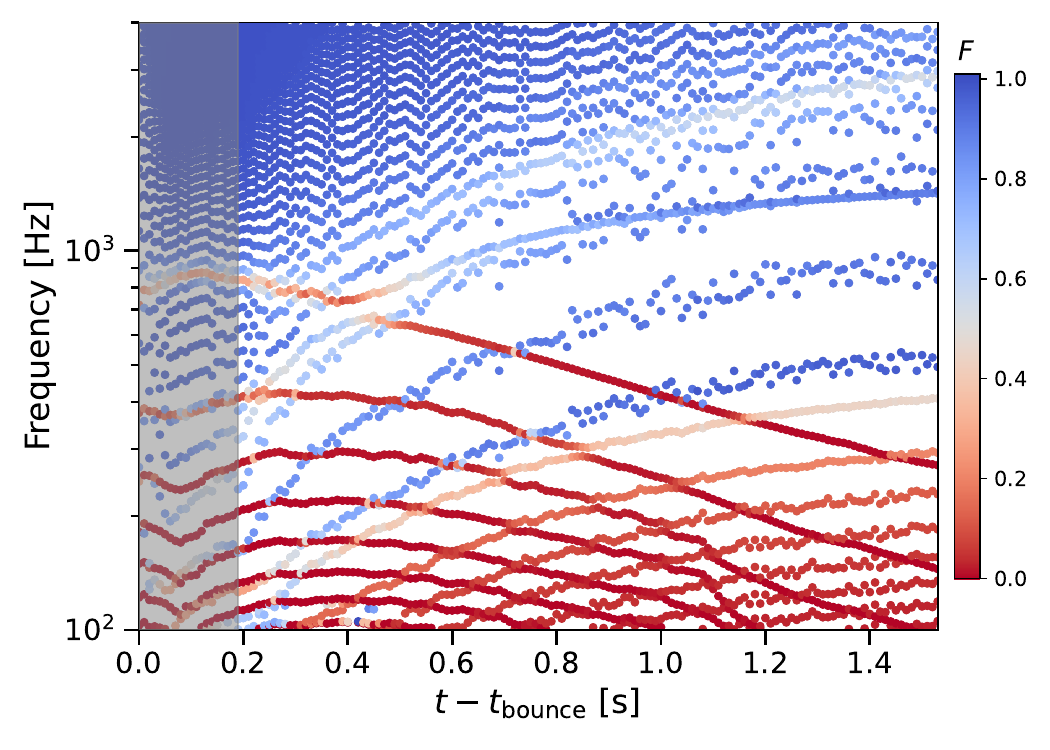}
  \caption{Spectrum of the modes from the reference model classified by the value of the energy fraction, $F$, contributing to f- and p-modes.  Values close to one (blue colors) correspond to the f- or p-modes, while values close to zero (red) correspond to the g-modes. The gray area covers the times when the shock has not stalled yet, and the linear analysis cannot be applied.
   }
  \label{fig:p_g_modes_energy_criterio}
\end{figure}

In Sec.~\ref{subsec:energyCalculation}, we presented the analytical framework for deriving the energy of the modes based on the work produced by their restoring forces (Eqs.~\eqref{eq:Mp}-\eqref{eq:Mg}). 
From the theory of stellar pulsations \cite{Unno1979, 1983_Cox}, it is well known that the restoring force of the p-modes is (the gradient of) pressure, whereas for the g-modes it is buoyancy. Thus, the energy $M_g$  will correspond to that of a g-mode and $M_p$ to that of a p-mode. For the case of an f-mode, the minimal requirement is that it has a free surface, i.e., the contribution of $M_{f}$. However, in the compressible fluid case, it is expected to have a large contribution of $M_{p}$ as well. In addition, we calculate $M_\alpha$, the energy of the gravitational potential perturbations (or metric perturbations).
For most of the modes, the energy of the metric perturbations, $M_\alpha$, is negligible compared to the total energy of the mode, $M_2$.  Only the f-modes (and some p-modes) have a significant contribution of $M_\alpha$ of at most $6.5\%$ of the mode energy, for the reference model.

Therefore, we can combine the contribution of $M_f$, $M_p$, and $M_\alpha$, to have an estimation of the contribution relevant for f- and p-modes. In order to identify these modes, we use the quantity $M_{f+p} = M_f + M_p + M_\alpha$ as a proxy for the energy of either of these modes, which we call f+p-modes, for short.
Since the integrals of Eqs. \eqref{eq:Mp}-\eqref{eq:Mf} contain second-order derivatives, which could lead to numerical inaccuracies, we calculate the energy of f- and p-modes according to, 
\begin{equation}\label{eq:Mf+p}
    M_{f+p} = M_2 - M_g.
\end{equation}

As a first step towards understanding the nature of PNS modes, we wish to categorize modes into two broad families: on the one hand, modes showing features of f- and/or p-modes, and on the other hand g-modes. For this purpose we define the energy fraction
\begin{equation}
    F = \frac{M_{f+p}}{M_2},
\end{equation}
which is a quantity ranging from $0$ to $1$ that tells us which fraction of the total mode energy corresponds to the combined work done by an f- or a p-mode, and the metric perturbations (mostly contributing to f-modes). Its complement, $1-F$, tells us the fraction corresponding to work done by a g-mode. 

Therefore, values of $F$ close to $1$ indicate that the mode is either an f- or a p-mode, and values close to zero (i.e. $1-F$ close to $1$) indicate that the mode is a g-mode.
The energy fraction $F$ is computed at each snapshot and for each eigenvalue, and we use $0.5$ as a threshold to distinguish between these two categories.

In Fig. \ref{fig:p_g_modes_energy_criterio}, we show the eigenfrequencies with respect to the post-bounce time for the reference model, color-coded according to the value of $F$. The frequencies have been obtained with \texttt{GREAT} using as an outer boundary condition a vanishing perturbation at the shock location, $\eta_1(r=r_s) = 0$. Blue/red colors show values of the energy fraction $F$ above/below the threshold $F = 0.5$ and thus indicate f- and p-modes/g-modes. Since we impose boundary conditions at the shock location and assume that the system is approximately in equilibrium, our eigenmode calculation is only valid if the shock is stalled. At times when the shock is moving fast, e.g. at early times before the shock gets stalled, or, in exploding models, when the shock expands rapidly, the approach is not valid and the calculations should not be considered. For the reference model this is the case for the first $\sim0.2$~s after bounce (shaded area in the plot).

Apart from the clear split between f+p-modes and g-modes, there are other interesting features in Fig.~\ref{fig:p_g_modes_energy_criterio}. The theory of linear oscillation modes \citep[see, e.g.,][]{1983_Cox,Unno1979} predicts that p- and g-modes appear in families, with modes ordered in each family according to the number of nodes in the radial eigenfunctions. For p-modes, the frequency increases as the number of nodes increases ($p_1$, $p_2$, $p_3 \dots$), while for g-modes the frequency decreases with an ascending number of nodes ($g_1$, $g_2$, $g_3 \dots$). The frequency of the f-mode has to be lower than the lowest order p-mode, and can be regarded, in some sense, as the lowest order (nodeless) mode of the family of the p-modes (although its nature is different). This is one of the reasons we treat the classification of f- and p-modes together. If one tries to use this standard prediction to our case, one can see that some of the g-modes seem to cross each other (see the increasing and decreasing red features), and similarly, some of the f+p-modes cross each other  
(see the smooth and the fluctuating blue features).  Since modes of the same angular degree cannot cross, these apparent crossings require them to be localised in physically separated regions. This points to at least two families of g-modes and two families of f+p-modes.
This raises the question whether all the modes come from the same region of the PNS or whether there could be separate families of modes located in different areas. We explore this in the following sections.

The second phenomenon worth exploring is avoided crossings. As the background evolves continuously, in our case due to contraction and accretion, two modes from different families can develop the same or very similar frequencies.  A perfect example can be observed in Fig.~\ref{fig:p_g_modes_energy_criterio} at around 0.4 s post-bounce and a frequency of about 700 Hz. Before that time, the rising track of a p-mode (blue) is on a course to meet the falling track of the highest g-mode (red). Instead of intersecting, they avoid the crossing, leaving a small frequency gap between the two tracks. During the interaction, our classification indicates that the two modes have mixed character. According to \cite{Unno1979, 1983_Cox} and based on a classification by the number of nodes, they interchange character, with the former g-mode continuing as a p-mode and vice versa. However, our energy-based classification shows that the g-mode continues to decrease after the avoided crossing, while the p-mode keeps increasing in frequency, as if their tracks crossed. 
Avoided crossings are also observed between the two families of g-modes (see, e.g., features at about $0.85$~s and $300$ Hz). In this case, the interchange of character can be observed in the shape of the eigenfunctions, but not so much in $F$. The strength of the interaction between two modes determines how large the frequency gap between the tracks is. A strong crossing (e.g., the first one mentioned) would indicate that both modes overlap in the same region of the star, with a strong interaction. A weak interaction (e.g., between the two g-modes) would indicate weakly interacting modes in physically separated regions. 

Closing this section with the first results of our new classification scheme, many questions are raised. Could the g-modes with various characteristics correspond to more than one family? Would the oscillatory p-modes hide more smooth ones? And if so, would these be distinct families of modes arising from different areas of the PNS? To address these, the following section analyzes different regions of the PNS-shock system.

%%%%%%%%%%%%%%%%%%%%%%%%%%%%%%%%%%%%%%%%%%%%%%%%%%%%%%%%%%%%%
\subsection{Stably stratified regions}
\label{subsec:BuoyancyRegions}
\begin{figure}[t]
\centering
\includegraphics[width=\linewidth]{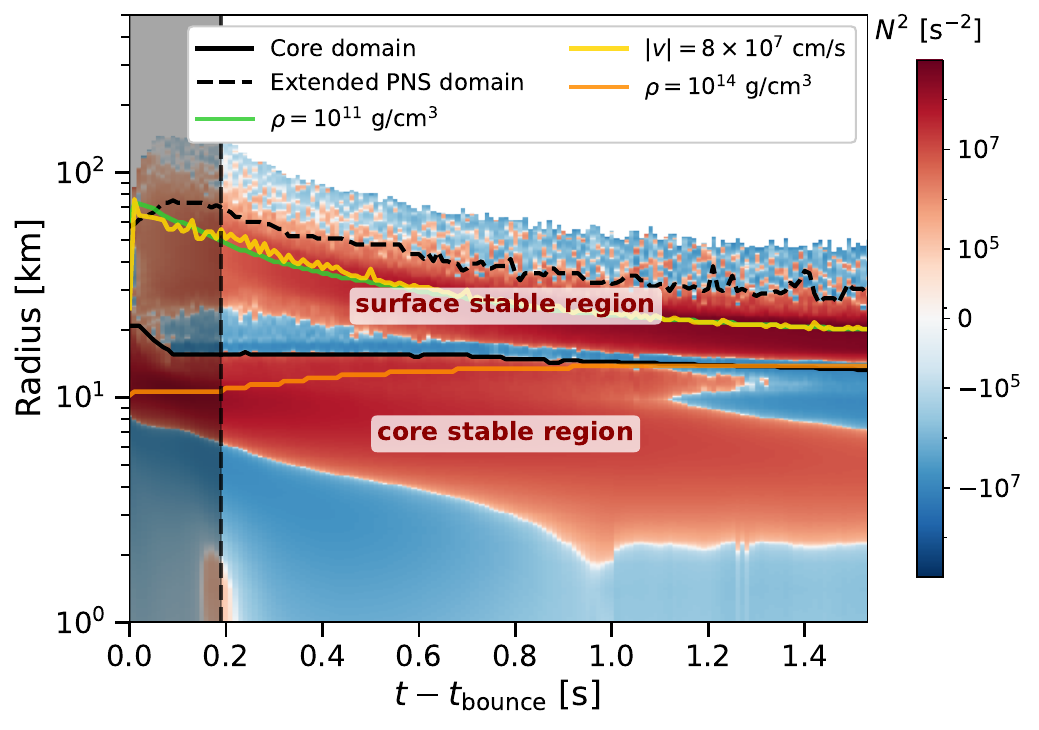}
\caption{Evolution of the squared BV frequency, $\mathcal{N}^2$, as a function of time and radius. The gray shaded area marks the early post-bounce phase before the shock stalls. Green, orange, and yellow curves denote the isocontours of density at $10^{11}\text{ g cm}^{-3}$, at $10^{14}\text{ g cm}^{-3}$,  and radial velocity at $|\upsilon|=8\times10^{7}\text{ cm s}^{-1}$, respectively. The solid and dashed black lines indicate the outer boundaries adopted for the core and extended PNS domains, respectively.}
\label{fig:BV_evolution}
\end{figure}

%%%%%%%%%%%%%%%%%%%%%%%%%%%%%%%%%%%%%%%%%%%%%%%%%%%%%%%%%%%%%
In all the simulations we consider in this paper, we observe the development of two distinct shells with a positive value of the square of the BV frequency ($\mathcal{N}^2 > 0$) that are stable against convection.
This structure has been reported previously \cite{Cerda-Duran2007,Murphy:2009, Morozova:2018, OConnor2018,Torres_et_al_II, Jakobus2024}. In these stable layers, buoyancy acts as the restoring force, enabling the propagation of gravity waves (g-modes).
Fig.~\ref{fig:BV_evolution} illustrates the time–radius evolution of the BV frequency for the reference model, where these two regions can be identified (red areas). We will refer to the inner stably stratified region, extending below $20$ km, as the  {\it core stable region}, and  the outer stably stratified region, extending between $\sim 20$ and $90$~km, as the {\it surface stable region}.

The central blue region corresponds to $\mathcal{N}^2<0$: the angle-averaged background is convectively unstable there, consistent with the Ledoux-unstable layer that separates the two stable regions in the multidimensional simulations.

The {\it PNS surface} is characterized by a steep density gradient and is identified with the location where the density is $10^{11}$~g~cm$^{-3}$
(green curve) \cite[see e.g.][]{Morozova:2018,Torres_et_al_II}. Accreting matter settles onto the PNS surface and hence a strong decrease in radial velocity is observed at the surface. The yellow line shows the radius at which the radial velocity drops below $8\times 10^7$~ cm/s (the exact value of the threshold velocity has been found empirically). Note that both definitions of the surface coincide almost perfectly. Given that the velocity criterion is motivated by the physics of accretion (as opposed to the empirical relation set by the density threshold), we chose it as our standard criterion in this work. In any case, the PNS surface is not a sharp surface, but rather a region extending over a few km that lies well within the surface stable region. 

From the PNS surface to the shock, densities are much lower than in the PNS interior, and the radial velocity is not negligible, producing an inflow of matter from the shock toward the PNS surface, where it accumulates. This {\it post-shock region}, bounded by the density gradient at the PNS surface and the sonic point at the shock, is typically convectively unstable and includes the gain layer, where neutrinos can deposit energy.

Motivated by this structure, we define three restricted computational domains to isolate the physical origin of the observed eigenmodes. 
\begin{itemize}
    \item The {\it core domain}, extending from the stellar center to the outer boundary of the core stable region (solid black line in Fig.~\ref{fig:BV_evolution}). This  domain contains the entire core stable region, additionally to a region with $\mathcal{N}^2<0$ located below.    
%    it is not assumed to be convectively stable at all radii. In particular, it also includes the inner region with $\mathcal{N}^2<0$ visible below the core stable region in Fig.~\ref{fig:BV_evolution}.}
    \item The {\it PNS domain}, extending from the center to the PNS surface, as defined above, using either the density or the velocity criterion. This domain contains the entire core stable region and part of the surface stable region.
    \item The {\it extended PNS domain}, extending from the center to the upper limit of the surface stable region (dashed black line in Fig. \ref{fig:BV_evolution}).  This domain fully contains both the core stable region and the surface stable region.  
\end{itemize}
The upper limit of the core and surface stable regions are determined as the first and second radius, respectively, at which $\mathcal{N}^2$ drops below $7\times 10^{5}~\rm{s}^{-2}$; this value is chosen to cover the entire region while avoiding an excessively oscillatory boundary.  For the core domain, we use a constant radius after $1$~s to avoid a sudden change in the domain radius. We will refer to the domain extending up to the shock, as discussed in the previous section, simply as {\it full domain}. The aim is to try to understand the nature of the eigenmodes computed in the full domain by comparing to the eigenmodes computed using the three restricted domains. We emphasize that all three regions start at the center and form a nested sequence of domains. More detailed spatial decompositions of the GW emission have been performed by Andresen et al.~\cite{Andresen:2017}, Mezzacappa et al.~\cite{Mezzacappa2023}, and Murphy et al.~\cite{Murphy2025}. In the nomenclature of the latter studies, our core and PNS domains broadly encompass Region 0 and Regions 0--3, respectively.

At the outer boundaries of the three restricted domains, we impose a vanishing Lagrangian pressure perturbation, $\Delta p = 0$ (see Section~\ref{subsec:NumerSolution}).

\begin{figure}[t]
  \centering
  \includegraphics[width=\linewidth]{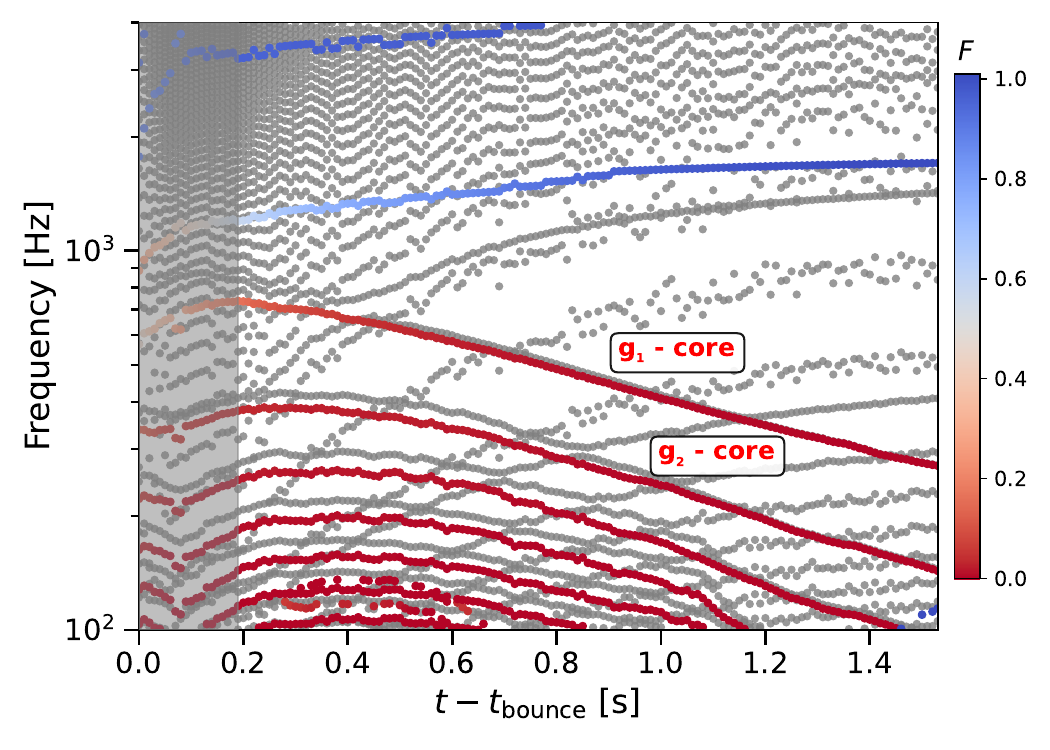}
  \caption{Comparison of the time evolution of mode frequencies for the reference simulation computed using the core domain (colored dots) and the full domain (gray dots). Modes are classified according to the value of $F$ as f+p modes (blue) or g-modes (red). The core g-modes are visible as red features that drift to lower frequencies with time. The gray-shaded area corresponds to the time period before the shock stalls.}
  \label{fig:freq_comparison_region1}
\end{figure}
Fig. \ref{fig:freq_comparison_region1} displays the eigenfrequencies computed by restricting the calculation to the core domain. To isolate which features of the full system originate within the core stable region, we overplot the modes of the core domain (colored) to the frequencies obtained when using the full domain (gray marks). The comparison shows that a family of low-frequency g-modes, specifically those that exhibit a systematic decrease in frequency over time, coincides with the g-modes obtained in the core domain (in red). This implies that this family of modes is generated within the core stable region of the PNS. Given their origin in a zone of positive squared BV frequency, and because the energy classification labels them as g-modes, we call them {\it core g-modes}, hereafter. The two sets of blue dots at frequencies of around 1000 Hz and higher are f+p-modes associated with the restricted domain and have no counterparts in the full domain calculation,
because they arise from the artificial truncation of the computational domain.

This analysis also sheds light on the avoided crossings discussed in Section~\ref{subsec:Class_p_g_modes}. When considering only the core domain, the highest-frequency g-mode does not interact with other modes, and no avoided crossing occurs. Compared with the analysis of the full domain (Fig.~\ref{fig:p_g_modes_energy_criterio} and gray dots in Fig.~\ref{fig:freq_comparison_region1}), this mode passes directly through the region where the most noticeable avoided crossing occurs, around $0.4$ s after bounce.

\begin{figure}[t]
  \centering
  \includegraphics[width=\linewidth]{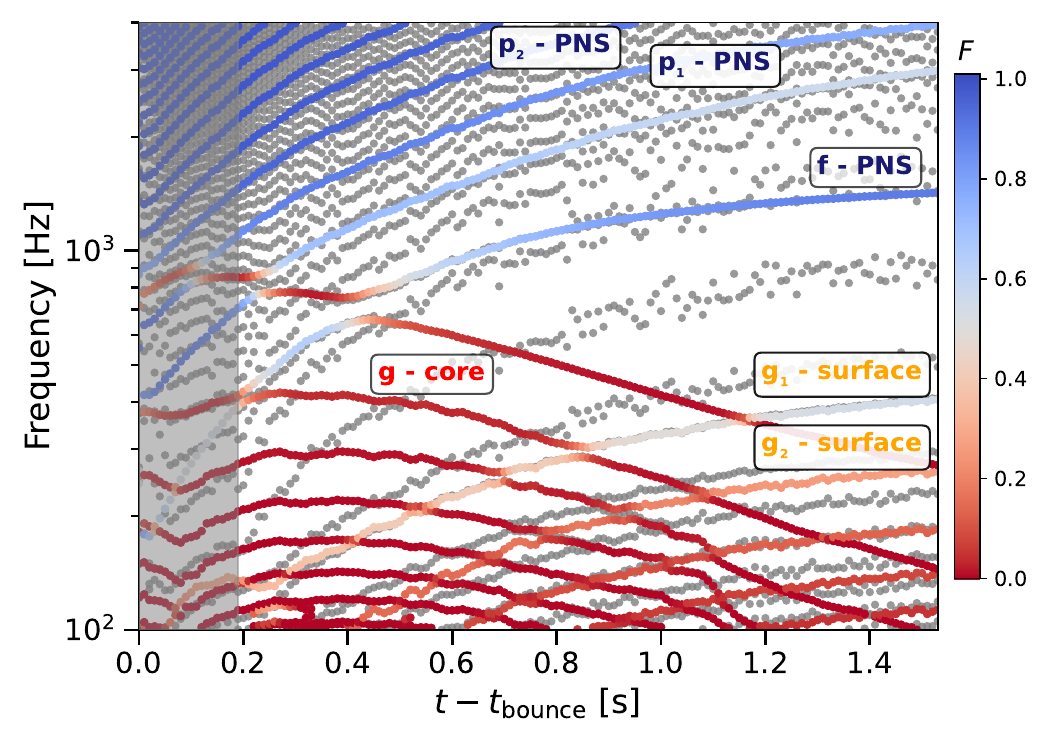}
  \caption{Comparison of the time evolution of mode frequencies for the reference simulation computed using the PNS domain (colored dots) and the full domain (gray dots). See Fig.~\ref{fig:freq_comparison_region1} for details. Two families of g-modes are visible, core-g-modes (downward drifting red features) and surface g-modes (upward drifting red features), along with a family of PNS f+p-modes (upward drifting blue features).}
\label{fig:freq_comparison_surface}
\end{figure}

In Fig.~\ref{fig:freq_comparison_surface}, we compare the results for the reference model computed with the full domain with those obtained with the PNS domain (using the density criterion). The comparison shows several interesting features. First, the low-frequency core g-modes identified in the analysis of the core domain remain consistent.

Second, one can clearly distinguish another family of g-modes (red) with increasing frequencies that was not present in the calculations restricted to the core domain. 
This family of g-modes is close to a set of g-modes with increasing frequency appearing in the full domain case. Therefore, we hypothesize that the origin of these modes is the surface stable region, but, because the PNS domain leaves part of the surface stable region outside (recall that this region extends outside the PNS surface), the frequency of the modes is not perfectly captured. We label this family as {\it surface g-modes}.
Note that the value of $F$ for the highest-frequency surface g-mode is around $0.5$ at times larger than $\sim 0.8$~s, meaning that the contributions of $M_g$ and $M_{f+p}$ are approximately equal. We attribute this to an artifact of the restriction imposed in the domain. If we consider the full domain, the mode is classified as a g-mode (see Fig. \ref{fig:p_g_modes_energy_criterio}). 

Third, the lowest frequency smoothly evolving f+p-mode (upward sloping blue feature, whose track ends at $\sim 1000$~Hz; see also Section~\ref{subsec:Class_p_g_modes}), coincides perfectly with the corresponding feature in the full domain calculation. We note that this is the same mode that is dominantly excited in the GW signal, visible as a track in Fig.~\ref{fig:GW_modes}. 
This mode does not appear when the core domain is used, only becoming visible when we extend the domain at least up to the PNS surface. Because it is the lowest frequency mode of the family and appears only when the PNS surface is included in the domain, we classify it as the {\it PNS f-mode}. We discuss this classification in more detail in the next sections, related to its global character inside the PNS. Higher frequency modes of the same family are likely {\it PNS p-modes}.  Except for the first PNS p-mode, the rest are not visible in full domain calculation, most probably buried by other modes in the same frequency range (see Fig.~\ref{fig:p_g_modes_energy_criterio}).
The appearance of this PNS f-mode produces a clear avoided crossing with a downward sloping core g-mode at about $0.4$~s. 

\begin{figure}[t]
  \centering
  \includegraphics[width=\linewidth]{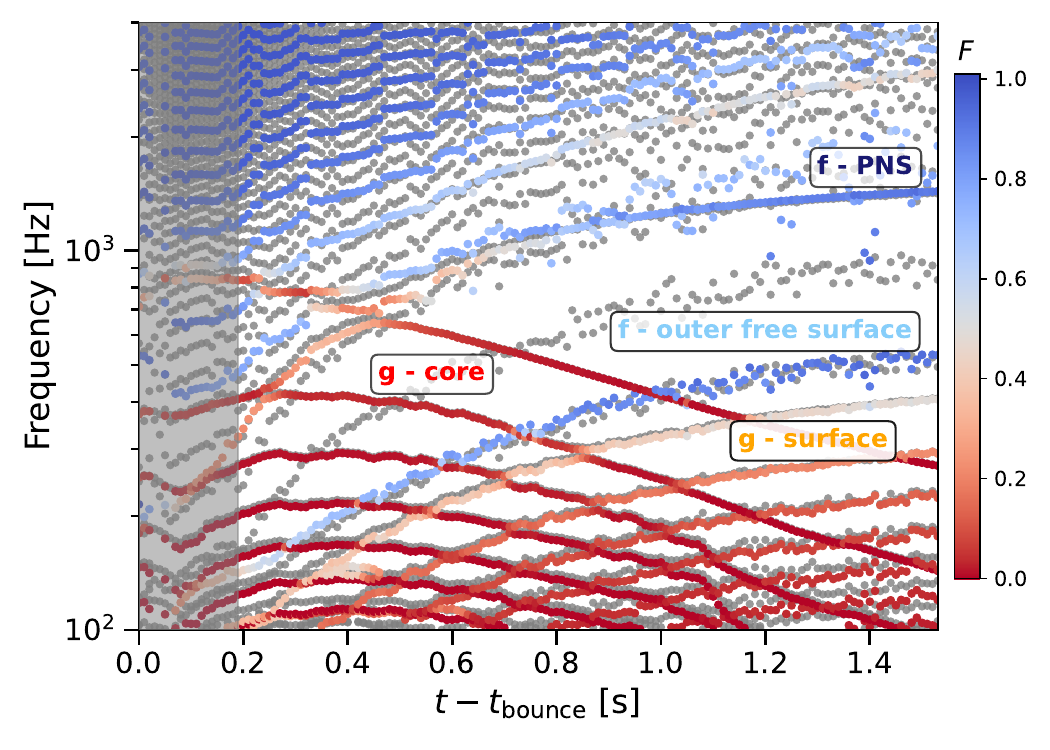}
  \caption{Comparison of the time evolution of mode frequencies from the reference simulation imposing the boundary condition at the shock location (gray dots) and considering the computational domain from the center to the upper limit of the surface BV region (colored dots). The gray area corresponds to the time period before the shock stalls. The frequency of the modes is classified as p-modes (blue) or g-modes (red) according to the value of $F$.}
\label{fig:freq_comparison_region2}
\end{figure}

Finally, we investigate the effect of considering the extended PNS domain. In Fig.~\ref{fig:freq_comparison_region2}, we compare the frequencies obtained using the extended PNS domain with the reference frequencies computed using the full domain.  In this case, in addition to the core g-modes, the surface g-modes, and the PNS f-mode, which were already present in more restricted domains, a new family of f+p-modes appears, which exhibits a fluctuating behavior. The lowest frequency component of this new f+p family of modes (fluctuating blue feature increasing from $\sim 100$~Hz up to $500$~ Hz) can be related to the presence of a free surface at the outer boundary of the domain (where $\Delta p=0$ has been imposed), and we call it {\it outer free-surface f-mode}.
The fluctuating nature of this mode is likely produced by variations in the location of the outer boundary and in the physical properties of the fluid there, which are determined numerically from the simulation (see fluctuating dashed black line in Fig.~\ref{fig:BV_evolution}).
The frequency of this mode follows very closely that of the corresponding f+p mode in the full domain (lowest frequency blue feature in Fig.~\ref{fig:p_g_modes_energy_criterio}). We refer to the analogous feature in the full domain calculation, related to the presence of a freely moving outer boundary (in that case the shock), as the {\it shock f-mode}. Therefore, there are two modes of similar nature, the outer free-surface f-mode (observed in the extended PNS domain calculations) and the shock f-mode (appearing in the full domain calculations), which have almost identical frequencies, despite being associated with outer boundaries at different radii. The reason for this coincidence is explored in section~\ref{sec:domain_dependence}.
Higher-order p-modes of the same family (fluctuating blue features above the shock f-mode frequency) do not coincide with the features in the full domain calculation. This is a consequence of truncating the domain below the shock, which strongly modifies the frequency of these p-modes living in the post-shock region.

With the increase of the outer boundary to the extended PNS domain, we now have two f-modes present at the same time (similarly to the case of the full domain). Only one of them is actually a proper f-mode, the one related to the outer free surface (the outer free-surface f-mode or the shock f-mode, depending on the domain considered). What we call PNS f-mode, related to the existence of a PNS surface, can be seen as an interface mode associated to the strong density gradient between the PNS and the low-density post-shock region 
\cite[see discussion in][]{Mueller:2013}. In practice, the low-density region can be replaced with good accuracy by a free surface boundary, giving almost identical results for the frequency of this mode (see results for the PNS domain).

Finally, we discuss the g-mode frequencies in the extended PNS domain in comparison to the ones obtained in domains considered previously. Including the full surface stable region rather than truncating it, as in the case of the PNS domain, leads to a much better agreement of the surface g-modes (upward sloping red features) with the corresponding ones in the full domain.
However, the first mode in this family of surface g-modes (upward sloping red-gray feature between $100$ and $400$~Hz), still has a value of $F$ around $0.5$ at later times ($>1$~s), albeit higher than in the case of using the PNS domain. This improvement shows that moving the boundary domain outwards helps in the classification of this particular mode. Actually, in the calculation with the full domain the situation improves even more, reaching a value of $F$ clearly above $0.5$ (see Fig.~\ref{fig:p_g_modes_energy_criterio}), which allows for an unambiguous classification of the mode. The reasons for this behavior are explored in Section~\ref{sec:domain_dependence}, while its impact on the classification is discussed in Section~\ref{subsec:BCs_surface_vs_shock}.

%%%%%%%%%%%%%%%%%%%%%%%%%%%%%%%%%%%%%%%%%%%%%%%%%%%%%%%%%
\subsection{Mode Propagation and Radial Energy Distribution}
\label{sec:EnergyLocalization}
%%%%%%%%%%%%%%%%%%%%%%%%%%%%%%%%%%%%%%%%%%%%%%%%%%%%%%%%%

In the previous sections, we have found indications that there are two families of g-modes (core and surface) and two families of f- and p-modes (PNS and shock). Up to this point, the classification of modes into these four families has been based on their qualitative behavior. 
In this and the following sections, we treat this qualitative classification as a hypothesis to be tested. Since the restricted-domain and energy-distribution diagnostics share the energy functional that defines the classification, we complement them with propagation diagrams and analytic estimates.

To test this hypothesis, we first examine whether our classification is consistent with the standard description of g- and p-modes based on propagation diagrams \cite{Unno1979, 1983_Cox, Aerts:2010}. For each angular degree $l$, the propagation diagram compares the squared angular frequency of a given mode, $\sigma^2$, with the squared relativistic BV and Lamb frequencies, $\mathcal{N}^2$ and $\mathcal{L}^2$, defined in Eqs.~(\ref{eq:BVfreq}) and (\ref{eq:Lamb_freq}), respectively. This comparison determines the local propagation character of the perturbation.

For g-modes, propagation is allowed where the squared angular frequency lies below both characteristic frequencies,
\begin{equation}
\sigma^{2} < \mathcal{N}^{2} \quad \text{and} \quad
\sigma^{2} < \mathcal{L}^{2}.
\end{equation}
In these regions, buoyancy acts as the restoring force. For p-modes, propagation is allowed where the squared angular frequency exceeds both characteristic frequencies,
\begin{equation}
\sigma^{2} > \mathcal{N}^{2} \quad \text{and} \quad
\sigma^{2} > \mathcal{L}^{2},
\end{equation}
and pressure provides the restoring force.

If $\sigma^2$ lies between $\mathcal{N}^2$ and $\mathcal{L}^2$, the perturbation is evanescent: its local radial behavior is exponential rather than oscillatory. Such layers inhibit wave propagation and can act as barriers between g- and p-mode propagation cavities.

In the post-bounce PNS and its surrounding post-shock region, the complex radial stratification produces multiple propagation cavities separated by evanescent layers. For each eigenfrequency, the propagation diagram maps where g- or p-mode propagation is locally allowed, while the radial energy distribution shows where the energy of the eigenmode is actually concentrated. Comparing these two diagnostics allows us to assess the proposed association of the modes with different parts of the system.

\begin{figure}[t]
  \centering
  \includegraphics[width=\linewidth]{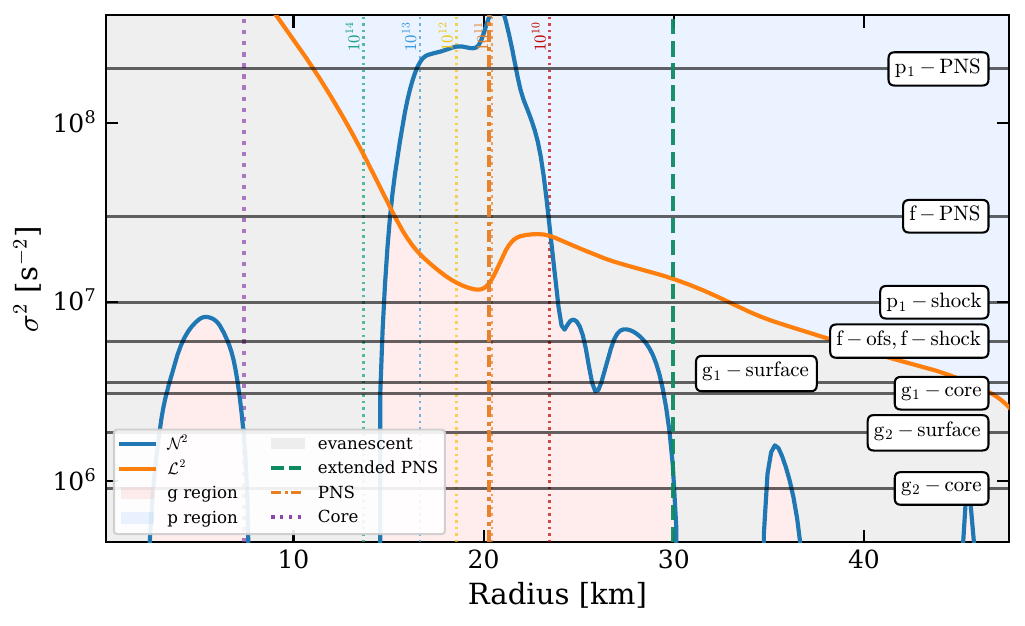}
  \caption{Example of a propagation diagram for the reference model at $t=1.4$ s after bounce. The blue and orange curves represent the squared Brunt--Väisälä and Lamb frequencies, $\mathcal{N}^2$ and $\mathcal{L}^2$, respectively. The p- and g-mode propagation cavities are shaded in red and blue, while the evanescent regions are shaded in gray. The horizontal lines mark the squared angular frequencies of the selected modes. The radial domain extends from the center to the shock. The vertical dashed lines mark selected density isocontours. The purple dashed, orange dash-dotted, and green dotted lines indicate the outer boundaries of the core, PNS, and extended PNS computational domains, respectively.}
  \label{fig:propagation_diagram}
\end{figure}

\begin{figure*}[t]
  \centering
  \includegraphics[width=\linewidth]{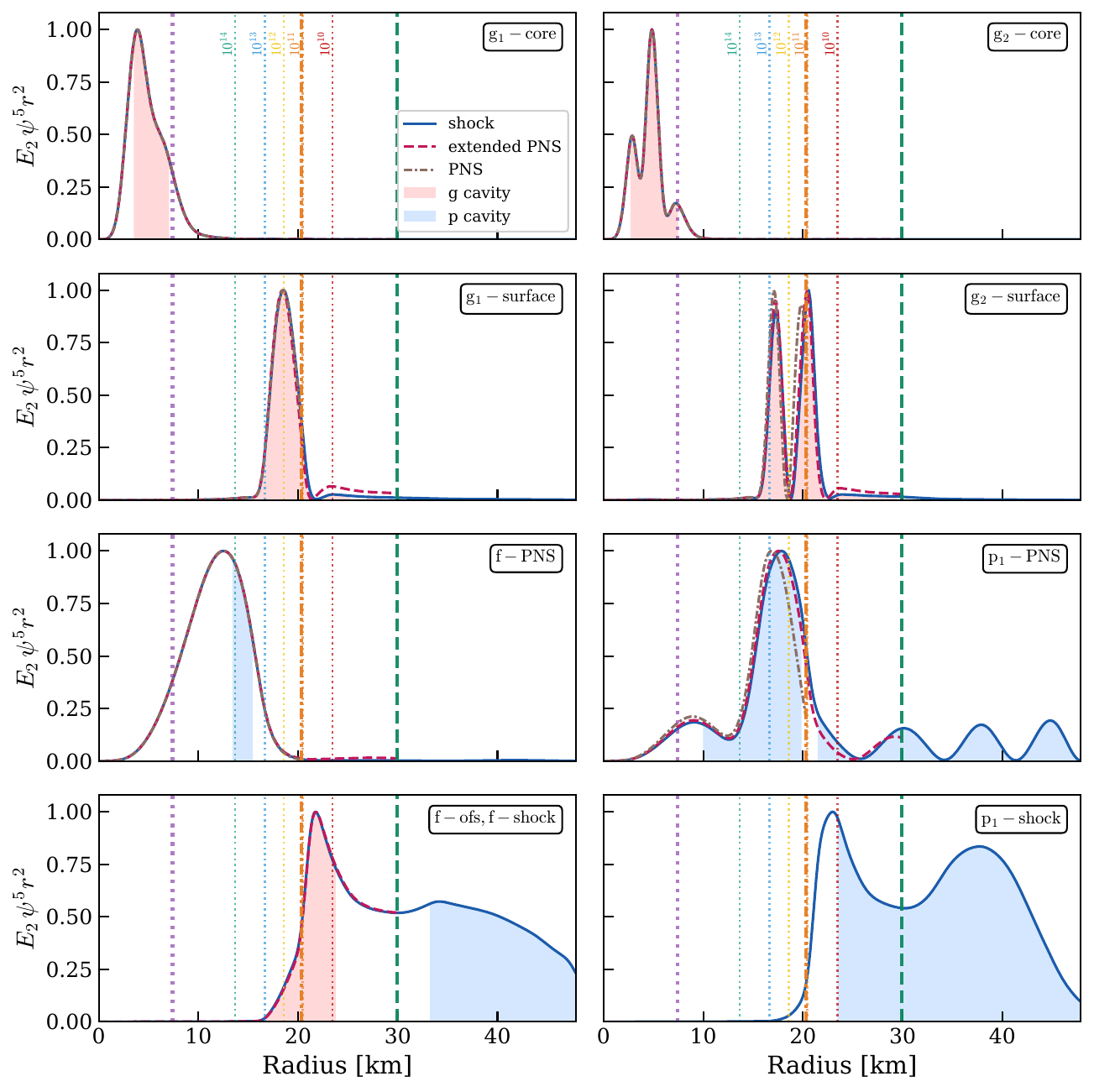}
  \caption{Radial distribution of the energy integrand for several selected modes at $t=1.4$ s after bounce. The solid blue line corresponds to the full computational domain, with the outer boundary condition imposed at the shock. The dashed red and dash-dotted brown lines correspond to the extended PNS and PNS domains, respectively. The thin vertical dashed lines mark selected density isocontours. The purple dashed, orange dash-dotted, and green dotted vertical lines indicate the outer boundaries of the core, PNS, and extended PNS computational domains, respectively.}
  \label{fig:modes}
\end{figure*}

Fig.~\ref{fig:propagation_diagram} shows the propagation diagram for the reference simulation at 1.4 s after bounce, considering the full radial domain up to the shock. At this time, several distinct g-mode and p-mode cavities can be identified. The horizontal lines correspond to the modes labeled in Fig. \ref{fig:modes}. Depending on their frequency, many of these modes cross different propagation and evanescent regions.

To determine whether modes computed using different domains represent the same physical mode, we compare their normalized radial energy distributions. Their similarity is quantified using the Bhattacharyya coefficient (BhC) \cite{Kailath1967}. Candidate pairs must satisfy $|f_B-f_A|\leq0.3f_A$. We then perform a one-to-one assignment in decreasing order of BhC, so that each mode is used at most once. We retain pairs with ${\rm BhC}\geq0.5$ and regard matches with ${\rm BhC}\geq0.7$ as robust identifications of the same physical mode.

\begin{table}[t]
\caption{Bhattacharyya coefficients comparing modes computed in the extended PNS and PNS domains with their candidate counterparts in the full domain at $t-t_{\rm b}=1.40$~s. A value of ${\rm BhC}=1$ indicates identical normalized radial energy distributions. A dash indicates that no pair satisfies the matching criteria.}
\label{tab:bc}
\begin{ruledtabular}
\begin{tabular}{lcc}
Mode & ${\rm BhC}_{\rm ext-PNS}$ & ${\rm BhC}_{\rm PNS}$ \\
\hline
$\mathrm{g}_1\text{-core}$    & 1.00 & 1.00 \\
$\mathrm{g}_2\text{-core}$    & 1.00 & 1.00 \\
$\mathrm{g}_1\text{-surface}$ & 0.97 & 0.93 \\
$\mathrm{g}_2\text{-surface}$ & 0.98 & 0.75 \\
$\mathrm{f}\text{-PNS}$       & 1.00 & 1.00 \\
$\mathrm{p}_1\text{-PNS}$     & 0.90 & 0.82 \\
$\mathrm{f}\text{-ofs}/\mathrm{f}\text{-shock}$ & 0.67 & -- \\
$\mathrm{p}_1\text{-shock}$   & 0.93 & -- \\
\end{tabular}
\end{ruledtabular}
\end{table}

To better understand the nature of the modes included in our analysis, Fig. \ref{fig:modes} shows the radial energy integrand for eight selected modes at $t=1.4$ s after bounce, with each panel corresponding to one mode. The locations of selected density isocontours and the outer boundaries of the restricted computational domains are also shown for reference.

The first row shows the first two core g-modes, identified in Sec. \ref{subsec:BuoyancyRegions} through the analysis of the core domain. Their energy is concentrated in the gravity-wave cavity of the core. The three curves compare the energy integrands obtained using the PNS, extended PNS, and full computational domains. Their close agreement indicates that, once the complete propagation cavity is included, the structure of these modes is largely insensitive to the location of the outer boundary. Consequently, the BhC values reported in Table \ref{tab:bc} are equal to unity to the precision shown for both comparisons. Within the core g-mode family, modes are ordered by decreasing frequency. The modes in the left and right panels are therefore labeled $\mathrm{g}_1\text{-core}$ and $\mathrm{g}_2\text{-core}$, respectively.

The second row of Fig. \ref{fig:modes} shows the g-modes associated with the surface-stable region. These panels support our initial interpretation: the energy of both modes is concentrated in the layer between $\rho=10^{12}$ and $10^{11}\,\mathrm{g\,cm^{-3}}$. The BhC values are high and increase when the restricted domain is extended outwards, from 0.93 to 0.97 for the $\mathrm{g}_1\text{-surface}$ and from 0.75 to 0.98 for the $\mathrm{g}_2\text{-surface}$. Following the same frequency-ordering convention, we label them as $\mathrm{g}_1\text{-surface}$ and $\mathrm{g}_2\text{-surface}$.

The third row shows the fundamental mode (left panel) and the first p-mode (right panel) of the PNS. Within each f+p-mode family, modes are ordered by increasing frequency: the lowest-frequency member is labeled as the f-mode, followed by $\mathrm{p}_1$, $\mathrm{p}_2$, and so on. The energy of the PNS f-mode is distributed throughout the PNS, with most of the contribution lying in an evanescent region and a smaller fraction in a p-mode cavity. This broad distribution, rather than confinement to a single g- or p-mode cavity, is consistent with its identification as the PNS f-mode. Its energy distributions agree across the three domains, with ${\rm BhC}=1$ to the precision shown. The energy of the $\mathrm{p}_1\text{-PNS}$ mode is concentrated in a p-mode cavity near the PNS surface, with a smaller contribution in the p-mode cavity associated with the shock. The agreement is slightly lower for this mode, particularly for the PNS domain, because this restricted domain does not encompass its full radial energy distribution. This results in ${\rm BhC}=0.90$ and $0.82$ for the extended PNS and PNS domains, respectively.

Finally, the last row shows modes associated with the region outside the PNS, for which no counterpart is found when the computational domain is truncated at the PNS surface. In the left panel, we compare the lowest-frequency member of the f+p-shock family, labeled as the shock f-mode, with the outer-free-surface f-mode obtained using the extended PNS domain. The energy of the shock f-mode extends across a g-mode and a p-mode cavity separated by an evanescent layer, supporting its identification as an f-mode. Although the two integrands appear similar over their common radial interval, their coefficient, ${\rm BhC}=0.67$, lies below the threshold for a robust match. This is consistent with treating them as distinct physical modes associated with outer boundaries at different locations. The right panel shows the next mode in the f+p-shock family, labeled as $\mathrm{p}_1\text{-shock}$. Its energy is concentrated in a p-mode cavity, consistent with its classification as a p-mode. It has a robust counterpart in the extended PNS domain, with ${\rm BhC=0.93} $, but no counterpart in the PNS domain.

%%%%%%%%%%%%%%%%%%%%%%%%%%%%%%%%%%%%%%%%%%%%%%%%%%%%%%%%%%
\subsection{Simple analytic estimates}
\label{sec:analytic_estimates}
%%%%%%%%%%%%%%%%%%%%%%%%%%%%%%%%%%%%%%%%%%%%%%%%%%%%%%%%%%

Before continuing with the analysis of the nature of the modes, and given that we have already gained some insight into which kind of modes are present in the system (core and surface g-modes, shock and PNS f+p-modes), we next formulate simple models for each of the mode families, based on analytic estimates. These models are not expected to match exactly the frequencies of the modes computed by solving the linear eigenvalue problem, but rather provide an estimate of how the different modes should scale with the properties of the system, and will allow in the next sections to test the consistency of the mode classification that we are proposing.

\subsubsection{PNS f- and p-modes}

To understand the behavior of the PNS f-mode frequency, we compare with that of a self-gravitating homogeneous fluid sphere as an approximation for the PNS interior. Under these assumptions, for a perturbation of spherical-harmonic degree $l$, the squared angular frequency is given by \citep[see e.g.][section 17.7]{1983_Cox}
\begin{equation}
    \sigma^2_l = \frac{2 l (l-1)}{2 l + 1}\frac{G M_{\rm PNS}}{R_{\rm PNS}^3},
\end{equation}
where $M_{\rm PNS}$ is the mass of the PNS enclosed within $R_{\rm PNS}$, computed as the radius where $\rho=10^{11}\text{ g cm}^{-3}$.
Note that this solution describes an incompressible perturbation even when the fluid is compressible, and coincides with Kelvin's classical result.

PNS p-modes can be estimated considering a homogeneous compressible fluid sphere with constant adiabatic index, $\Gamma_1$, and no buoyancy ($\mathcal{N}^2=0$). In the asymptotic limit ($n \gg 1$), the mode frequency reads  \citep[see, e.g., Sec.~17.7 of][]{1983_Cox}
\begin{equation}
\sigma^2_{ln} \simeq \left ( 2 \Gamma_1 n^2 + \frac{l(l+1)}{2\Gamma_1 n^2}
\right ) \frac{GM_{\rm PNS}}{R_{\rm PNS}^3},
\label{eq:pns:pmodes}
\end{equation}
where $n$ is the radial order and $\Gamma_1$ is the radial average of the adiabatic index over the PNS interior ($r < R_{\rm PNS})$.

\subsubsection{Outer free surface f- and p-modes}
The outer free surface f-mode is associated with the presence of a low-density region layer surrounding the PNS, extending from its surface to the outer boundary of the domain, that behaves as a free surface. This can be modeled following the work of Lamb (\cite{lamb1932}, see art. 264. Eq. (9)), adopting the classical framework for an incompressible fluid layer with constant density surrounding a spherical core of radius $R_{\text{PNS}}$ up to a radius $R_{\text{out}}$. The interior PNS is assumed to provide a fixed background gravitational field, while the self-gravity of the perturbed layer is neglected. Under these assumptions, the squared angular frequency is given by
\begin{equation}
\label{eq:f-mode}
\sigma_l^2 \simeq l(l+1) \frac{\left(\frac{R_{\text{out}}}{R_{\text{PNS}}}\right)^l-\left(\frac{R_{\text{PNS}}}{R_{\text{out}}}\right)^{l+1}}
{(l+1)\left(\frac{R_{\text{out}}}{R_{\text{PNS}}}\right)^l+l\left(\frac{R_{\text{PNS}}}{R_{\text{out}}}\right)^{l+1}}\,\frac{\mathcal{G}_{\rm out}}{R_{\rm out}},
\end{equation}
where $\mathcal{G}_{\rm out}\simeq\frac{GM_{\rm PNS}}{R_{\text{out}}^2}$ is the gravitational acceleration at the outer boundary.

The shock p-modes can be estimated using the same configuration as above, consisting of a core and a shell extending from $R_{\text{PNS}}$ to $R_{\text{out}}$, with a constant sound speed $c_s$ and no buoyancy ($\mathcal{N}^2=0$). In the WKB limit\footnote{Wentzel–Kramers–Brillouin approximation, which assumes that the wavelength of the mode is much shorter than the characteristic scale of variation of the background medium. }, 
the acoustic-gravity dispersion relation reduces to \citep[see, e.g., Sec.~15.2 of][]{Unno1979}
\begin{equation}
\sigma^2 \simeq c_s^2\left(k_r^2+k_h^2\right),
\end{equation}
where $k_r$ and $k_h$ are the radial and horizontal wavenumbers, respectively. The horizontal wavenumber describes the angular variation of the perturbation along a spherical surface and is determined by the spherical-harmonic degree $l$. We approximate the radial wavenumber by imposing a standing wave condition across the layer, 
\begin{equation}
    k_r \simeq \frac{n\pi}{R_{\text{out}}-R_{\text{PNS}}}, 
\end{equation} 
where $n$ is the radial order, and obtain the horizontal wavenumber at the mean cavity radius $\bar r=(R_{\text{out}}+R_{\text{PNS}})/2$, 
\begin{equation} 
    k_h^2 \simeq \frac{l(l+1)}{\bar r^2}. 
\end{equation}
This gives
\begin{equation}
\label{Eq:p-modes}
\sigma_{l n}^2 \simeq
c_{s,\rm eff}^2
\left[
\left(\frac{n\pi}{R_{\text{out}}-R_{\text{PNS}}}\right)^2
+
\frac{l(l+1)}{\bar r^2}
\right].
\end{equation}

The effective sound speed in Eq.~(\ref{Eq:p-modes}) is obtained from the sound
travel time across the cavity: for high-order acoustic modes the radial
quantization condition reads $\sigma \int dr/c_s = \pi n$
(see, e.g., Sec.~3.4 of \cite{Aerts:2010}).
\begin{equation}
c_{s,\rm eff} = \left( R_{\rm out} - R_{\rm PNS} \right)
\left[ \int_{R_{\rm PNS}}^{R_{\rm out}} \frac{dr}{\alpha(r)\, c_s(r)} \right]^{-1},
\label{eq:cs_eff}
\end{equation}
i.e.\ the harmonic mean of $\alpha c_s$ across the shell. Note that, since the
lapse is already included in the travel time, no additional redshift correction
is applied to Eq.~(\ref{Eq:p-modes}).

\subsubsection{Redshift correction}

The estimates above for f- and p-modes are computed in classical gravity. However, the eigenvalue problem that we are solving is formulated in general relativity, and the frequencies obtained correspond to those measured by a distant observer. Therefore, we should apply a gravitational redshift correction to these simple expressions,
\begin{equation}
\sigma_{l,\infty}=\alpha_{\rm out}\sigma_l,
\end{equation}
where, assuming a Schwarzschild metric outside the PNS, in isotropic coordinates it reads
%in the isotropic Schwarzschild approximation used here, 
%
\begin{equation}
\alpha_{\rm out}=\left ( 1-\dfrac{M}{2R_{\text{out}}} \right )
\left (
1+\dfrac{M}{2R_{\text{out}}}
\right )^{-1}.
\end{equation}

\subsubsection{g-modes}

For the case of g-modes, we can obtain the mode frequencies from the profile of the
Brunt-V\"ais\"al\"a frequency in the region hosting the mode. For this purpose, we detect the boundaries of the two stable regions (core and surface) using as a threshold  
$\mathcal{N}^2_{\rm thr} = 7\times10^{5}~\mathrm{s}^{-2}$. Note that, since $\mathcal{N}^2$ is a relativistic expression, it does not need any additional redshift correction.

In the asymptotic limit ($n \gg 1$), the periods of g-modes of consecutive radial order form a uniformly spaced sequence (see Sec.~3.4 of \cite{Aerts:2010}), which in terms of the angular frequency reads 
\begin{equation} 
\sigma_{ln} \simeq \frac{\sqrt{l(l+1)}}{\pi\, S\, (n+\varepsilon_{\rm g})} \int \frac{\mathcal{N}}{r}\, dr , 
\label{eq:gmodes} 
\end{equation} 
where the integral extends over the corresponding stable region, so that $\sigma \propto 1/n$. Here $\varepsilon_{\rm g}$ is a phase of order unity, set by the behavior of the eigenfunctions at the boundaries of the region, and $S$ is a normalization factor close to unity that absorbs higher-order corrections to the asymptotic expansion \citep{Tassoul1990} as well as the mild dependence of the integral on the adopted threshold. Both are constants; they depend neither on time nor on the radial order.

%%%%%%%%%%%%%%%%%%%%%%%%%%%%%%%%%%%%%%%%%%%%%%
\subsubsection{Comparison with numerical eigenmodes}
%%%%%%%%%%%%%%%%%%%%%%%%%%%%%%%%%%%%%%%%%%%%%%

\begin{figure}[t]
  \centering
  \includegraphics[width=0.49\textwidth]{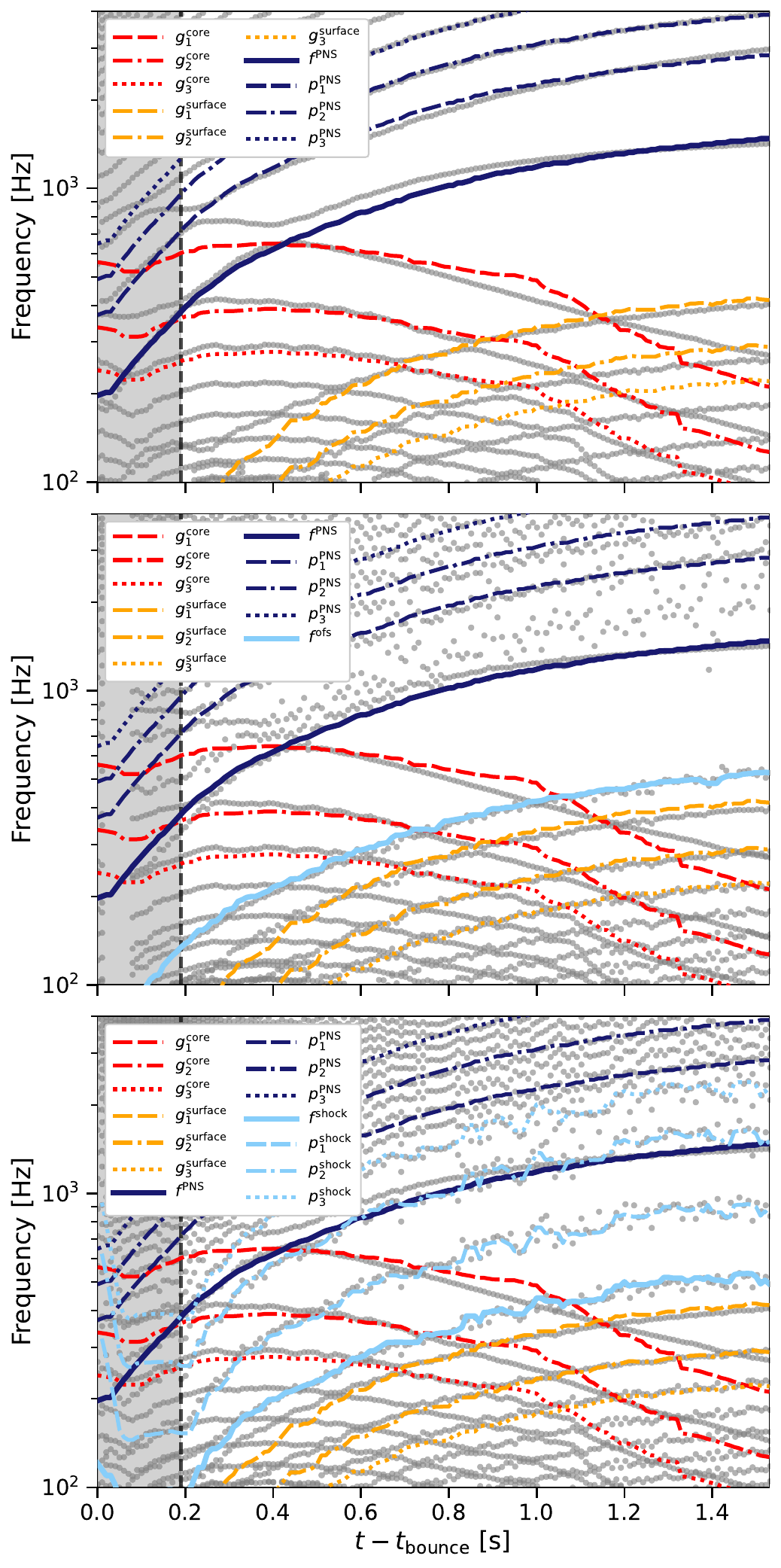}
  \caption{Results from the simple analytical estimates of section~\ref{sec:analytic_estimates} compared to the numerical ones when modes are computed in the PNS domain (upper panel), the extended PNS domain (middle panel) and full domain (lower panel).}
  \label{fig:comparison_to_analytical_model}
\end{figure}

In Fig.~\ref{fig:comparison_to_analytical_model} we compare these simple estimates with the numerically computed eigenmodes for three of the cases considered in Section~\ref{subsec:BuoyancyRegions}, from top to bottom, the PNS, the extended PNS and the full domain.

For the case of core g-modes, regardless of the domain considered, the analytical estimation (red lines) follows the trend of the modes that we identified as core g-modes. The reason is that in all three cases, the domain includes the core stable region completely where these modes originate.

For the surface g-modes, the simple analytic estimate (orange lines) gives qualitatively similar results to the numerical eigenmodes. However, only when the whole surface stable region is enclosed in the domain (extended PNS or full domain, middle and lower panel, respectively), the analytical and numerical frequencies show similar trends. In the case of the PNS domain (upper panel), part of the surface stable region remains outside the domain, so the numerical eigenmodes tend to have lower frequencies (particularly in the higher harmonics).

For the PNS f+p-modes, there is good agreement between the analytic model (dark blue lines) and the numerical results in the three spatial domains. The reason is that these modes, originating from the PNS, can be properly resolved within the three domains considered, as all three include the whole PNS.

In the case of the f+p-modes, corresponding to the low-density layer surrounding the PNS, the simple model cannot be applied when the PNS domain is considered, since by construction, this area is not incorporated. In contrast, in the extended PNS domain, the PNS surface is included within the computational domain, with an extra low-density layer extending to the outer (domain) boundary. However, this is not a distinct physical region of the CCSN, but an artifact of imposing a free surface boundary condition at a chosen radius. The simple estimate for the f-mode (light blue solid line in the middle panel) confirms the interpretation of an outer free-surface f-mode, corresponding precisely to the numerically computed f-mode associated with the free-surface at the outer boundary. Finally, in the full domain, the total region considered for the analytical estimates (light blue curves in the lower panel), is physical, since it extends from the PNS surface to the shock. The excellent matching is a good indication that our simple model captures the physics of the underlying f+p-modes. 

The fact that these simple analytic models can reproduce the trends of the numerically computed modes (in some cases with excellent agreement), is a strong indication that our interpretation of the different families of modes and their nature is accurate. We note that several of the analytical expressions that we use are nominally only valid in the asymptotic limit or the WKB approach, which effectively holds only if $n \gg 1$. Nevertheless, the analytical estimates are consistent with the numerical results, even if we apply these expressions to cases with $n\sim 1$.

%%%%%%%%%%%%%%%%%%%%%%%%%%%%%%%%%%%%%%%%%%%%%%%%%%%%%%%%%%
\subsection{Dependence on the radial domain boundary}
\label{sec:domain_dependence}
%%%%%%%%%%%%%%%%%%%%%%%%%%%%%%%%%%%%%%%%%%%%%%%%%%%%%%%%%%

In Section~\ref{subsec:BuoyancyRegions} we observed that, as we consider domains of progressively larger size, more physical regions are incorporated into the analysis, and new modes appear. Some modes seem to change very little once the domain is sufficiently large (e.g. core g-modes), indicating that those modes should be localized in certain regions of the PNS. This is consistent with the energy localization of the modes, as shown in Section~\ref{sec:EnergyLocalization}.
To investigate this effect more systematically, we study the dependence of the mode frequency and character (value of $F$) as a function of the location of the outer boundary of the computational domain, for a fixed simulation time ($t=$1.4 s post-bounce). We select the location of the outer boundary, $R_{\rm out}$, by selecting a density threshold, $\rho_{\rm out}$ in the range $\sim 10^{9} - 10^{14}$~g cm$^{-3}$, the lower limit corresponding to the shock location.
This test allows us to better characterize the structural properties of the system.

\begin{figure}[t]
  \centering
    \includegraphics[width=0.49\textwidth]{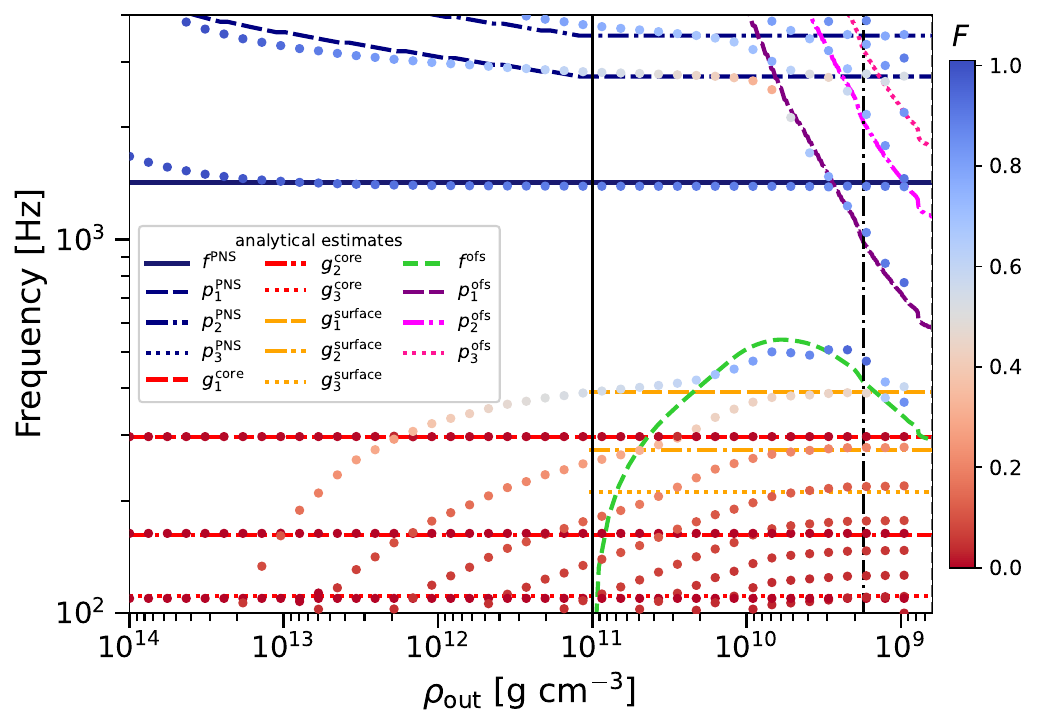}
    \hfill
    \includegraphics[width=0.49\textwidth]{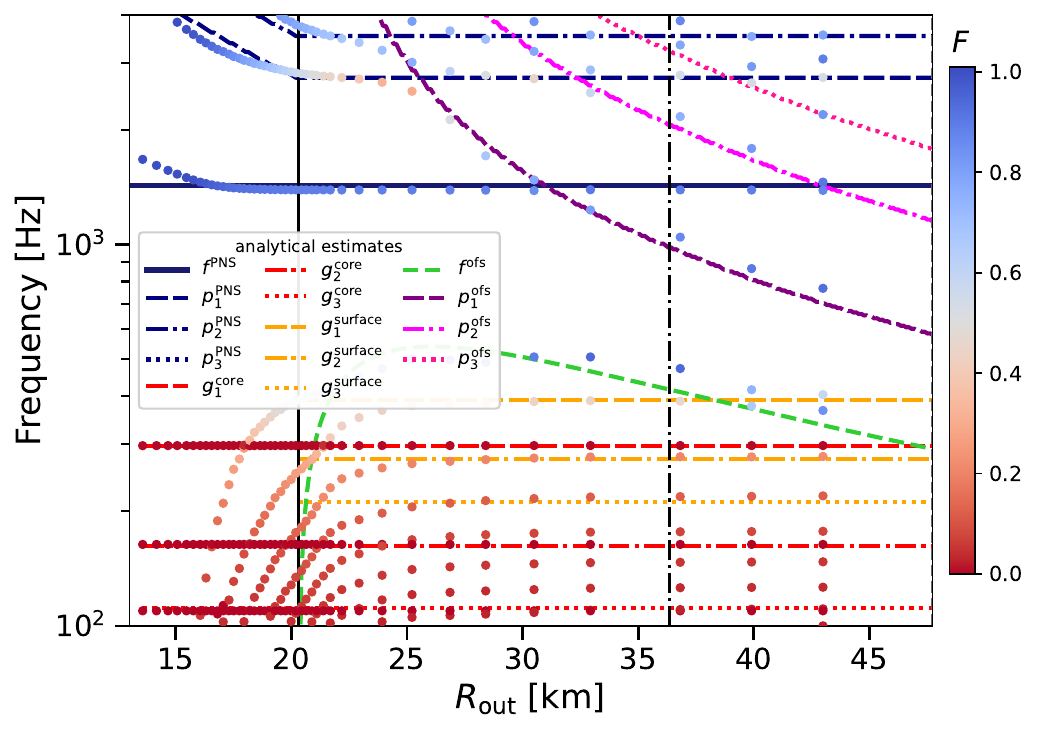}
  \caption{Spectrum of the modes for different density thresholds at a fixed time ($1.4$~s post-bounce), shown as a function of the outer cut density $\rho_{\rm out}$ (upper) and of the corresponding outer cut radius $R_{\rm out}$ (bottom). Modes are classified according to the value of $F$ as f+p modes (blue) or g-modes (red). Overlaid curves show the analytic modes: the PNS f-mode $f^{\rm PNS}$ and p-modes $p_n^{\rm PNS}$ (dark blue), the core g-modes $g_n^{\rm core}$ trapped in the core stable region (red), the surface g-modes $g_n^{\rm surface}$ of the surface stable region (orange), the free-surface mode $f^{\rm ofs}$ (green) of the cavity bounded by the cut, and the corresponding shell p-modes $p_n^{\rm ofs}$ (purple, magenta, and deep pink for $n=1,2,3$). Vertical lines mark, from left to right, the PNS surface (solid black) and the top of the surface-stable region (dot-dashed black).}
  \label{fig:freq-vs-density}
\end{figure}

Fig.~\ref{fig:freq-vs-density} shows the results as a function of $\rho_{\rm out}$ (upper panel) and $R_{\rm out}$ (lower panel).
We also apply the energy classification criterion to distinguish between p- and g-modes (blue and red, respectively).
Additionally, we compare the frequencies of the different mode families with the simple analytical models discussed in Section~\ref{sec:analytic_estimates}.
Because our analytical models are highly idealized, we do not expect an exact quantitative match between these analytical lines and the numerical simulation frequencies. However, the analytical curves capture the qualitative trends and slopes of the simulated frequencies remarkably well. To facilitate the comparison, we rescale each of the frequencies by a constant factor of order unity.
 
Let us focus first on the f-modes. When the domain extends outside of the PNS surface ($R_{\rm out}\gtrsim R_{\rm PNS} \simeq 20$~km) there are two such modes: the outer free-surface (ofs) f-mode (blue points at $400-500$~Hz) related to the domain boundary, and the PNS f-mode (blue marks right above $1000$~Hz), which is actually an interface f-mode originating from the existence of the PNS surface. 
However, for smaller domains ($R_{\rm out}<R_{\rm PNS}$), the PNS surface lies beyond the domain boundary, and only one f-mode appears, associated with the outer boundary. This mode is produced by the artificial truncation of the computational domain inside the PNS and should not be identified with the physical PNS f-mode. Nevertheless, its frequency connects continuously with that of the PNS f-mode at $\sim 1000$~Hz as $R_{\rm out}$ approaches $R_{\rm PNS}$.

The frequency of the PNS f-mode barely changes once $R_{\rm out} \ge R_{\rm PNS}$, consistent with the fact that this is a mode confined in the PNS interior. 
However, the frequency of the outer free-surface f-mode increases very rapidly as the boundary moves outwards across the steep density gradient near the PNS surface, where the sudden drop in mass-loading inside the cavity dominates over the geometry expansion. The simple analytical model for the outer free-surface f-mode (Eq.~\eqref{eq:f-mode}, green dashed line), qualitatively explains this behavior.
In particular, in the limit in which $R_{\rm out}$ is close to $R_{\rm PNS}$, its squared frequency scales as $l(l+1)\left(R_{\rm out}/R_{\rm PNS}-1\right)\mathcal{G}_{\rm out}/R_{\rm out}$, going to zero if $R_{\rm out}=R_{\rm PNS}$.
In the numerically computed eigenfrequencies, this behavior manifests as an avoided crossing between the outer free surface f-mode and the g-modes of the PNS, which modifies the frequencies and eigenfunctions of all participating modes due to their mutual coupling.

Similarly, the family of modes identified as outer free-surface p-modes (downward curving blue marks at high frequencies) follow qualitatively the behavior of the simple analytical model (purple dashed, dot-dashed magenta, and dotted deep pink curves). They appear when the boundary position is placed above the radius of the PNS, and their frequencies decrease with increasing radius. In the limit where $R_{\rm out}$ extends to the shock location (full domain), their frequencies approach those of the shock p-modes.
The reason for this behavior is that these modes live in the post-shock region, extending from the PNS radius to $R_{\rm out}$, so modifying the size of the domain has a big impact on their frequency.
Additionally, there are PNS p-modes (series of blue marks of approximately constant-frequency at $\sim 2000$~Hz and above), whose frequencies remain nearly constant for sufficiently large domains, including the full PNS. The simple analytical model (Eq.~\eqref{eq:pns:pmodes}, dark blue dashed and dot-dashed horizontal lines) explains this behavior.

The family of PNS f- and p-modes (solid dark blue and dark blue dashed and dot-dashed horizontal lines) intersects in frequency with the family of outer free-surface p-modes (purple dashed, dot-dashed magenta, and dotted deep pink curves). The presence of avoided crossings in the numerically calculated eigenfrequencies clearly evidences that both families of modes interact with each other. These avoided crossings are analogous to those discussed above, but the continuous parameter of the eigenvalue problem is now $R_{\rm out}$ (or, equivalently, $\rho_{\rm out}$), rather than time.
Thus, even though the simulation time is fixed, the mode frequencies vary continuously with the location of the outer boundary and can undergo avoided crossings.

Finally, we compare the two families of g-modes (red marks) with the simple analytic estimates for the core (red line) and surface (orange line) g-modes (see Section~\ref{sec:analytic_estimates}).  
The frequencies of core g-modes (computations and analytic model) are constant and independent of $R_{\rm out}$ (horizontal series of red marks at $300$~Hz and below).
This is expected and confirms the findings of the previous sections: once the region containing the mode energy (core stable region) is fully enclosed by the computational domain, both the frequency and the eigenfunction remain unaltered by the boundary location. 
For surface g-modes, the comparison with the analytical model is more quantitative. For $R_{\rm out}$ larger than the extended PNS domain boundary (at $\sim$36 km, at this time), the frequencies of both the analytical and the numerical modes are constant, since the whole surface stable region is contained in the domain. For smaller values of $R_{\rm out}$, the surface stable region is progressively truncated, so the frequency decreases with decreasing $R_{\rm out}$, until the region is completely excluded at about $15$~km, at which point the mode disappears.
%%%%%%%%%%%%%%%%%%%%%%%%%%%%%%%%%%%%%%%

%%%%%%%%%%%%%%%%%%%%%%%%%%%%%%%%%%%%%%%%%%%%%%%%%%%%%%%%%%%%
\section{Mode classification}\label{sec:ModeClass}
%%%%%%%%%%%%%%%%%%%%%%%%%%%%%%%%%%%%%%%%%%%%%%%%%%%%%%%%%%%%
Our next task is to develop a flexible methodology to classify the modes into the four families that we have identified in the preceding sections, without requiring an extended analysis for each simulation. Our ultimate goal is to achieve an automatic classification of the modes, in order to compare with GW spectrograms and extract information about the system through universal relations. For the rest of this section, for all the simulations, the outer boundary condition is imposed at the shock location, except if mentioned otherwise.

%%%%%%%%%%%%%%%%%%%%%%%%%%%%%%%%%%%%%%%%%%%%%%%%%%%%%%%%%%%%%%%%%
\subsection{Automatic classification}\label{subsec:AllFamilies}
%%%%%%%%%%%%%%%%%%%%%%%%%%%%%%%%%%%%%%%%%%%%%%%%%%%%%%%%%%%%%%%%%
Our classification scheme is based on both the energy of the modes and their region of origin. Hence, we first perform the analysis of Sec.~\ref{subsec:Class_p_g_modes} to identify whether a mode is an f+p- or a g-mode based on the work done by their restoring forces. Next, we calculate the energy of the mode in different domains identifying the following distinct families of modes (see Sec.~\ref{subsec:BuoyancyRegions}): 
\begin{itemize}
    \item Core and surface g-modes are originated from two localized stable regions at the core and the surface of the PNS, respectively. Based on this spatial distribution, we propose the following classification criterion:
    first we select all modes with $F<0.5$, i.e. those with g-mode character. Then we compute the energy of the g-mode, $M_g$, integrating separately the energy at densities higher and lower than a density threshold, $\rho_{\rm thr} = 10^{14}$ $\rm g/cm^3$. This threshold separates the interior region (from the stellar center to $\rho_{\rm thr}$) and the exterior (from $\rho_{\rm thr}$ to the outer boundary of the computational domain). At late times, this threshold coincides approximately with the outer boundary of the core stable region (see Sec. \ref{subsec:BuoyancyRegions}), so it naturally separates the core and surface stable regions. By comparing the energy below and above this threshold the mode is classified as a core g-mode ($\rho>\rho_{\rm thr}$) or as a surface g-mode (otherwise).
    
    \item PNS and shock f+p-modes inhabit the PNS interior and the post-shock region, respectively. Therefore, we can also classify the modes according to their spatial localization. First we select all modes with $F\ge$ 0.5, i.e. with f+p character. Then we compute $M_{f+p}$, integrating in two different areas: the PNS (from the stellar center to its surface) and the post-shock region (from the surface of the PNS to the shock). If $M_{f+p}$ is larger in the PNS than in the post-shock region, then the mode is attributed to the PNS f+p-modes. Otherwise, it is labeled as a shock f+p-mode. For the estimation of the PNS surface we use the velocity-based criterion described in section~\ref{subsec:BuoyancyRegions}).

\end{itemize}

Once the families of the modes are identified correctly, then the modes within each family should be properly ordered with respect to their number of nodes (see section~\ref{subsec:BuoyancyRegions}). For that reason, we consider the lowest frequency of the f+p-PNS modes, as the f-mode of the PNS, the next one as the p$_1$-PNS and so on. In other words, the order of the p-modes will increase as the frequency increases. The f+p-shock modes are ordered in a similar fashion. On the contrary, the g-modes have an ascending order with descending frequency, thus the highest frequency of the g-core modes will be the g$_1$-core. The same applies for the g-surface modes, too. 

\begin{figure}[t]
  \centering
  \includegraphics[width=\linewidth]{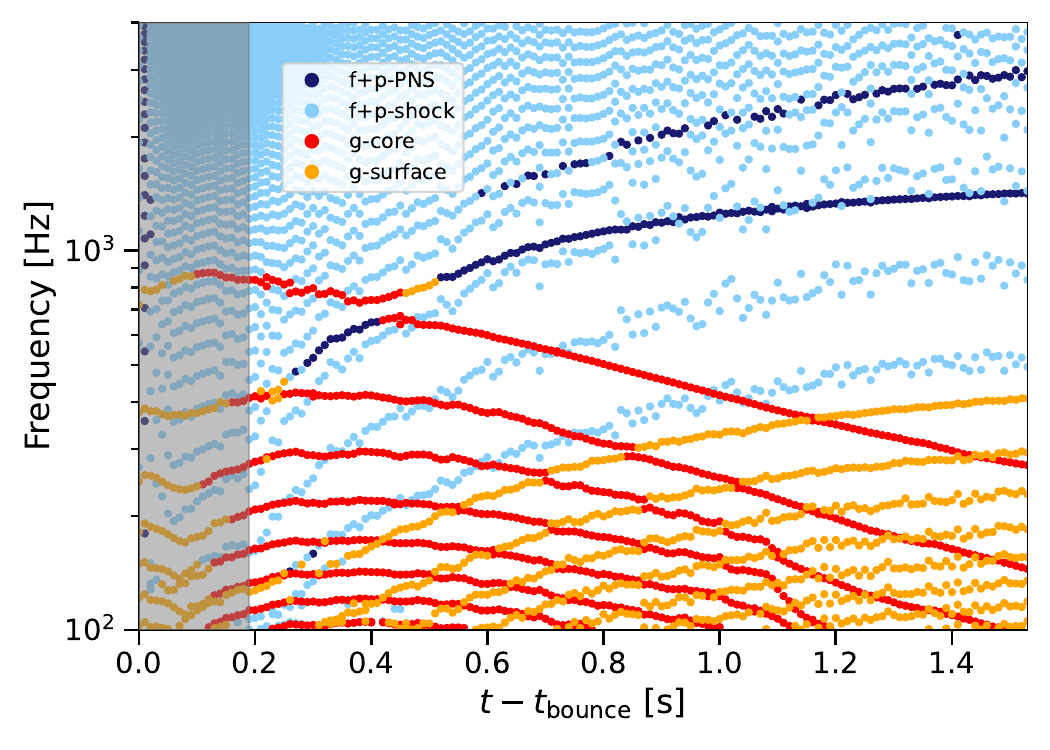}
  \caption{Automatic classification of the four mode families for our reference CCSN model: p-PNS (dark blue), p-shock (light blue), g-core (red), and g-surface (orange).}
  \label{fig:p_g_modes_energy_criterio_4families}
\end{figure}

Fig.~\ref{fig:p_g_modes_energy_criterio_4families} displays the four mode families for the reference CCSN model. The first f+p-PNS mode that appears around $\sim 1000$ Hz is the f-PNS, while the following one is the p$_1$-PNS mode. Here, we cannot identify higher order p-PNS modes, as they are highly interacting with the p-shock modes. The f-shock mode has the lowest frequency of the f+p-shock family. The modes labeled here as core g-modes agree with the modes identified to originate from the core domain in Fig.~\ref{fig:freq_comparison_region1} of Sec.~\ref{subsec:BuoyancyRegions}. Similarly, the g-surface modes correspond to the ones of Fig.~\ref{fig:freq_comparison_region2}.

In addition, we have tested how sensitive our results are to the definition of the PNS surface. In addition to the velocity-based criterion of section~\ref{subsec:BuoyancyRegions}, we have tested a criterion based on the value of the density, consisting in defining the surface of the star at a fraction of the central density $\rho_c = \rho(r=0)$. We find that $0.01\%$ of the central density is able to locate the PNS surface properly. Since both criteria perform similarly we present only the results obtained using the velocity criterion.

To test the abilities of the automatic classification, we apply it to the eigenmodes computed for the full set of $28$ CCSN simulations used in this work (cf.~Section~\ref{sec:CCSNeSimulations}). Given that the procedure allows individual modes to be labeled, each mode can be unambiguously identified across all simulations, enabling a direct comparison of their frequencies. For each mode of each simulation we end up with a series of mode frequencies as a function of time that typically contains a few misclassified modes. This arises from the difficulty of classifying modes properly near the avoided crossings (especially the strong ones) where the value of $F$ is $\sim 0.5$. To eliminate those misclassified outliers, we apply a filter, as described in Appendix~\ref{app:Outliers}.

In Fig.~\ref{fig:modes_all_sims} we present such a comparison for the first modes of each of the families. The color scale represents the different EoS, from the softest (SFHo) to the stiffest (GShen-NL3), while the symbols indicate the dimensionality of the simulation and the two codes, CoCoNuT and Aenus-ALCAR.
For all models, the frequency of the f-PNS mode grows with time, from $\sim 500$~Hz at $\sim 0.2$~s up to $1000$-$1500$~Hz at $1.5$~s post-bounce. This range is similar to that of the GW signal HFF reported in the simulations of~\cite{Cerda-Duran2025}. Variability can be attributed to the EoS, the progenitor structure and the code (different gravity and neutrino treatment).
The f-shock mode manifests as the most fluctuating one, as it depends on the shock location. The shock may undergo oscillations, or in some simulations not be completely stalled and keep expanding at a lower rate. Moreover, in the case of exploding models, the f-shock mode frequency will decrease. The g$_1$-core mode appears at about $800$-$1000$~Hz, at $0.2$~s and, for most models, decreases monotonically as the value of $\mathcal{N}^2$ in the core stable region decreases. In cases where this region disappears ($\mathcal{N}^2$ becomes negative) the mode frequency approaches  zero.
Lastly, the g$_1$-surface mode appears at about $100$-$200$~Hz at $0.2$~s and, for most models, increases monotonically with time. In this case, even after filtering, there is a small cluster of misclassified modes above the main branch at $\sim 0.25$~s. The reason is that the classification becomes less reliable in the vicinity of the strong avoided crossing between the first core g-mode and the PNS f-mode.

\begin{figure*}
    \centering

    \includegraphics[width=0.47\textwidth]{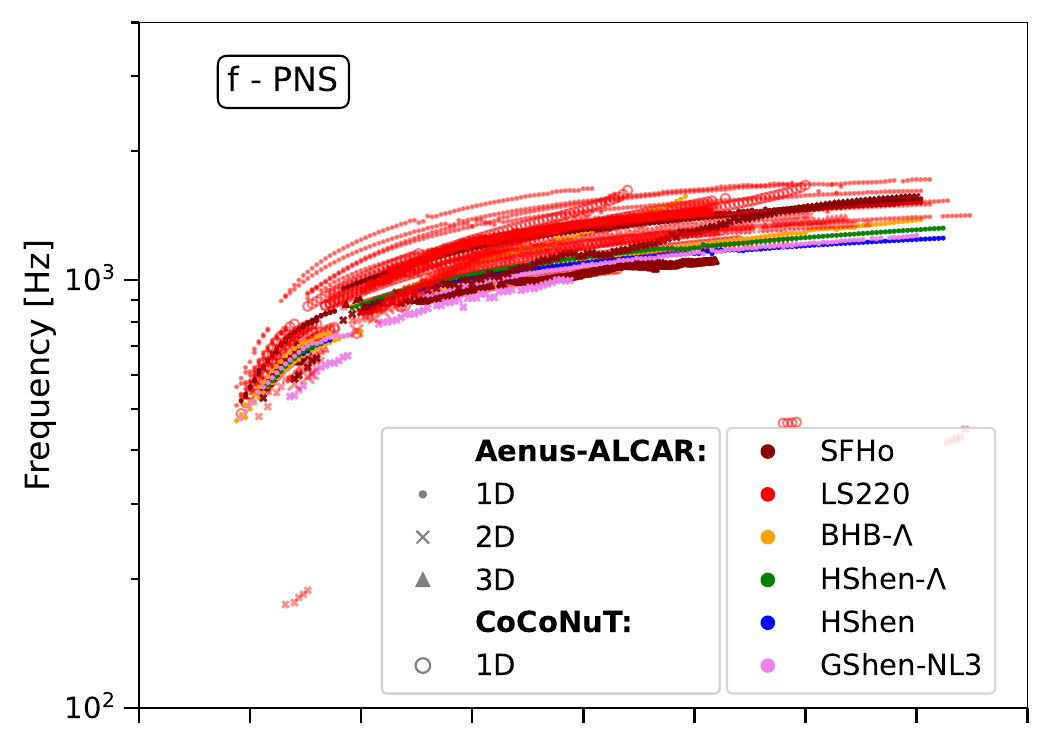}
    \hfill
    \includegraphics[width=0.48\textwidth]{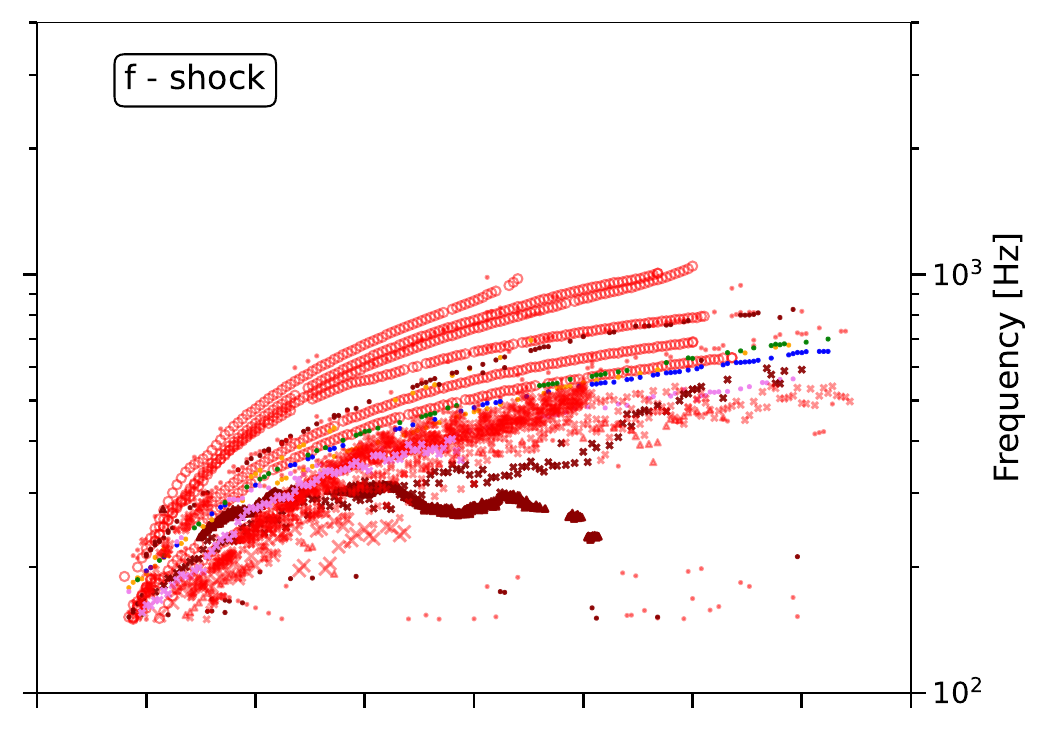}

    \vspace{0.2cm}

    \includegraphics[width=0.49\textwidth]{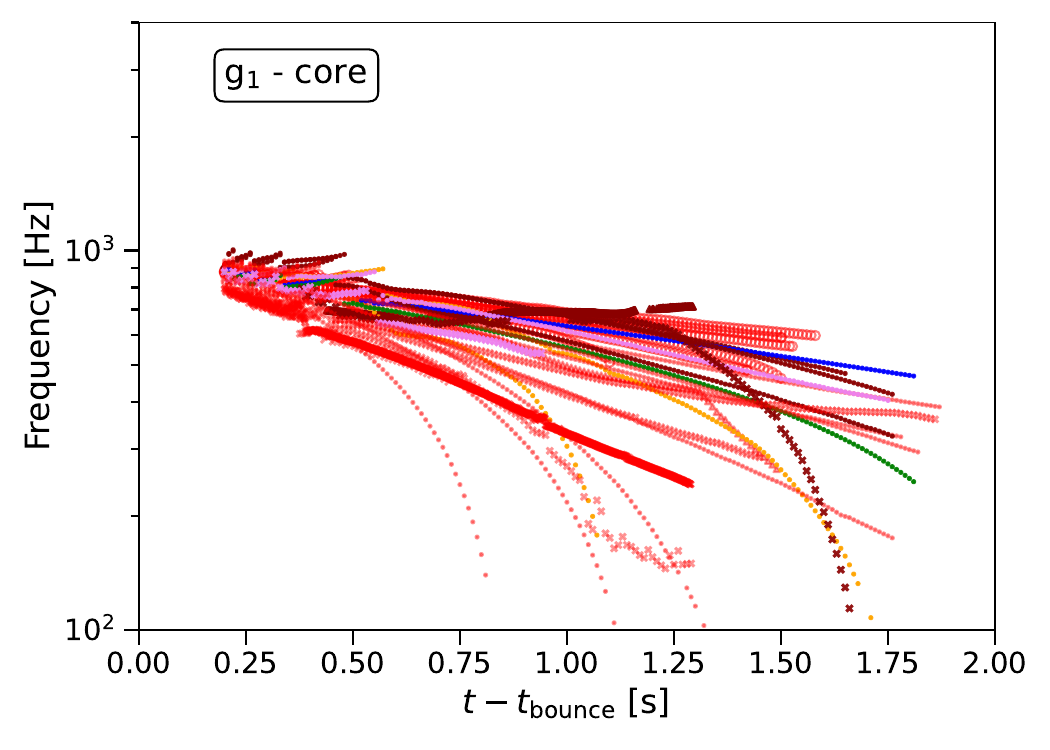}
    \hfill
    \includegraphics[width=0.49\textwidth]{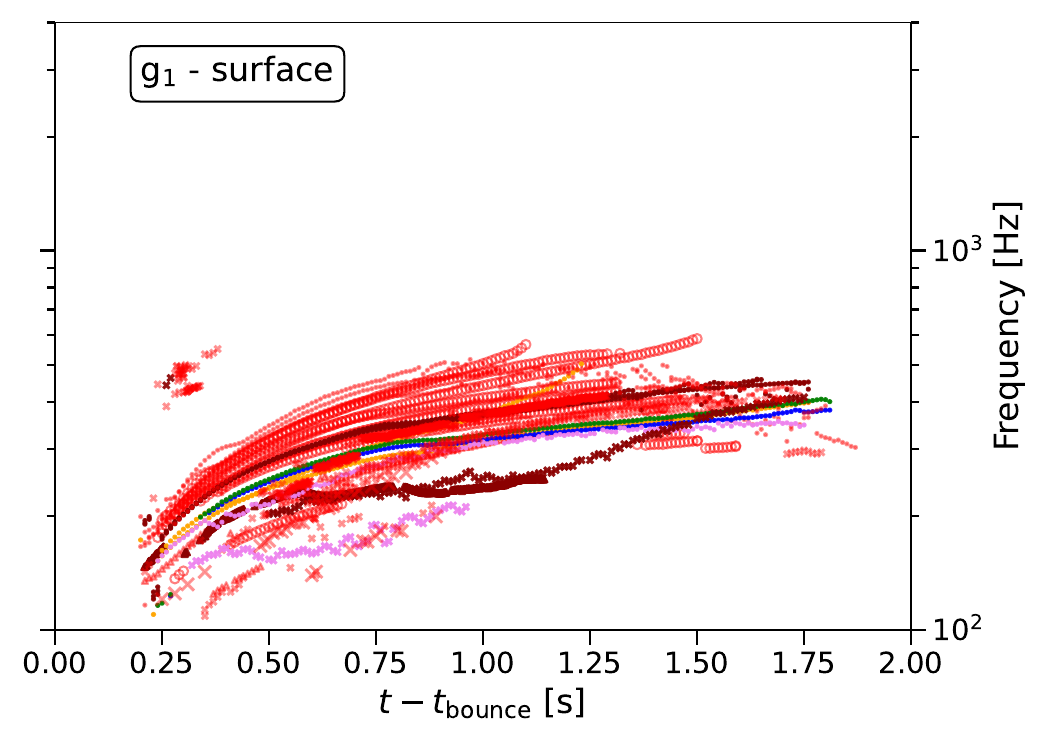}

    \caption{From left to right: upper plots: f-PNS and f-shock, bottom plots: g$_1$-core and g$_1$-surface for all the models of Table \ref{tab:simulations}. The color represents the EoS, while the different symbols depict the dimensionality; $\cdot$ : 1D, x : 2D, $\blacktriangle$ : 3D, and the code; $\bullet$ : Aenus-ALCAR, $\circ$ : CoCoNuT.
    }
    \label{fig:modes_all_sims}
\end{figure*}

Our suite of CCSN simulations contains one model that explodes (namely model 2D s25 LS220; illustrated in 
Fig.~\ref{fig:shockposition} by the rapid increase in the shock radius).
For this model, Fig.~\ref{fig:modes_all_sims} only shows the mode frequencies up to the time at which the shock starts expanding rapidly. The reason is that our analysis is not suitable if the background changes in a timescale comparable or faster than the mode period. However, applying our method despite the limitations still yields reasonable results for the g-modes and the PNS f+p-modes. This is expected because those modes do not depend significantly on the shock location. Nevertheless, since the results are based on one single model, we prefer to be cautious about the capabilities of our procedure when an expanding shock is present. The main difficulty we foresee is that, when using the full domain, the frequency of the shock f+p-modes will decrease drastically and the domain will be filled with high-order shock p-modes, which will make the classification more difficult. An alternative approach would be to perform the analysis using the PNS domain or the extended PNS domain for exploding models, since those consider only the PNS and its surroundings. However, in some of the simulations of non-exploding models, this choice leads to misclassifications of the f-PNS and the g$_1$-surface modes (see following section). Last, one could also take into account a combination of two different boundaries, e.g., the shock and the extended PNS domain. Investigating those possible solutions is out of the scope of this work and will be reported elsewhere, considering more exploding models.

%%%%%%%%%%%%%%%%%%%%%%%%%%%%%%%%%%%%%%%%%%%%%%%%%%%%
\subsection{Surface vs shock boundary conditions}
\label{subsec:BCs_surface_vs_shock}
%%%%%%%%%%%%%%%%%%%%%%%%%%%%%%%%%%%%%%%%%%%%%%%%%%%%

\begin{figure}[t]
    \centering
    \includegraphics[width=0.47\textwidth]{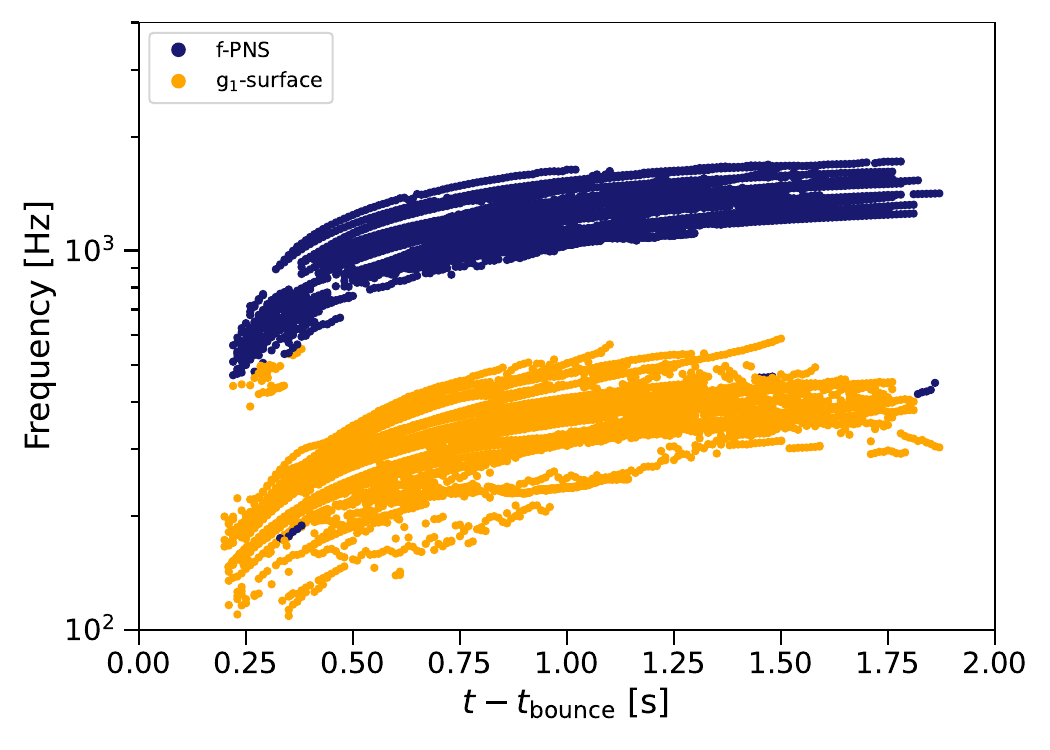}
    \vspace{0.2cm}
    \includegraphics[width=0.47\textwidth]{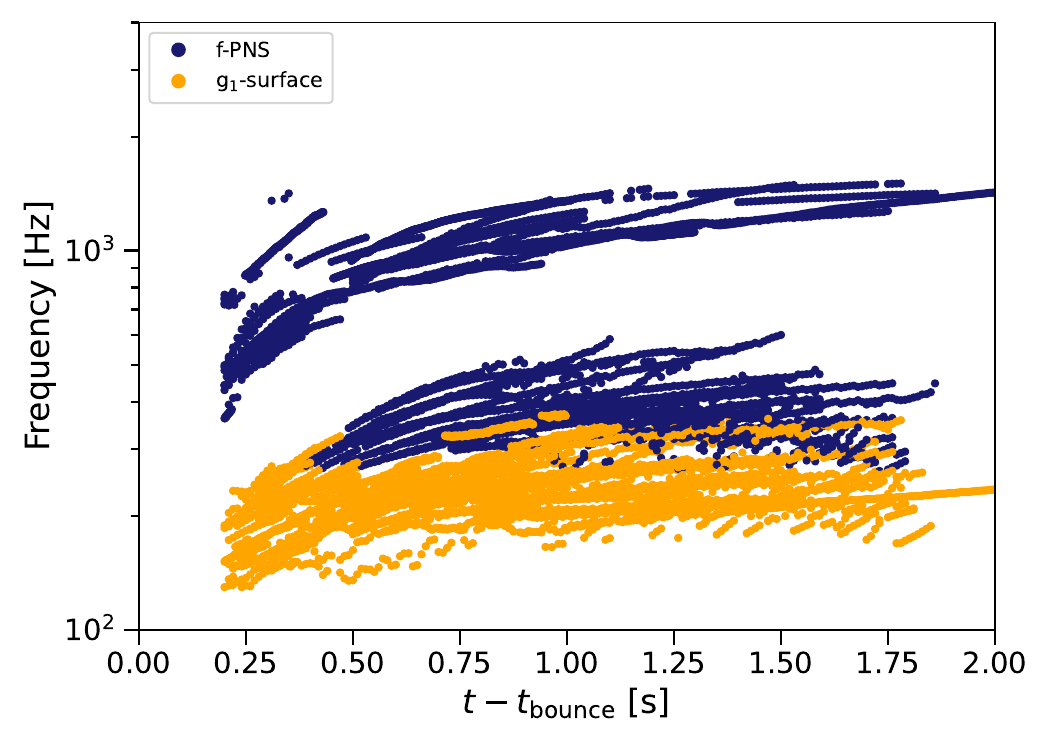}
    \caption{The f-mode of the PNS and the $\rm g_1$-surface for all the simulations have been obtained imposing BC at the shock location (upper plot) and at the surface of the PNS (lower graph).}
    \label{fig:comparison_f_PNS_g_surf}
\end{figure}

Before closing this section, it is worth pointing out the importance of the location of the outer boundary for a correct classification of the various modes of the system. 
Fig.~\ref{fig:comparison_f_PNS_g_surf} displays the time evolution of the frequencies of the f-PNS and the g$_1$-surface modes
for all models. The top plot shows the frequencies computed for an outer BC imposed at the shock location (full domain), while the bottom one places the BC at the surface of the PNS (PNS domain). It is evident that when the BC is set at the shock, those two modes are classified properly, as they are grouped in two distinct areas for all the simulations. However, when the outer BC is the surface of the star, many g$_1$-surface modes are labeled as f-PNS. Since the first mode of the family of the f+p-modes has been misclassified, then the rest of them will be too, leading to the lack of f-PNS modes in the lower plot. Notice that in the bottom graph, the modes change nature without the appearance of avoided crossings. This is further evidence that this change is artificial and not physical, as it does not stem from any interaction between the modes.

%%%%%%%%%%%%%%%%%%%%%%%%%%%%%%%%%%%%%%%
%%%%%%%%%%%%%%%%%%%%%%%%%%%%%%%%%%%%%%%
\section{Conclusions}
\label{sec:conclusions}
%%%%%%%%%%%%%%%%%%%%%%%%%%%%%%%%%%%%%%%
We have presented a novel scheme to automatically classify the spectrum of oscillation modes of a newly born PNS surrounded by a stalled accretion shock. To do so, we have made use of a comprehensive suite of CCSN simulations, showing that our procedure yields accurate mode classification across diverse CCSN models, involving different numerical codes, progenitors, equations of state, and treatments of gravity and neutrino physics. We have investigated the nature of the modes by considering the different regions of the CCSN where they originate. Our study has revealed different families of f-, p- and g-modes in the system, stemming from different areas of the PNS-shock system. The accurate characterization and analysis of these modes is essential for GW asteroseismology of PNS.

Our classification procedure is physically motivated, as it is based on the energy of the restoring forces of specific modes. Moreover, our new approach removes some of the classification uncertainties present in previous studies. Specifically, we have found four mode families: f+p-PNS modes, originating in the PNS interior, f+p-shock modes, arising from the cavity enclosed by the PNS surface and the shock, and g-modes. The latter are further classified as g-surface modes and g-core modes, stemming, respectively, from either the stably stratified region at the PNS surface or the stable PNS core region. The existence of these four families has been further confirmed through simple analytical estimates (one for each family) able to describe the behavior of the mode frequencies when modifying the background.

%Our analysis settles the issue of the nature of the dominant GW emission mode in CCSNe (a.k.a.~the High Frequency Feature). We have shown that the HFF is the f-mode of the PNS, associated with the presence of a strong density gradient at the PNS surface. Given that matter is continuously accreting onto the PNS, this mode can also be seen as an interface mode between the PNS and the post-shock region. Our study also clarifies the nature of the frequency break in the GW power spectral density (or power gap) at about $0.4$~s post-bounce, observed in a number of numerical simulations (e.g.~\cite{Morozova:2018, Bruel2023, Vartanyan2023, EggenbergerAndersen2021,Zha2024}). The interpretation of this gap as an avoided crossing in~\cite{Torres_et_al_I,Morozova:2018, Sotani2020} is consistent with the results of this work, which highlights a prominent avoided crossing between the PNS-f mode and the first core g-mode. We note that core g-modes have been also reported in~\cite{Torres_et_al_II,Jakobus2023, Jakobus2024} and our identification as a core g-mode of one of the two modes involved in the avoided crossing  coincides with the analysis of \cite{Jakobus2024}.

Our analysis settles the issue of the nature of the dominant GW emission mode in CCSNe (a.k.a.~the High Frequency Feature). We have shown that the HFF is the f-mode of the PNS, associated with the presence of a strong density gradient at the PNS surface. Given that matter is continuously accreting onto the PNS, this mode can also be seen as an interface mode between the PNS and the post-shock region. Our analysis supports previous results identifying an important $f$-mode contribution to the dominant GW emission in CCSNe 
\cite{Morozova:2018,Sotani2019,Sotani2020,Vartanyan2023,Murphy2025}. Within our energy-based classification, the HFF is identified with the PNS $f$-mode, since it has $f+p$ rather than buoyancy character and is the lowest-frequency member of the PNS $f+p$ family, with its energy distributed throughout the PNS. This differs from classifications based on nodal structure such as the GCN scheme adopted in \citet{Murphy2025}, in which the mode tracking the HFF is labeled successively as a g-mode and then  an f-mode. We interpret this difference as arising from the distinct criteria used to classify the modes: our identification is based on their energetic/restoring-force character rather than on their nodal properties.
Our study also clarifies the nature of the frequency break in the GW power spectral density (or power gap) at about $0.4$~s post-bounce, observed in a number of numerical simulations (e.g.~\cite{Morozova:2018, Bruel2023, Vartanyan2023, EggenbergerAndersen2021,Zha2024}). The interpretation of this gap as an avoided crossing in~\cite{Torres_et_al_I,Morozova:2018, Sotani2020} is consistent with the results of this work, which highlights a prominent avoided crossing between the PNS-f mode and the first core g-mode. We note that core g-modes have been also reported in~\cite{Torres_et_al_II,Jakobus2023, Jakobus2024} and our identification as a core g-mode of one of the two modes involved in the avoided crossing  coincides with the analysis of \cite{Jakobus2024}.

It is worth pointing out that the identification of oscillation modes in isolated (cold) neutron stars is far simpler than in the case of PNS following a CCSN event. In the former case the low values of the BV frequency keep g-modes strictly below the f-mode. This allows to easily identify the nodeless f-mode and order the rest by increasing frequency. CCSNe are far more complex. Intense neutrino activity creates strong entropy and lepton gradients, boosting the BV frequency above the Lamb frequency. As a result, g-modes often exceed the frequencies of low-order f- and p-modes. Additionally, the system's structure introduces multiple mode-generating regions: two zones for g-modes, two interfaces (shock and surface) for f-modes, and two acoustic cavities for p-modes. This structural complexity creates a dense, overlapping spectrum that had hindered, up to this study, clear mode classification.

Simple classification methods based on node counting  (Cowling classification) (see e.g.~\cite{1983_Cox,Torres_et_al_I, Morozova:2018,Sotani2020,Zha2024})  fail to follow the modes across avoided crossings where the mode character changes. There have been some efforts to solve this issue by applying generalizations of the Cowling criterion~\cite{Torres_et_al_II, Rodriguez2023}. While those produce  cleaner mode identification they do not solve the problem of the avoided crossings nor provide a framework to handle four coexisting families of modes. A partial solution is to use methods based on the tracking of eigenfunctions across avoided crossings~\cite{Torres_et_al_II}; however,  this approach is not reliable enough to be used in an automated way. As an alternative, \cite{Rodriguez2023} presented a method based on the location of the kinetic energy of the mode in the star, that allows one to separate the f-PNS and the core g-mode, yielding a correct classification across this particular avoided crossing. However, the extension of this approach to classify all families of modes remains uncertain.

One of the most important results from this work has been to highlight the importance of taking the full domain of the system into account, i.e.~up to the location of the shock, to characterize the complete mode spectrum. Previous studies have used the PNS surface as the outer boundary for the eigenvalue calculations \cite{Morozova:2018, Westernacher-Schneider2019, Westernacher-Schneider2020, Sotani2020, Wolfe2023}. This allows for a clean identification of the f-PNS mode, without the contamination of shock f+p-modes that may coincide in frequency after the onset of the explosion. Moreover, the frequency of this particular mode is well captured even if the domain is limited to the PNS interior, as shown by~\cite{Sotani2019} and corroborated by our results. However, using such a restricted domain leads to the misclassification of the surface g-mode, which partially lives outside the PNS surface. In addition, the p-modes originating in the post-shock region and the shock f-mode are, by construction, impossible to identify when the full domain of the PNS-shock system is not considered.

All CCSN models used in this investigation have revealed two regions stable against convection, where the BV frequency is positive, separated by an unstable layer $(\mathcal{N}^2<0)$. The two families of g-modes we have identified are associated with each of those areas. Such convectively stable regions have been reported in previous 2D and 3D CCSN simulations~\cite{Pan2019,EggenbergerAndersen2021,Mezzacappa2023,Jakobus2024}. 
Jakobus et al.~\cite{Jakobus2024} investigated how entropy and electron fraction (\(Y_{e}\)) gradients contribute to \(\mathcal{N}^{2}\), reporting an inner stable region and a stratified PNS surface. They showed that the entropy gradient dominates \(\mathcal{N}^{2}\), while the \(Y_{e}\) gradient exerts a stabilizing effect. Similarly, Mezzacappa et al. \cite{Mezzacappa2023} identified two stable regions separated by a Ledoux convection zone. Despite a sign difference in our definitions of \(\mathcal{N}^{2}\), our qualitative findings align with those of~\cite{Mezzacappa2023}. Eggenberger \& Andersen~\cite{EggenbergerAndersen2021} found two positive \(\mathcal{N}^{2}\) regions; however, instead of a negative \(\mathcal{N}^{2}\) layer, the intermediate unstable convective zone is identified via the turbulent energy flux. Additionally, Pan et al.~\cite{Pan2019} showed a positive \(\mathcal{N}^{2}\) across the entire domain, with only a faint hint of separation between the two regions.
These discrepancies stem from differing definitions of \(\mathcal{N}^{2}\), neutrino/gravity treatments, numerical codes, progenitors, and EoS. To our knowledge, a systematic analysis of how these two stable regions form is still lacking. While this formation process can heavily influence the existing mode families, such an investigation remains outside the scope of this work.

In the near future, the automatic classification framework developed in this investigation will enable systematic studies of PNS mode dependencies on key physical parameters, such as the progenitor star and EoS. We will focus on modes prominently observed in simulated GW spectrograms, specifically the f-PNS, g-core, and SASI modes. For the latter, we will adopt the formalism from~\cite{Tseneklidou:2025} to account for the quasi-stationary accretion flow in the post-shock region, which drives the associated p-modes and the shock f-mode unstable~\cite{Foglizzo2002}. Ultimately, this line of work aims to advance quasi-universal relations between mode frequencies and PNS properties by extending current frameworks~\cite{Sotani2017,Torres_et_al_letter,Sotani2019}. These relations are essential for interpreting future GW observations, probing the CCSN explosion engine, and constraining the physical parameters of the system.

\section*{Acknowledgements}
We are grateful to Prof. Bernhard Müller for fruitful discussions. This work is supported by the Spanish Agencia Estatal de Investigación (grants PID2024-159689NB-C21, PID2021-127495NB-I00 and PID2025-171322NB-C22) funded by MICIU/AEI/10.13039/501100011033 and by FEDER/EU and by the European Union “NextGenerationEU", and by the Generalitat Valenciana (Prometeo grants CIPROM/2022/49 and CIPROM/2022/13).  We also acknowledge support from the Astrophysics and High Energy Physics programme of the Generalitat Valenciana ASFAE/2022/026 funded by MCIN and the European Union NextGenerationEU (PRTR-C17.I1). The CCSN simulations have been performed on the server Lluisvives of the Servei d'Informàtica de la Universitat de València and on the Red Española de Supercomputación (RES) on MareNostrum (grants AECT-2024-2-0038 and AECT-2024-3-0014).

\appendix

\section{Lagrangian displacement}\label{app:LagrangianDispl}
%%%%%%%%%%%%%%%%%%%%%%%%%%%%%%%%%%%%%%%%%%%%%%%%%%%%%%%%%%%%%
The Lagrangian displacement in the orthonormal basis of the flat 3-metric, $\hat{e}_i=\{\partial_r, \frac{1}{r}\partial_{\theta}, \frac{1}{r\sin{\theta}}\partial_{\phi} \}$, is given by Eq. \eqref{eq:xi_vector_harmonics}, where the radial and transverse vector spherical harmonics with polar parity are, respectively,
\begin{align}
    \vec{Y}^{R,lm} &= \vec{n} \, Y^{lm}   \label{eq:YR} \\
    \vec{Y}^{E,lm} &= \left[ l(l+1) \right]^{-1/2}r \, \vec{\nabla}Y^{lm} ,  \label{eq:YE}
\end{align}
with $\vec{n}$ being the unit radial vector and $\vec{\nabla}$ the gradient operator with respect to the flat metric. For further information on the (vector) spherical harmonics and their properties the interested reader is referred to \cite{Thorne:1980}. In spherical coordinates, the unit radial vector is $\vec n = \hat e_r$, and thus the displacement operator can be expressed as
\begin{align}\label{eq:xi_orthonormalBasis}
    \vec{\xi} = &\eta_{{}_R} Y^{lm} \hat{e}_r  \nonumber \\
    &+ \frac{\eta_{{}_E}}{\sqrt{l(l+1)}}\left( \partial_{\theta}Y^{lm}\hat{e}_{\theta} + \frac{1}{\sin{\theta}}\partial_{\phi}Y^{lm}\hat{e}_{\phi}\right).
\end{align}
However, Eqs.~\eqref{eq:xi_r}-\eqref{eq:xi_phi} are expressed in the coordinate basis $\vec e_i=\{\partial_r, \partial_{\theta}, \partial_{\phi}\}$. The orthonormal basis is linked to the coordinate one, according to the following relations,
\begin{align}
    \hat{e}_r &= \vec e_r \,, \label{eq:e_r}  \\
    \hat{e}_{\theta} &= \frac{1}{r}\, \vec e_{\theta} \,,  \label{eq:e_theta}  \\
    \hat{e}_{\phi} &= \frac{1}{r\sin{\theta}} \vec e_{\phi}  \label{eq:e_phi} \, . 
\end{align}
Taking into account the above transformations we can now compare Eq.~\eqref{eq:xi_orthonormalBasis} to \eqref{eq:xi_r}-\eqref{eq:xi_phi}. By doing so, we find the following relations among the radial functions,
\begin{align}
    \eta_{{}_R} &= \eta_{{}_1}, \label{eq:etas_radial}  \\
    \eta_{{}_E} &= \sqrt{l(l+1)}\, \eta_{{}_2}  . \label{eq:etas_ang}
\end{align}

\section{Calculation of the perturbation of the relativistic energy density}\label{app:CalcKineticEnergy}
%%%%%%%%%%%%%%%%%%%%%%%%%%%%%%%%%%%%%%%%%%%%%%%%
This appendix presents the detailed derivation of the perturbations of the total energy.
For aesthetic reasons, we shall denote as $q_0$ any scalar quantity in the background, $q_1$ its first order Eulerian perturbation and $q_2$ its second order perturbation. Wherever there is need to distinguish between the Eulerian and the Lagrangian perturbations, the letters $\delta$ and $\Delta$ will be used, respectively. We point out that we consider second-order perturbations only for the total energy calculation, while the system of equations being solved to obtain the frequencies is linearized. For most of the appendix we neglect metric perturbations (Cowling approximation) and only at the end we evaluate the impact of including those.

The first step is to calculate the perturbation of the relativistic energy density from Eq.~\eqref{eq:total_E}. Keeping up perturbative terms up to second order it follows that,
\begin{align}\label{eq:dE}
    E_0 &+ E_1 + E_2 = \nonumber \\
    &\bigg[ (\rho h)_0 + (\rho h)_1 + (\rho h)_2
    \bigg]\left( W_0 + W_1 + W_2 \right)^2 \nonumber \\
    &- \left( p_0 + p_1 + p_2 \right),
\end{align}
where for a background solution in hydrostatic equilibrium, as the velocity is zero, the Lorentz factor, $W_0=1$, and, since $W_1=0$, it has only second order perturbations
\begin{equation}\label{eq:W_2}
    W_2 = \frac{\upsilon_1^2}{2}  = \frac{\alpha^{-2}}{2}\upsilon_1^{*2},
\end{equation}
where $\upsilon_1^2=g_{ij}\upsilon_1^i\upsilon_1^j$.
The first order perturbation of $\rho h$ can be written as
\begin{equation}\label{eq:drhoh}
    \left( \rho h \right)_1 = \left( 1 + \frac{1}{c_s^2} \right)p_1 - \left(\rho h\right)_0 \xi^i \mathcal{B}_{0i},
\end{equation}
as shown in \cite{Torres_et_al_I}. Taking into account the definitions of enthalpy and energy density, its second-order perturbation will be given by
\begin{equation}\label{eg:rhoh2}
    (\rho h)_2 = e_2 + p_2.
\end{equation}

Substituting the above relations in the perturbation of the relativistic energy density, Eq.~\eqref{eq:total_E}, the 0th, 1st and 2nd perturbative order components can be expressed, respectively, as
\begin{align}
    E_0 &= \left( \rho h\right)_0 - p_0 \,,  \label{eq:E0} \\
    E_1 &= \left( \rho h\right)_1 - p_1  \,, \label{eq:E1} \\
    E_2 &= \left( \rho h\right)_0\upsilon_1^2 +  e_2 \,. \label{eq:E2}
\end{align}
The next step is to derive the second order perturbations of the energy density, $e_2$. From the definition of the energy density, $e\equiv \rho(1+\varepsilon)$, the second order perturbation will be
\begin{equation}\label{eq:De2_generic}
    \Delta e_2 = \Delta \rho_2 + \rho \Delta \epsilon_2 + \epsilon \Delta \rho_2 + \Delta \rho_1 \Delta \epsilon_1  .
\end{equation}
In order to find relations among the fluid variables, we need to re-examine the first law of thermodynamics and derive relations for second order variations. In the adiabatic case, i.e.~assuming there is no energy exchange ($s=$const.) or composition change ($Y_e=$const.), the first law of thermodynamics reads
\begin{equation}\label{eq:1stLaw_generic}
    \Delta U = - p \Delta V,
\end{equation}
where $U$ is the internal energy and $V$ the volume. The baryon number density, $n$, is related to $N$, the number of particles, according to $n = \frac{N}{V}$. If we perturb the latter, for a constant number of particles, then we find after some calculations that the volume variation is related to the variation of the baryon number density according to
\begin{align}\label{eq:DV2}
     \Delta V = -\frac{N\Delta n}{n^2}\left( 1 -\frac{\Delta n}{n} \right),
\end{align}
where third order or higher perturbative terms are omitted. The internal energy is related to the volume element through $\rho \epsilon = U/V \xrightarrow[]{\rho=mn} U = mN\epsilon$. The variation of this last relation will be $\Delta U = mN\Delta \epsilon$ and therefore Eq. \eqref{eq:1stLaw_generic} will be recast into
\begin{align}
    m N\Delta \epsilon = p \frac{N\Delta n}{n^2}\left( 1-\frac{\Delta n}{n} \right) ,
\end{align}
where we have also used Eq.~\eqref{eq:DV2}. Simplifying the above relation and expressing the number density $n$ in terms of the density $\rho = mn$, we arrive at the following
\begin{align}
    \Delta \epsilon = p\frac{\Delta \rho}{\rho^2} - \frac{p}{\rho^3}(\Delta \rho)^2 \,.
\end{align}
If we expand the latter equation up to second order terms, $\Delta \epsilon = \Delta\epsilon_1 +\Delta\epsilon_2$, then we get the following for the perturbations of the energy density,
\begin{align}
    \Delta \epsilon_1 &= \frac{p}{\rho^2}\Delta \rho_1  \,, \label{eq:Deps1}  \\
    \Delta \epsilon_2 &= \frac{p}{\rho^2}\Delta \rho_2 - \frac{p}{\rho^3}(\Delta \rho_1)^2  \,. \label{eq:Deps2}
\end{align}
If we substitute Eqs.~\eqref{eq:Deps1} and \eqref{eq:Deps2} into Eq.~\eqref{eq:De2_generic}, then we get the following relation for the second order Lagrangian perturbations of the energy density and the rest-mass density,
\begin{equation}\label{eq:De2_Drho2}
    \Delta e_2 = h \Delta \rho_2  .
\end{equation}

At this point, let us define the relation between the Lagrangian and Eulerian perturbations up to second order terms. The Eulerian perturbation of any scalar quantity, e.g. $e$, is given by 
\begin{equation}\label{eq:Eulerian2}
    \delta e(\tilde{r}) = e(\tilde{r}) - e_0(\tilde{r}) = e_1 + e_2 \,,
\end{equation}
and the Lagrangian perturbation up to second order perturbative terms as,
\begin{align}\label{eq:Lagrangian2}
    \Delta e(\tilde{r}) &= e(\tilde{r} + \vec{\xi}) - e_0(\tilde{r}) \rightarrow \nonumber \\
    \Delta e(\tilde{r}) &= e(\tilde{r}) + {\xi^i}\nabla_i e(\tilde{r}) + \frac{1}{2}\xi^i\xi^j\nabla_i\nabla_j e_0 - e_0(\tilde{r}),
\end{align}
where $\nabla_i$ is defined with respect to the 3-metric $\gamma_{ij}$.
Considering \eqref{eq:Eulerian2}, the Lagrangian perturbations of first and second order are, respectively,
\begin{align}
    \Delta e_1 &= e_1 + {\xi^i}\nabla_i e_0 \,,  \label{eq:De1} \\ %= e_1 + \xi^i\partial_i e_0 
    \Delta e_2 &= e_2 + {\xi^i}\nabla_i e_1 + \frac{1}{2}\xi^i\xi^j\nabla_i\nabla_j e_0  \,, \label{eq:De2}
\end{align}
where for a scalar quantity it holds that ${\xi^i}\nabla_i e_0 = \xi^i\partial_i e_0$. The above relations can actually be used for any scalar quantity and, since they are written in a generic form, they can also be used for vector quantities.
Now, Eq.~\eqref{eq:De2_Drho2} can be expressed in terms of the Eulerian perturbations by means of Eq.~\eqref{eq:De2}, 
\begin{align}\label{eq:delta_e2}
    &\delta e_2 +\vec{\xi}\cdot\vec{\nabla} \delta e_1 + \frac{1}{2}\xi^i\xi^j\partial_i\partial_j e_0 =  \nonumber \\
    &h\left( \delta\rho_2 +\vec{\xi}\cdot \vec\nabla \delta \rho_1 + \frac{1}{2}\xi^i\xi^j\partial_i\partial_j \rho_0 \right).
\end{align}
In Sec.~\eqref{subsec:StandingWave}, we showed that in order to have standing wave solutions, the Lagrangian displacement, as well as the other first order perturbations of scalar quantities, are purely real and their time dependence is $\cos{\sigma t}$. However, we do not know the time dependence of the second order perturbations. Eq.~\eqref{eq:E2} contains the kinetic term of $\upsilon_1^2$, which because of Eq.~\eqref{eq:TimeDependence}, has a phase defined by $\sin^2{\sigma t}$. Since the kinetic and potential energy are out of phase, then the potential terms pulsate according to $\cos^2{\sigma t}$. Following this logic, it is easily proven that Eq.~\eqref{eq:delta_e2} can be written as
\begin{equation}\label{eq:delta_e2_only_kinetic}
    \delta e_2 = h \delta \rho_2 \: + \: \text{potential terms},
\end{equation}
and consequently, the relativistic energy density, Eq. \eqref{eq:E2}, results in
\begin{equation}\label{eq:E2_pot_terms}
    E_2 = \left( \rho h\right)_0\upsilon_1^2 + h \delta \rho_2 \: + \: \text{potential terms}.
\end{equation}

Before continuing to the calculation of the term $\delta\rho_2$, let us introduce the connection of the perturbed advective velocity to the Lagrangian displacement, that we have already implicitly used for the above result.
The Lagrangian perturbation of the velocity is linked to the displacement vector \cite{Tseneklidou:2025} according to,
\begin{equation}\label{eq:Du_dxi}
    \Delta \upsilon^{*i} = \frac{d\xi^i}{dt},
\end{equation}
where $\frac{d\xi^i}{dt} = \partial_t \xi^i + \upsilon^{*j}\nabla_j\xi^i$ and $\Delta  \upsilon^{*i} = \delta  \upsilon^{*i} + \xi^{j} \nabla_j  \upsilon^{*i}$. Expanding the above relation keeping up to non-zero second order perturbative terms and using Eq.~\eqref{eq:De2}, we can get the Eulerian perturbations of first and second order, respectively,
\begin{align}
    \upsilon^{*i}_1  &= \partial_t \xi^i  \label{eq:dv1} \\
    \upsilon^{*i}_2  &= \upsilon^{*j}_1 \nabla_j \xi^i - \xi^j\nabla_j \upsilon^{*i}
    \overset{\eqref{eq:dv1}}{=} \partial_t \xi^j\nabla_j \xi^i  - \xi^j\nabla_j \partial_t \xi^i . \label{eq:dv2}
\end{align}

Our last challenge to the path of the energy calculation is the derivation of the second-order perturbation of the density. For that purpose, we will turn to the continuity equation and perturb it keeping up to second-order terms. Recall that the continuity equation is given by
\begin{equation}\label{eq:continuity}
    \frac{1}{\sqrt{\gamma}}\partial_t [\sqrt{\gamma} D] + \nabla_i[  D\upsilon^{*i} ] = 0,
\end{equation}
where $D=\rho W$. For further details on the derivation of the continuity equation, the interested reader is referred to \cite{Torres_et_al_II} and references therein.
The components of the continuity equation including first and second-order perturbations are, respectively, 
\begin{align}
    &\partial_t \rho_1 + \nabla_i \left( \rho_0 \upsilon_1^{*i} \right) =0  \rightarrow \nonumber \\
    &\rho_1 = -\frac{1}{\sqrt{\gamma}}\nabla_i \left( \rho_0 \xi^i \right), \label{eq:contin1} \\
    &\partial_t \rho_2 = -\psi^4\rho\frac{\alpha^{-2}}{2}\partial_t \dot{\xi}^2 - \nabla_i\left[ \rho_0 \upsilon_2^{*i} + \rho_1 \upsilon_1^{*i}  \right] \label{eq:contin2}.
\end{align}

Taking into account Eqs.~\eqref{eq:dv2} and \eqref{eq:contin1}, and after some calculations, the latter second-order equation becomes
\begin{align}\label{eq:contin2_subs}
    \partial_t \rho_2 = &-\rho\frac{\alpha^{-2}\psi^4}{2}\partial_t \dot{\xi}^2 
    -\nabla_i \bigg[ \rho_0 \left( \partial_t \xi^j\nabla_j\xi^i \right) \nonumber \\
     &-\nabla_j\left( \rho_0 \xi^j \partial_t\xi^i \right)  \bigg],
\end{align}
where the last two terms of the right-hand side can be written as
\begin{align}
    &\nabla_i \left[ \rho_0 \left( \partial_t \xi^j\nabla_j\xi^i \right) \right] = \frac{1}{2}\partial_t \left[  \partial_i \left( \rho_0 \xi^j\nabla_j\xi^i \right)\right]  \label{eq:term1_cont2} \\
    &\nabla_i \left[ \partial_j \left( \rho_0 \xi^j \partial_t \xi^i \right) \right] = \frac{1}{2}\partial_t \left\{ \nabla_i \left[ \nabla_j \left( \rho_0 \xi^i \xi^j \right) \right] \right\}  , \label{eq:term2_cont2}
\end{align}
where $\rho_0$ belongs to the static background solution and thus is independent of time.
Eq.~\eqref{eq:contin2_subs} can be integrated in time, resulting in
\begin{align}\label{eq:cont2_pot_terms}
    &\rho_2 = -\rho_0\frac{\alpha^{-2}\psi^4}{2}\dot{\xi}^2 \nonumber \\ &-\frac{1}{2}\left\{ \partial_i \left[ \rho_0 \left(  \xi^j\nabla_j\xi^i \right)
     -\nabla_j\left( \rho_0 \xi^j \xi^i \right)  \right] \right\},
\end{align}
with the last two terms being potential as they scale with $\cos^2{\sigma t}$, according to our previous analysis following Eq.~\eqref{eq:TimeDependence}. In conclusion, we can express the second-order perturbation of the density as
\begin{equation}\label{eq:cont2_final}
    \rho_2 = -\rho_0\frac{\alpha^{-2}\psi^4}{2}\dot{\xi}^2 \; + \; \text{potential terms} .
\end{equation}
Combining the latter equation with Eq. \eqref{eq:E2_pot_terms}, the 
relativistic energy density of the fluid motion is
\begin{equation}\label{eq:E2_final}
    E_2 = \frac{1}{2}\alpha^{-2}\psi^4\left( \rho h \right)_0 \dot{\xi}^2  \; + \; \text{potential terms}.
\end{equation}

Before closing this appendix, let us relax the Cowling approximation by including the perturbations of the metric up to second order. All terms including $\psi_1$ and $\alpha_1$ lead to potential terms, so one needs not pay much attention to them. Terms including $\alpha_2$ are not present because they originate from the Lorentz factor and contribute only to orders higher than quadratic. The only remaining terms are those including $\psi_2$. After straightforward but lengthy calculations we find that
\begin{equation}\label{eq:E2_dpsi2}
    E_2 = \frac{1}{2}\alpha^{-2}\psi^4\left( \rho h \right)_0 \dot{\xi}^2  + \left( \rho h \right)_0 \frac{6}{\psi_0} \psi_2 \; + \; 
    \substack{\text{potential}\\\text{terms}}.
\end{equation}

To evaluate if $\psi_2$ is a kinetic or a potential term, we use an alternative expression for the ADM mass \citep{Gourgoulhon:2012} 
\begin{equation}
    M=-\frac{1}{2\pi}\int \bar \nabla_i \bar \nabla^i \psi \,  \sqrt{f} d^3x.
\end{equation}
Therefore, it is trivial to get the second-order perturbation of the mass as 
\begin{equation}
    M_2=-\frac{1}{2\pi}\int \bar \nabla_i \bar \nabla^i \psi_2 \,  \sqrt{f} d^3x.
\end{equation}

We have already established that $M_2$ is constant; then, since $\psi_2$ admits separation of variables for the time component, it necessarily has to be constant in time. Therefore, the $\psi_2$ terms add just a constant shift in the mode energy. This term represents a non-spherical constant deformation of the background metric induced by the presence of the perturbation, and does not contribute to the potential nor the kinetic energy of the mode.

%%%%%%%%%%%%%%%%%%%%%%%%%%%%%%%%%%%%%%%%%%%%%%%%%%%%%%%%%
\section{Verification that the force is conservative}\label{app:ConservativeForce}
%%%%%%%%%%%%%%%%%%%%%%%%%%%%%%%%%%%%%%%%%%%%%%%%%%%%%%%%%
The acceleration of a fluid element, Eqs.~\eqref{eq:mom1_Drho}-\eqref{eq:mom2_Drho}, 
is a linear differential operator acting on $\xi^i$, $\delta \hat \rho$ and $\delta \hat \alpha$. However, both $\delta \hat \rho$ and $\delta \hat \alpha$ can be expressed in terms of $\xi^i$ (see~\cite{Torres_et_al_II}). Therefore, the force density, which is proportional to the acceleration, can be written in a compact form as
\begin{equation}
    f_i = \mathcal{K}_{ij} \xi^j,
\end{equation}
where $\mathcal{K}_{ij}$ is a differential-integral\footnote{Note that, according to \cite{Torres_et_al_II}, $\delta \hat \alpha$ can be written as a second spatial integral of $\xi^i$.}
operator depending on the location, $\vec r$, but not on $\xi^i$ itself.
In that case the work density can be computed from Eq.~\eqref{eq:work1} as 
\begin{equation}
w = \int f_i  \, d\xi^j = 
\mathcal{K}_{ij} \int \xi^j \, d\xi^i,
\end{equation}
If $\mathcal{K}_{ij}$ is symmetric ($\mathcal{K}_{ij} =\mathcal{K}_{ji}$) then the integral results in 
\begin{equation}
w = \frac{1}{2} \mathcal{K}_{ij} \xi^i \xi^j = \frac{1}{2} f_i \xi^{i}. 
\end{equation}
In that case, since the work only depends on the displacement, the force density, $\vec f$, can be regarded as conservative. 
However, since $\mathcal{K}_{ij}$ is a differential-integral operator in our case, it is not a trivial task to prove that it is symmetric.

Instead, we show here that the force is conservative for the harmonic time dependence of the normal modes obtained by \cite{Torres_et_al_II}. With this purpose in mind, 
let us write the displacement as $\vec \xi = \vec \xi_0 \cos \sigma t$. In that case 
\begin{align}
w &=  
\mathcal{K}_{ij} \int \xi^j \, \frac{d\xi^i}{dt} dt
= -\sigma \, \mathcal{K}_{ij} \;\xi^j_0\;\xi_0^i  \, \int  \cos\sigma t\; \sin\sigma t \;dt , \nonumber
\\ &=  \frac{1}{2} \, \mathcal{K}_{ij} \;\xi^j_0\;\xi_0^i \;\cos^2 \sigma t = \frac{1}{2} \mathcal{K}_{ij} \xi^i \xi^j = \frac{1}{2} f_i \xi^{i},
\end{align}
which gives the same result as above, i.e. that the force is conservative, at least along the path of the oscillation.

%%%%%%%%%%%%%%%%%%%%%%%%%%%%%%%%%%%%%%%%%%%%%%%%%%%%%%%%%
\section{Outliers elimination}\label{app:Outliers}
%%%%%%%%%%%%%%%%%%%%%%%%%%%%%%%%%%%%%%%%%%%%%%%%%%%%%%%%%

In order to detect misclassified modes (outliers), we use a random sample consensus (RANSAC) fitting model \cite{RANSAC}. RANSAC repeatedly fits random subsets of the data to an nth-order polynomial and looks for the fit that is supported by the largest number of points. This procedure effectively eliminates the impact of outliers in the polynomial fit and at the same time allows us to identify those outliers, which is our main interest here. We apply this approach through the Python implementation available in Scikit-learn \cite{scikit-learn} (\texttt{sklearn.linear\_model.RANSACRegressor}).

\begin{figure}
    \centering
    \includegraphics[width=0.98\linewidth]{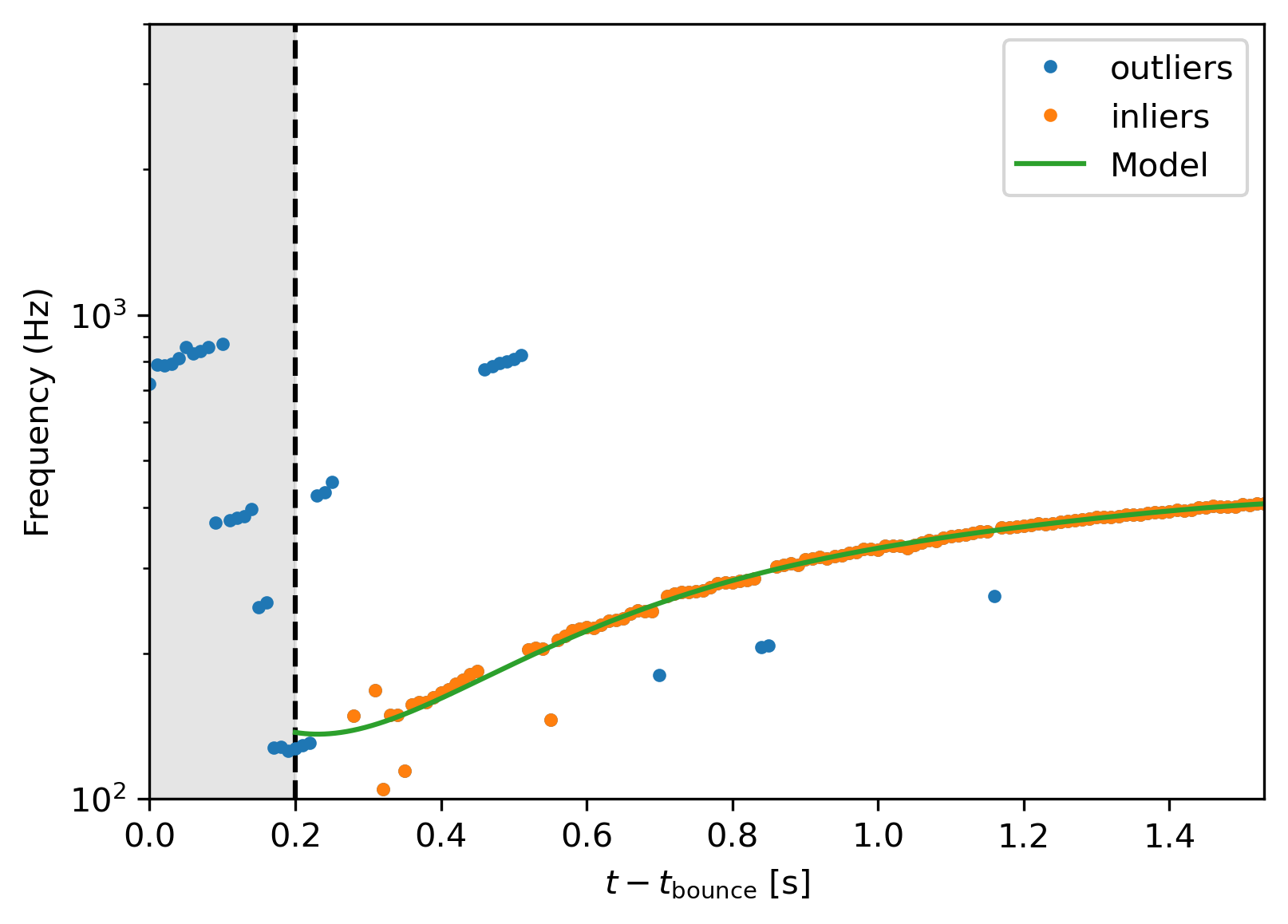}
    \caption{Outlier elimination using the RANSAC fitting model, applied to the first surface g-mode of the reference CCSN simulation. We plot the fitting model (green line), the modes marked as outliers (blue dots), and the remaining modes once outliers are disregarded (orange dots).}
    \label{fig:fit_and_clean}
\end{figure}

The threshold for outlier removal is based on the median absolute deviation (as implemented in the Python library) and we use 5th-order polynomials. Fig.~\ref{fig:fit_and_clean} shows an example of how the procedure works for the first surface g-mode of the reference CCSN model. Following the procedure laid out in the main text, we discard the modes at early times ($<0.2$~s for this model) for which the shock is still rapidly expanding. The RANSAC model is applied to the remaining data, resulting in a fitting model (green curve) that is used to identify the outliers (blue marks).  The remaining modes (orange marks) constitute the set of filtered modes used in our analysis. We note that this particular mode is shown as an example because it presents a handful of outliers, but in most of the cases (in particular for core g-modes and the PNS f-mode) outliers are either very few or even absent.

%%%% Bibliography %%%%%
\bibliographystyle{apsrev}
\bibliography{references}

\newpage

\end{document}